\documentclass{article}
\usepackage{times}
\usepackage{empheq}
\usepackage{amsmath, amssymb}
\usepackage{amsthm}
\usepackage{bbm}
\usepackage[T1]{fontenc}
\usepackage[english]{babel}
\usepackage[utf8]{inputenc}
\usepackage{hyperref}
\usepackage{mathrsfs}
\usepackage{mathabx}
\usepackage{color}
\usepackage{anysize}
\usepackage{apacite}
\usepackage{natbib}
\marginsize{2cm}{2cm}{1.5cm}{1.5cm}
\usepackage[toc,page]{appendix}
\usepackage{url}
\usepackage{bm}
\usepackage{mathtools}
\usepackage[font={small,sc}]{caption}
\usepackage{ragged2e}
\usepackage[shortlabels]{enumitem}
\usepackage{algorithm}
\usepackage{algpseudocode}
\usepackage{algorithmicx}
\usepackage{comment}    
\usepackage{lineno}
\usepackage{tensor}
\usepackage{subcaption}

\usepackage{makecell}
\usepackage{multirow}
\usepackage{array}

\algnewcommand\algorithmicinput{\textbf{Input:}}
\algnewcommand\Input{\item[\algorithmicinput]}
\algnewcommand\algorithmicdefine{\textbf{Define}}
\algnewcommand\Define{\State\algorithmicdefine \ }
\algnewcommand\algorithmicverify{\textbf{Verify}}
\algnewcommand\Verify{\State\algorithmicverify \ }

\numberwithin{equation}{section}

\newtheorem{Def}{Definition}[section]
\newtheorem{Prop}{Proposition}[section]

\AtBeginDocument{}

\DeclareMathOperator{\logit}{logit}

\newcommand{\Rd}{\mathbb{R}^{d}}

\newcommand{\BorelRd}{\mathcal{B}(\mathbb{R}^{d})}

\newcommand{\Cov}{\mathbb{C}ov}
\newcommand{\Var}{\mathbb{V}ar}

\newcommand{\R}{\mathbb{R}}

\title{The evolving categories multinomial distribution: introduction with applications to movement ecology and vote transfer}

\author
{Ricardo Carrizo Vergara$^1$, Marc Kéry$^1$, Trevor Hefley$^2$\\
     \normalsize{$^{1}$Population Biology Group, Swiss Ornithological Institute, 6204 Sempach, Switzerland.}\\
	\normalsize{$^{2}$Department of Statistics,
Kansas State University, Manhattan, KS 66506, USA.}\\
}

\date{}

\begin{document}
 
\maketitle

\maketitle

\noindent {\bf Abstract} \quad We introduce the evolving categories multinomial (ECM) distribution for multivariate count data taken over time. This distribution models the counts of individuals following iid stochastic dynamics among categories, with the number and identity of the categories also evolving over time. We specify the one-time and two-times marginal distributions of the counts and the first and second order moments. When the total number of individuals is unknown, placing a Poisson prior on it yields a new distribution (ECM-Poisson), whose main properties we also describe. Since likelihoods are intractable or impractical, we propose two estimating functions for parameter estimation: a Gaussian pseudo-likelihood and a pairwise composite likelihood. We show two application scenarios: the inference of movement parameters of animals moving continuously in space-time with irregular survey regions, and the inference of vote transfer in two-rounds elections. We give three illustrations: a simulation study with Ornstein-Uhlenbeck moving individuals, paying special attention to the autocorrelation parameter; the inference of movement and behavior parameters of lesser prairie-chickens; and the estimation of vote transfer in the 2021 Chilean presidential election.

\bigskip


\section{Introduction}
\label{Sec:Intro}

Count data are ubiquitous in statistics. Probability distributions for modeling integer-valued data have been developed since the very beginning of probability theory and are used in virtually every discipline that applies probability and statistics. One particular scenario often present in applications is when count data are taken over time. In addition, often counts refer to entities (e.g., particles, animals, people, or \textit{individuals} for using a general term), whose attributes experience a dynamics over time. In this context, a certain multivariate distribution arises as a conceptually simple model for such counts: we call it here the evolving categories multinomial (ECM) distribution. It considers individuals moving from category to category over time, with the additional feature that \textit{the categories may also evolve over time}. The categories are abstract and they may mean whatever is required in an application context where individuals have to be categorized and counted dynamically: e.g., spatial volumes for gas particles; survey landscapes for animal abundance; political preference, age group or professional activity sector of humans; seasonal consumption goods preferred by costumers; health state of agents during pandemics, to mention but a few examples. The ECM distribution is the joint distribution followed by the counts per category and time of individuals following independently the same dynamics across the evolving categories. It can be used to infer movement properties using the counts as sole information.

While the construction of this multivariate discrete distribution is simple and general, we are unaware of any work focusing on its general abstract form or mathematical properties. Technically, any application with counts of independently moving individuals has used this probabilistic structure, so in that sense the distribution is not new. Our contribution is to pay special attention to this mathematical object, giving it a name, studying its basic properties, developing methods for statistical inference and demonstrating its utility by addressing important questions in scientific fields such as ecology and sociology.

One important aspect of this abstract construction is that the exact individual ``position'' at each time is lost, and an aggregate becomes the sole available information. This key issue is present in every application where the individual information is unavailable because it is too expensive, sensitive or even physically impossible to collect. The classical example of statistical mechanics \citep{gallavotti2013statistical,pathria2017statistical} for studying broad scale properties of many-particles systems is perhaps the most common example. In the social sciences, some aggregate information about people's behavior can be available for some institutions (e.g., market and educational institutions, social media, governments) or even publicly available (e.g., vote counts in transparent democracies), but sensitive individual information is often protected. In ecology, both kinds of information are often available: animal count data from surveys \citep{seber1986review} and telemetry data of animal's positions from tracking devices \citep{hooten2017animal}. But tracking animals has inconveniences such as biased sampling or expensive tracking technology, while count data are very widely available, usually cheaper to collect, and can be obtained from citizen-science data collection initiatives \citep{sullivan2009ebird}.

In this work we study the minimal mathematical properties of this distribution so we can apply it to model important phenomena and provide fitting techniques. We aim for two application scenarios from ecology and sociology. In the ecological scenario, we infer animal movement parameters from survey data (spatio-temporal counts) taken at irregular times in continuous space-time. There is a substantial literature on the modeling of animal movement \citep{turchin1998quantitative,grimm2013individual} and of abundance \citep{andrewartha1954distribution,thorson2024spatio} separately, but the two concepts have hardly ever been united in a single framework, either in practice with integrated data scenarios or in theory by studying the statistical properties issued from the conceptual connection between movement and count. We refer to \citep{chandler2022modeling,roques2022spatial,potts2023scale,buderman2025integrated} for some recent works using, either explicitly or implicitly, the connection between movement and broad-scale count. An important and hitherto neglected aspect is the space-time autocorrelation induced by movement. Here we exploit it to infer movement parameters which do not occur on the average (intensity) field. We illustrate with a simulation study with steady-state Ornstein-Uhlenbeck moving individuals. Also, we show how the ECM distribution is useful in classification problems, where the proportions of animals following different movement behaviors must be estimated. We illustrate with an analysis of translocated lesser prairie-chickens (\textit{Tympanuchus pallidicinctus}) in the U.S. \citep{berigan2024lesser}. In the sociological scenario, we study vote transfer between election rounds, which is an \textit{ecological inference problem} \citep{schuessler1999ecological,king2004ecological}. Here, the ECM distribution offers a different approach to existing methodology commonly found in the literature. We illustrate by inferring vote transfer during the 2021 Chilean presidential election.

Our work is organized as follows. In Section \ref{Sec:ECM} we introduce the ECM distribution with its main properties: characteristic function, one-time and two-times marginal distributions, and first and second order moments. In Section \ref{Sec:ECMPoisson} we \textit{Poissonize} the ECM distribution in order to consider cases with unknown total number of individuals, describing thus a new distribution here called ECM-Poisson. Since likelihoods are intractable, in Section \ref{Sec:Inference} we propose two ad-hoc estimating functions for inference which allow the estimation of parameters occurring in the mean and the autocorrelation of the counts. The first is a Gaussian pseudo-likelihood respecting mean and covariance, justified by the central limit theorem and conceived for cases with huge number of individuals. The second is a pairwise composite likelihood, conceived for cases where the number of individuals is too low for the Gaussian pseudo-likelihood to be reliable. The distributions of the pairs are related to the bivariate binomial and Poisson distributions \citep{kawamura1973structure}. Applications are shown in Section \ref{Sec:Applications}. In Section \ref{Sec:Comparison} we discuss similarities and differences with Poisson processes and Markov models, which can be retrieved as particular cases. We conclude in Section \ref{Sec:Perspectives} with ideas to explore more about the ECM distribution and further potential applications.

Most part of mathematical results are explained intuitively in the main body of this paper. Mathematical proofs, which rely mainly on characteristic functions, are given in Appendix \ref{App:Proofs}. Reproducible \texttt{R} routines are openly available at the Github repository \url{https://github.com/CarrizoV/Introducing-the-ECM-distribution}.

\section{The ECM distribution}
\label{Sec:ECM}

We consider an individual following a stochastic dynamics across evolving categories: at each time the individual belongs to a single category among many possible ones, and the categories also may change with time. Let $n \in \mathbb{N} $ be the number of time steps\footnote{The construction of this distribution is done in discrete \textit{index} time. These index times may not necessarily refer to \textit{equally spaced chronological time points} or similar. What the index time refers to depends on the application scenario. For example, index times may represent moments in a continuous-time interval (possibly with irregular lag between them), long periods of time  (days, years, centuries, etc.), or particular events (election rounds, selling seasons, concerts, etc.).}. For each $k = 1 , ... , n$ we consider $m_{k} \in \mathbb{N}$ exhaustive and exclusive categories (or classes, or boxes, or states, or bins, etc.) to which an individual may belong at time $k$. The subsequent stochastic dynamics is completely determined by the full-path probabilities:
\begin{equation}
\label{Eq:full-pathProbs}
    p_{l_{1} , ... , l_{n}}^{(1,...,n)} := \hbox{``probability of belonging to categories $l_{1},\ldots , l_{n}$ at times $1,\ldots , n$ respectively''}, 
\end{equation}
for every $(l_{1} , \ldots , l_{n}) \in \lbrace 1 , \ldots , m_{1} \rbrace \times \ldots \times \lbrace 1 , \ldots , m_{n} \rbrace$. There are $(\prod_{k=1}^{n}m_{k}) - 1$ free probabilities to determine, since they must sum to one due to the exhaustiveness of the categories at each time:
\begin{equation}
\label{Eq:ProbSum1}
\sum_{l_{1}=1}^{m_{1}} ... \sum_{l_{n}=1}^{m_{n}} p_{l_{1} , ... , l_{n}}^{(1,...,n)} = 1.
\end{equation}
With the full-path probabilities one can obtain, by marginalization, the sub-path probabilities, indicating the probability of belonging to some categories at some times, not necessarily considering all the $n$ time steps. In this work we shall focus mainly on the one-time and two-times probabilities: $p_{l}^{(k)}$ denotes the probability of belonging to category $l$ at time $k$; $p_{l,l'}^{(k,k')}$ denotes the probability of belonging to category $l$ at time $k$ and to category $l'$ at time $k'$. As explained further, these probabilities will be enough to describe important properties and develop ad-hoc fitting methods.

We now consider $N$ individuals following independently this stochastic dynamics across the evolving categories. Our focus is on the number of individuals counted in each category at each time. For each $k \in \lbrace 1 , ... , n \rbrace $ and $l \in \lbrace 1 , ... , m_{k} \rbrace$, we define the random variable
\begin{equation}
\label{Eq:DefQComponent}
    Q_{l}^{(k)} := ``\hbox{number of individuals belonging to category } l \hbox{ at time } k". 
\end{equation}
We assemble the variables \eqref{Eq:DefQComponent} into a single \textit{random arrangement}
\begin{equation}
\label{Eq:DefQArrangment}
    \bm{Q} := \big(  Q_{l}^{(k)} \big)_{k \in \lbrace 1 , ... , n \rbrace , l \in \lbrace 1 , ... , m_{k} \rbrace}. 
\end{equation}

\begin{Def}
\label{Def:ECMDistribution}
    Under the previously defined conditions, we say that the random arrangement $\bm{Q}$ follows an evolving categories multinomial distribution (abbreviated ECM distribution).
\end{Def}

There are many possible ways to arrange the variables in $\bm{Q}$. One useful manner is to see $\bm{Q}$ as a \textit{list or collection of random vectors}, by writing $\bm{Q} = ( \vec{Q}^{(k)} )_{k \in \lbrace 1 , ... , n \rbrace}$, with
\begin{equation}
\label{Eq:DefVectorQk}
    \vec{Q}^{(k)} = ( Q_{1}^{(k)} , ... , Q_{m_{k}}^{(k)} ), \quad \forall k \in \lbrace 1 , ... , n \rbrace.
\end{equation}
Note that $\bm{Q}$ is not a random matrix unless $m_{1} = ... = m_{n}$. The arrangement $\bm{Q}$ could be wrapped into a traditional random vector with $\sum_{k=1}^{n}m_{k}$ components, but for ease of presentation we maintain the indexing scheme in \eqref{Eq:DefQArrangment}. We describe the distribution of an ECM random arrangement through its characteristic function.

\begin{Prop}
\label{Prop:CharFunctionQ}
The characteristic function $\varphi_{\bm{Q}} : \mathbb{R}^{m_{1} + ... + m_{n}} \to \mathbb{C}$ of $\bm{Q}$ is
    \begin{equation}
    \label{Eq:CharFunctionQ}
        \varphi_{\bm{Q}}(\bm{\xi} ) = \left[  \sum_{l_{1}=1}^{m_{1}} ... \sum_{l_{n}=1}^{m_{n}}p_{l_{1}, ... , l_{n}}^{(1 , ... , n)} e^{i( \xi_{l_{1}}^{(1)} + ... + \xi_{l_{n}}^{(n)} )}  \right]^{N},
    \end{equation}
    with $ \bm{\xi} = \left( \xi_{l}^{(k)} \right)_{k \in \lbrace 1 , ... , n \rbrace, l \in \lbrace 1 , ... , m_{k} \rbrace} \in \mathbb{R}^{m_{1} + ... + m_{n}}$.\footnote{For ease of exposition, we consider the arrangements of the form $(q_{l}^{(k)})_{k \in \lbrace 1 , ... , n \rbrace, l \in \lbrace 1 , ... , m_{k} \rbrace}$ as members of the space $\R^{m_{1} + ... + m_{n}}$, although technically their indexing differs from the traditional one used for vectors. In any case, the vector space of such arrangements is isomorphic to  $\R^{m_{1} + ... + m_{n}}$.}
\end{Prop}

The following Proposition \ref{Prop:OneTimeDistQ} describes the one-time marginals, that is, the distribution of the random vectors $\vec{Q}^{(k)}$ as defined in \eqref{Eq:DefVectorQk}. The word \textit{multinomial} in the name ``ECM'' comes from this result.

\begin{Prop}
\label{Prop:OneTimeDistQ}
    Let $k \in \lbrace 1 , ... , n \rbrace$. Then,
    \begin{equation}
        \vec{Q}^{(k)} \sim Multinomial\left( N , (p_{1}^{(k)} , ... , p_{m_{k}}^{(k)})
 \right)
    \end{equation}
\end{Prop}

This result can be seen intuitively: for a given time $k$ we assign $N$ individuals, independently and with common probabilities, to the $m_{k}$ categories, so $\vec{Q}^{(k)}$ is a multinomial random vector. The two-times conditional distribution is also related to the multinomial distribution. Let us introduce the two-times conditional probabilities for  $k \neq k'$:
\begin{equation}
\label{Eq:ConditionalPathProbs}
    p_{l' | l}^{(k' | k)} := p_{l,l'}^{(k , k')} / p_{l}^{k}, \quad \forall (l,l') \in \{ 1 , \ldots , m_{k} \} \times \{ 1 , \ldots , m_{k'} \}
\end{equation}
\begin{Prop}
\label{Prop:TwoTimesDistQ}
    Let $k,k' \in \lbrace 1 , ... , n \rbrace$, $k \neq k'$. Then, the random vector $\vec{Q}^{(k')}$ conditioned on $\vec{Q}^{(k)}$ has the distribution of a sum of $m_{k}$ independent $m_{k'}$-dimensional multinomial random vectors:
    \begin{equation}
        \vec{Q}^{(k')} \mid \vec{Q}^{(k)} \sim \sum_{l=1}^{m_{k}} Multinomial\left( Q_{l}^{(k)} , ( p_{1 \mid l}^{(k'\mid k)} , \ldots , p_{m_{k'} \mid l}^{(k'\mid k)} ) \right) \quad \hbox{(indep. sum)}.
    \end{equation}
\end{Prop}

Proposition \ref{Prop:TwoTimesDistQ} can also be interpreted intuitively: the individuals in each category at time $k$ generate a multinomial vector when moving to the categories at time $k'$, with corresponding conditional probabilities. Since individuals move independently, the aggregate of these independent random vectors yields the final count at time $k'$. This structure with a sum of independent multinomial random vectors is a particular case of the Poisson-multinomial distribution  \citep{daskalakis2015structure,lin2023computing}, which is constructed as the sum of independent multinomial random vectors of size $1$ with different bins probabilities.

Propositions  \ref{Prop:OneTimeDistQ} and \ref{Prop:TwoTimesDistQ} allow us to compute the first and second moment structures of $\bm{Q}$. We use the Kronecker delta notation $\delta_{l,l'} = 1$ if $l=l'$, and $\delta_{l,l'} = 0$ if $l \neq l'$.

\begin{Prop}
\label{Prop:MomentsECM}
    The mean and covariance structures of $\bm{Q}$ are given respectively by
    \begin{equation}
    \label{Eq:MeanECM}
        \mathbb{E}\left( Q^{(k)}_{l}  \right) = Np_{l}^{(k)},
    \end{equation}
    \begin{equation}
\label{Eq:CovQ}
    \Cov\left( Q_{l}^{(k)}  ,   Q^{(k')}_{l'}  \right)  = \begin{cases}
        N\left( p_{l,l'}^{(k,k')} - p_{l}^{(k)}p_{l'}^{(k')}  \right) \quad & \hbox{ if } k \neq k', \\
        N\left(   \delta_{l,l'}p_{l}^{(k)}    - p_{l}^{(k)}p_{l'}^{(k')}   \right) \quad & \hbox{ if } k = k'.
    \end{cases} 
\end{equation}
    for every $k,k' \in \lbrace 1 , ... , n \rbrace$ and $(l,l') \in \lbrace 1 , ... , m_{k} \rbrace \times \lbrace 1 , ... , m_{k'} \rbrace.$
\end{Prop}

\section{Unknown $N$: the ECM-Poisson distribution}
\label{Sec:ECMPoisson}

In many applications, the total number of individuals $N$ will be unknown. Methods are then needed to use the ECM distribution in such scenarios. Here we explore the option of assuming $N$ is random following a Poisson distribution. This provides a connection with Poisson counts-type models.

Let $\lambda > 0$ be a parameter (the \textit{size rate}) and let $N \sim Poisson(\lambda)$. We use the same notations and indexing as in Section \ref{Sec:ECM} for number of time steps, number of categories, and path probabilities. Let $\bm{Q}$ be a random arrangement whose distribution is, when conditioned on $N$, an ECM distribution with $N$ individuals. Then $\bm{Q}$ follows a new distribution which we call the \textit{ECM-Poisson distribution}. 

\begin{Prop}
\label{Prop:CharFunctionQPoisson}
    The characteristic function $\varphi_{\bm{Q}}$ of an ECM-Poisson random arrangement $\bm{Q}$ is
    \begin{equation}
    \label{Eq:CharFunctionQPoisson}
        \varphi_{\bm{Q}}(\bm{\xi} ) = \exp \Big\lbrace \lambda \left(  \sum_{l_{1}=1}^{m_{1}} ... \sum_{l_{n}=1}^{m_{n}}  p_{l_{1}, ... , l_{n}}^{(1 , ... , n)} e^{i( \xi_{l_{1}}^{(1)} + ... + \xi_{l_{n}}^{(n)} )} \ - 1  \right)  \Big\rbrace  ,
    \end{equation}
    for every $ \bm{\xi} = \left( \xi_{l}^{(k)} \right)_{k \in \lbrace 1 , ... , n \rbrace, l \in \lbrace 1 , ... , m_{k} \rbrace} \in \mathbb{R}^{m_{1} + ... + m_{n}}$.
\end{Prop}

The following result for the one-time marginal distribution derives immediately from the connection between the Poisson and multinomial distributions. We use the tensor product notation: if $\vec{X} = (X_{1} , \ldots , X_{\tilde{n}})$ is a random vector, we note $\vec{X} \sim \bigotimes_{j=1}^{\tilde{n}}\mathscr{L}_{j}$ to say that it consists of independent components, each $X_{j}$ following the distribution $\mathscr{L}_{j}$.

\begin{Prop}
\label{Prop:OneTimeDistQPoisson}
    Let $k \in \lbrace 1 , ... , n \rbrace$. Then, $\vec{Q}^{(k)}$ has independent Poisson components:
    \begin{equation}
    \label{Eq:QkECMPoissonIndep}
        \vec{Q}^{(k)} \sim \bigotimes_{l=1}^{m_{k}} Poisson(  \lambda p_{l}^{(k)} ).
    \end{equation}
\end{Prop}

The ECM-Poisson case contrasts thus with the ECM case where $\vec{Q}^{(k)}$ has correlated binomial components. For the two-times distribution, there is a particular subtlety to take into account. Since the categories are exhaustive at each time, knowing $\vec{Q}^{(k)}$ implies that we know the value of $N = \sum_{l=1}^{m_{k}}Q_{l}^{(k)}$. Therefore, the distribution of $\vec{Q}^{(k')}$ given $\vec{Q}^{(k)}$ for $k' \neq k$ is simply the same as in Proposition \ref{Prop:TwoTimesDistQ}. However, it is more enlightening to study the case where only some components of $\vec{Q}^{(k)}$ are known, implying that we do not know $N$.

For every $k$, denote $\tilde{m}_{k} = m_{k} - 1$ (assuming $m_{k} \geq 2$) and denote $\vec{\tilde{Q}}^{(k)} =  (  Q_{1}^{(k)} , \ldots , Q_{\tilde{m}_{k}}^{(k)} )$ the sub-vector of $\vec{Q}^{(k)}$ considering the categories up to $\tilde{m}_{k}$. Suppose that we know $\vec{\tilde{Q}}^{(k)}$ for a given $k$, and take $k' \neq k$. Then, each quantity $Q_{l}^{(k)}$ is redistributed in the $m_{k'}$ categories at time $k'$ similarly to Proposition \ref{Prop:TwoTimesDistQ}, producing a sum of independent multinomial contributions to the vector $\vec{Q}^{(k')}$. In addition, the unknown quantity $Q_{m_{k}}^{(k)}$ generates independent Poisson counts, similarly as in Proposition \ref{Prop:OneTimeDistQPoisson}, which are added to $\vec{\tilde{Q}}^{(k')}$. This intuition is precisely stated in Proposition \ref{Prop:TwoTimesDistQPoisson}. 

\begin{Prop}
\label{Prop:TwoTimesDistQPoisson}
    Let $k,k'\in \{ 1 , \ldots , n \}$, $k \neq k'$. Then, $\vec{Q}^{(k')}$ conditioned on $\vec{\tilde{Q}}^{(k)}$ has the distribution of a sum of $\tilde{m}_{k}$ independent $m_{k'}$-dimensional multinomial random vectors plus an independent random vector of $m_{k'}$ independent Poisson components:
    \small
    \begin{equation}
\vec{Q}^{(k')} \mid \vec{\tilde{Q}}^{(k)} \sim \bigotimes_{l'=1}^{m_{k'}}Poisson( \lambda p_{m_{k},l'}^{(k,k')} )\ + \ \sum_{l=1}^{\tilde{m}_{k}} Multinomial\left( Q_{l}^{(k)} ,  ( p_{1 \mid l}^{(k' \mid k )} , \ldots , p_{m_{k'} \mid l}^{(k' \mid k )}  ) \right) \quad \hbox{(indep. sums)}. 
    \end{equation}
    \normalsize
\end{Prop}

Similar distributions have been worked out in the context of evolving Poisson point processes from moving particles. See \citep[Section 4]{roques2022spatial}, where a future point process conditioned on a past Poisson process is described as the union of an inhomogeneous Poisson process and independent thinned binomial point processes. Proposition \ref{Prop:TwoTimesDistQPoisson} is the resulting distribution of the count of such processes over different spatial areas (see Section \ref{Sec:MovingIndividuals}). 

The second order moments of an ECM-Poisson random arrangement are simpler than in the ECM case.

\begin{Prop}
\label{Prop:MeanCovQPoisson}
 The mean and covariance structures of $\bm{Q}$ are given respectively by
    \begin{equation}
    \label{Eq:MeanECMPoisson}
        \mathbb{E}\left( Q^{(k)}_{l}  \right) = \lambda p_{l}^{(k)},
    \end{equation}
    \begin{equation}
    \label{Eq:CovECMPoisson}
        \mathbb{C}ov\left( Q^{(k)}_{l} \ , \ Q^{(k')}_{l'}  \right) = \begin{cases}
        \lambda p_{l}^{(k)} \delta_{l,l'} \quad & \hbox{ if } k = k', \\ 
             \lambda p_{l,l'}^{(k,k')} \quad & \hbox{ if } k \neq k',
        \end{cases} 
    \end{equation}
        for every $k,k' \in \lbrace 1 , ... , n \rbrace$ and $(l,l') \in \lbrace 1 , ... , m_{k} \rbrace \times \lbrace 1 , ... , m_{k'} \rbrace.$
\end{Prop}

\section{Inference methods}
\label{Sec:Inference}

Consider the problem of computing the probability $\mathbb{P}(\bm{Q} = \bm{q})$ for an ECM random arrangement $\bm{Q}$ and a consistent arrangement $\bm{q}$. For the case $n = 2$, one could do
\begin{equation}
    \mathbb{P}(\bm{Q} = \bm{q}) =\mathbb{P}(\vec{Q}^{(2)} = \vec{q}^{(2)} \mid \vec{Q}^{(1)} = \vec{q}^{(1)} )\mathbb{P}( \vec{Q}^{(1)} = \vec{q}^{(1)} ), 
\end{equation}
and use Propositions \ref{Prop:OneTimeDistQ} and \ref{Prop:TwoTimesDistQ}. $\mathbb{P}( \vec{Q}^{(1)} = \vec{q}^{(1)} )$ is a simple multinomial probability, and $\mathbb{P}(\vec{Q}^{(2)} = \vec{q}^{(2)} \mid \vec{Q}^{(1)} = \vec{q}^{(1)} )$ is a Poisson-Multinomial probability which can be obtained as a convolution of multinomial probabilities. Under this approach, one must convolve $m_{1}$ arrays of dimension $m_{2}$, the $l$-th array having values ranging from $0$ to $q_{l}^{(1)}$. For $m_{1}$ and $m_{2}$ of the order of $10^{2}$ (which is common in ecological applications with a domain split in many subdomains), this convolution becomes intractable both in terms of computation time and memory use, getting worse if the values in $\bm{q}$ are high (likely when $N$ is high). A computing method using fast Fourier transforms from the characteristic function is explored in \citep[Section 4.4]{lin2023computing}. The study is done only for  $m_{2} \leq 5$ since the algorithm computes arrays of dimension $(N+1)^{m_{2}-1}$ (in their setting $N = m_{1}$), becoming impractical even for small values of $m_{2}$. Besides the convolutional formulation, no closed-form expression for the Poisson-Multinomial part is known. The case $n \geq 3$ is even more intricate, since the Poisson-Multinomial distribution is not present and the cost of Fourier transforms of characteristic function \eqref{Eq:CharFunctionQ} are more sensitive to $m_{1},m_{2},N$ as when $n=2$. The use of the full-path probabilities \eqref{Eq:full-pathProbs} requires to storage $\prod_{k=1}^{n}m_{k}$ values, which in certain scenarios are not provided but must be computed (see Section \ref{Sec:MovingIndividuals}).

The explicit ECM and ECM-Poisson likelihoods are thus considered currently intractable. Fitting methods requiring likelihood evaluation cannot be applied. Due to this, we explore two suitable estimating functions for inference: a Gaussian pseudo-likelihood and a pairwise composite likelihood. Only the one-time and two-times probabilities are required for these techniques. From our developments, these methods are immediately appliable and can be used to infer parameters occurring in the autocorrelation structure of the count.

The basic parameters of the ECM distribution are the full-path probabilities \eqref{Eq:full-pathProbs}. Since there are $(\prod_{k=1}^{n}m_{k})-1$ of them but only $\sum_{k=1}^{n}(m_{k}-1)$ informative data, it is impossible to obtain precise estimates with a single realization. One option, which we exploit in Section \ref{Sec:VoteTransfer}, is to consider a sufficient number of independent replicates. The other option, explored in Section \ref{Sec:MovingIndividuals} and thought for the single-realization case, is to express the path probabilities as functions of a much smaller set of parameters for which inference is possible. The methods here presented can help for both scenarios.

\subsection{Maximum Gaussian Likelihood Estimation (MGLE)}
\label{Sec:MGLE}

If $\bm{Q^{(N)}}$ is an ECM random arrangement of $N$ individuals, it can be expressed as a sum of $N$ iid ECM random arrangements with $1$ individual. Note that the characteristic function of $\bm{Q^{(N)}}$ in Proposition \ref{Prop:CharFunctionQ}, is the $N$-power of a given characteristic function. This allows us to invoke a multivariate central limit theorem argument and claim that $\bm{Q^{(N)}}$ is asymptotically Gaussian as $N \to \infty$. Precisely, let $\bm{m_{Q^{(1)}}}$ and $\bm{\Sigma_{Q^{(1)}}}$ be respectively the mean arrangement and the covariance tensor of an ECM arrangement with $1$ individual. Then, when $N \to \infty$
\begin{equation}
    \label{Eq:QasymptGaussian}
    \frac{\bm{Q^{(N)}} - N\bm{m_{Q^{(1)}}} }{\sqrt{N}} \stackrel{l}{\to} \mathcal{N}(  \bm{0} , \bm{\Sigma_{Q^{(1)}}} ), 
\end{equation}
where $\stackrel{l}{\to}$ denotes the convergence in distribution. The idea is simply to replace the original likelihood with a Gaussian one with identical mean and covariance (Proposition \ref{Prop:MomentsECM}) when $N$ is large. For the ECM-Poisson case, the classical Gaussian approximation of the Poisson distribution for large rates can also be applied for large $\lambda$, replacing $N$ with $\lambda$ in formula \eqref{Eq:QasymptGaussian} and using the corresponding mean and covariance given in Proposition \ref{Prop:MeanCovQPoisson} (with $\lambda = 1$).

Replacing the exact likelihood with a Gaussian one through the use of central limit theorems is common practice in statistics. See \citep{daskalakis2015structure,lin2023computing} for the Poisson-Multinomial distribution. Here, this method requires the one-time and two-times probabilities. If they are not available, they must be computed and then used to construct the mean vector and the covariance matrix. Then, the computational cost of the Gaussian likelihood depends on the number of counts in $\bm{Q}$, as is typical for Gaussian vectors.

\subsection{Maximum Composite Likelihood Estimation (MCLE)}
\label{Sec:MCLE}

Composite likelihoods are estimating functions constructed from likelihoods of subsets of the variables. They are used for parameter estimation of models whose likelihood is too costly to evaluate \citep{lindsay1988composite,varin2011overview}. MCLE provides less efficient estimates than proper maximum likelihood estimates, but for which an asymptotic theory is well developed. In particular, consistency and asymptotic Gaussianity (when adding proper information) are common properties of these estimates  \citep{godambe1960optimum,kent1982robust,cox2004note}.  The most commonly used composite likelihood is the pairwise composite likelihood \citep{varin2006pairwise,caragea2007asymptotic,padoan2010likelihood,bevilacqua2015comparing}. Let $(X_{1} , \ldots,  X_{\tilde{n}})$ be a random vector with $\tilde{n}$ components whose distribution depends on a parameter $\theta$. The estimating function is given by the product of the likelihoods of the pairs:
\begin{equation}
    c\mathcal{L}_{X_{1} , \ldots , X_{\tilde{n}}}(\theta) = \prod_{j=1}^{\tilde{n}-1}\prod_{k=j+1}^{\tilde{n}} \mathcal{L}_{X_{j},X_{k}}(\theta).
\end{equation}
This function is rich enough to provide estimates for parameters occurring in the bivariate marginals, particularly those in the covariance structure. This is the approach we explore here.

Consider an ECM or ECM-Poisson random arrangement $\bm{Q}$. We express the pairwise composite log-likelihood of $\bm{Q}$ for a given parameter $\theta$ as
\begin{equation}
\label{Eq:ECMCompositeLoglikelihood}
    c\ell_{\bm{Q}}(\theta) = \sum_{k=1}^{n} \sum_{l=1}^{m_{k}-1}\sum_{l' = l+1}^{m_{k}}  \ell_{Q_{l}^{(k)}, Q_{l'}^{(k)}}(\theta) \ + \ \sum_{k=1}^{n-1} \sum_{k'=k+1}^{n} \sum_{l=1}^{m_{k}}\sum_{l'=1}^{m_{k'}} \ell_{Q_{l}^{(k)}, Q_{l'}^{(k')}}(\theta).
\end{equation}
The triple sum in \eqref{Eq:ECMCompositeLoglikelihood} includes pairs at the same time, while the quadruple sum includes the pairs at different times. The results in Sections \ref{Sec:ECM} and \ref{Sec:ECMPoisson} allow us to obtain the  log-likelihoods of the pairs $\ell_{Q_{l}^{(k)}, Q_{l'}^{(k')}}(\theta)$. For the ECM case, the pairs at same time $k=k'$ are sub-vectors of a multinomial vector, so their distribution is simple. When $k \neq k'$ a special structure appears in $(Q_{l}^{(k)} , Q_{l'}^{(k')})$. In the case with just $1$ individual, it is a random binary vector following thus a bivariate Bernoulli distribution  \citep[Section 2.1]{kawamura1973structure}. For $N$ individuals, $(Q_{l}^{(k)} , Q_{l'}^{(k')})$ is the addition of $N$ independent bivariate Bernoulli vectors, obtaining a \textit{bivariate binomial vector} \citep[Section 2.2]{kawamura1973structure}.

\begin{Prop}
\label{Prop:PairsECMBivariateBinomial}
Let $\bm{Q}$ be an ECM random arrangement. Then, for $k \neq k'$,  $(Q_{l}^{(k)} , Q_{l'}^{(k')} )$ follows a bivariate binomial distribution of size $N$ with joint success probability $p_{l,l'}^{(k,k')}$ and respective marginal success probabilities $p_{l}^{(k)}$ and $p_{l'}^{(k')}$.
\end{Prop}

For the ECM-Poisson case with $k=k'$, $(Q_{l}^{(k)} , Q_{l'}^{(k')})$ has independent Poisson components. For $k\neq k'$, each component can be written as the sum of two independent Poisson: $Q_{l}^{(k)}$ is the count of those belonging to $l$ at $k$ \textit{and} to $l'$ at $k'$ plus those belonging to $l$ at $k'$ \textit{and not} to $l'$ at $k'$ ( $Q_{l'}^{(k')}$ is analogue). The count of those belonging to $l$ at $k$ and to $l'$ at $k'$ occurs in both sums. The resulting law is known as the \textit{bivariate Poisson distribution} \citep[Section 3]{kawamura1973structure}.

\begin{Prop}
\label{Prop:ECMPoissonPairBivariatePoisson}
    Let $\bm{Q}$  be an ECM-Poisson random arrangement. Then, for $k \neq k'$,  $( Q_{l}^{(k)} , Q_{l'}^{(k')}  )$ follows a bivariate Poisson distribution with joint rate $\lambda p_{l,l'}^{(k,k')}$ and respective marginal rates $\lambda p_{l}^{(k)}$ and  $\lambda p_{l'}^{(k')}$.
\end{Prop}

In Appendices \ref{Proof:PairsECMBivariateBinomial}, \ref{Proof:ECMPoissonPairBivariatePoisson} we give more details on the bivariate binomial and Poisson distributions, including formulae to compute the pair likelihoods  (\eqref{Eq:BivBinomialProb}, \eqref{Eq:BivPoissonProb}). For MCLE the one-time and two-times probabilities must be known, similarly as in MGLE. Bivariate binomial and Poisson distributions require extra computations of sums, whose lengths increase with the values in $\bm{Q}$, getting larger for large $N$ or $\lambda$. MCLE is thus more costly than MGLE, but, as seen further in Section \ref{Sec:Applications}, it can provide better estimates when $N$ or $\lambda$ are too low for the Gaussian pseudo-likelihood to be reliable.

\section{Applications}
\label{Sec:Applications}

We propose two standalone application scenarios with count data: one for inferring movement parameters in ecology, and the other for inferring vote transfer in sociology. Readers interested in one field are not required to read the application on the other field. The models in our illustrations are rather simplistic proofs of concept kind, and their specific assumptions may not be entirely biologically or sociologically realistic. The conclusions should not be seen as definite.

\subsection{Counting moving individuals}
\label{Sec:MovingIndividuals}

An underlying continuous space-time movement of individuals induces particular statistical properties on space-time survey counts. Here we show how use the ECM distribution to analyze those properties and to learn about movement parameters from aggregated count data. Let $t_{0}$ be an initial time. Consider $N$ individuals moving continuously and randomly over $\Rd$. We denote $X_{j} = (X_{j}(t))_{t \geq t_{0}}$ the $\Rd$-valued stochastic process which describes the trajectory of individual $j$. We define the \textit{abundance random measure} as the space-time (generalized) random field identified with the family of random variables $ \Phi = (\Phi_{t}(A) )_{A \in \BorelRd, t \geq t_{0} }$ which represent the population abundance:
\begin{equation}
\label{Eq:DefPhit(A)Concept}
\Phi_{t}(A) := \hbox{``number of individuals in region $A$ at time $t$''}.
\end{equation}
Note that $A$ represents a (Borel) survey region $A \subset \Rd$, not a single point of $\Rd$. We use the term \textit{random measure} since for every $t$ the application $A \mapsto \Phi_{t}(A)$ is a random measure \citep{kallenberg2017random}, more particularly a spatial point process \citep{daley2006introduction}. Let us assume the $N$ individuals follow iid trajectories, that is, the processes $(X_{j})_{j}$ are iid with the distribution of a reference process $X = (X(t))_{t \geq t_{0}}$. Let $t_{1} , ... , t_{n} \geq t_{0}$ be survey time points (possibly continuously irregularly sampled). For each time $t_{k}$, let $A_{1}^{(k)} , ... , A_{m_{k}}^{(k)}$ be a collection of $m_{k} \geq 1$ survey regions forming a partition of $\Rd$. The categories to be used are simply given by
\small
\begin{equation}
    \hbox{individual $j$ belongs to category $l$ at time $k$} \Longleftrightarrow \hbox{individual $j$ is in $A_{l}^{(k)}$ at time $t_{k}$,} 
\end{equation}
\normalsize
which are exhaustive and exclusive since the survey regions form a partition. We define the random variables
\begin{equation}
    Q_{l}^{(k)} := \Phi_{t_{k}}(A_{l}^{(k)}), \quad l \in \{ 1 , \ldots , m_{k} \},k \in \{ 1 , \ldots , n \}. 
\end{equation}
The arrangement $\bm{Q} = ( Q_{l}^{(k)} )_{k,l} $ contains the aggregated counts of $N$ individuals following iid movement dynamics across evolving categories: $\bm{Q}$ follows the ECM distribution. The path probabilities are determined by the distribution of $X$:
\begin{equation}
\label{Eq:PathProbPHI}
p_{l_{1} , ... , l_{n}}^{(1, ... , n)} = \mathbb{P}\left( X(t_{1}) \in A_{l_{1}}^{(1)} , \ldots , X(t_{n}) \in A_{l_{n}}^{(n)}  \right).
\end{equation}
The sub-path probabilities can be obtained by marginalization, replacing some of the $A_{l_{k}}^{(k)}$ sets in \eqref{Eq:PathProbPHI} by $\Rd$. In particular, one-time probabilities are determined by the one-time marginal distributions of the process $X$, and two-times probabilities by the bivariate two-times marginals. 

That the path probabilities are completely determined by the distribution of $X$ has two consequences. First, the parameters governing the distribution of $\bm{Q}$ are now the same as those in the distribution of $X$, which are likely much smaller in number than the total amount of path probabilities in a general ECM case. This may allow inference even in the single-realization scenario. Second, we can analyze a variety of trajectory models with relative freedom. For instance, if $X$ is any Gaussian process and the sets $A_{l}^{(k)}$ are rectangles, many available libraries in various programming languages allow computation of the path probabilities \citep{genz2009computation,botev2017normal,cao2022tlrmvnmvt}. Note also that, in principle, no Markov assumption is required on the process $X$, contrarily to the common case of interacting particle systems in statistical mechanics \citep{martin1976limit,liggett1985interacting}.

The case where $N$ is unknown provides a particular connection with Poisson point processes. If $N \sim Poisson(\lambda)$ and if we define $\Phi$ considering $N$ independent moving individuals as above, then $\bm{Q}$ is ECM-Poisson. At each time $t_{k}$, the counts $( \Phi_{t_{1}}(A_{1}^{(k)}) , \ldots , \Phi_{t_{k}}(A_{m_{k}}^{(k)}))$ are independent Poisson variables with respective rates $\lambda \mathbb{P}(X(t_{k}) \in A_{l}^{(k)})$ (Proposition \ref{Prop:OneTimeDistQPoisson}), and this structure holds for any possible choice of disjoint survey regions $(A_{l}^{(k)})_{l}$. In other words, for every $t$, $\Phi_{t}$ is a spatial inhomogeneous Poisson process with intensity $ A \mapsto \lambda \mathbb{P}( X(t) \in A ) $. Thus, $\bm{Q}$ contains the counts of an \textit{evolution of inhomogeneous Poisson processes}, with a particular temporal dependence induced by the movement, involving for instance the bivariate Poisson distribution.

\subsubsection{Illustration 1: simulation studies on steady-state Ornstein-Uhlenbeck trajectories}
\label{Sec:SimuStudyOU}

To show we can infer movement parameters with count data, we qualitatively investigate both the non-asymptotics and asymptotics properties of estimates (as the number of individuals grows) using a simulation study with individuals moving according to a steady-state Ornstein-Uhlenbeck process (OU), which is a widely used trajectory model \citep[Section 3.8]{uhlenbeck1930theory,gardiner2009stochastic}. One parameter (here $\sigma$) only occurs in the covariance of the movement, thus the space-time autocorrelation of the counts, but not in the mean. It is therefore impossible to estimate it using only the intensity information. For instance, an inhomogeneous space-time Poisson process model for the counts cannot be used to estimate it. Here we prove that with our approach even such kind of parameters can be estimated from the mere count data.

A $2$-dimensional (isotropic) steady-state OU process is a Gaussian process $X = (X(t))_{t \geq t_{0}}$ with values in $\R^{2}$, say $X(t) = (X_{1}(t),X_{2}(t))$, such that
\begin{equation}
    \mathbb{E}(  X(t) ) = z \quad ; \quad \Cov(X_{j}(t) , X_{k}(s) ) = \delta_{j,k}\tau^{2} e^{-\frac{1}{2}\frac{\sigma^{2}}{\tau^{2}} |t-s| }, \quad \forall t,s \geq t_{0}, 
\end{equation}
where $z = (z_{1} , z_{2}) \in \R^{2}$ and $\sigma , \tau > 0$. In other words, $X$ is a stationary Gaussian process with constant mean and exponential covariance function, as widely used in time-series modeling \citep{,allen2010introduction,hamilton2020time}. OU processes have been used in ecology to model animal movement around an activity center \citep{fleming2014fine,eisaguirre2021multistate,gurarie2017correlated}. The OU parameters can be interpreted biologically: since $X_{j}(t) \sim \mathcal{N}( z_{j} ,  \tau^{2} )$, $z$ is the activity center and $\tau$ controls the extent of the home range \citep{borger2008there,powell2012home}; $\sigma$ controls the instantaneous quadratic variation of the process, which can be interpreted as how ``fast'' an individual moves:
\begin{equation}
\label{Eq:OUQuadraticvariationSigma}
    \lim_{\Delta t \to 0^{+}} \frac{\Var\left( X(t + \Delta t) - X(t) \right)}{\Delta t} = \sigma^{2}.
\end{equation}
In Section \ref{Sec:AppChickens} we shall also use the non-steady-state version of an OU process. The simplest manner of defining such a process is as a steady-state OU process conditioned on $X(t_{0})=x_{0}$ for a initial time $t_{0}$ and a initial position $x_{0} \in \R^{2}$. Then, the process describes an individual starting at $x_{0}$ and traveling towards $z$ in a sort of direct manner (exponentially fast on average), to finish in an approximately steady-state OU process around $z$ (convergence to the steady-state when $t \to \infty$). Another common equivalent way of defining an OU process is through a stochastic differential equation and using a different parametrization,  setting $\theta = \frac{1}{2}\frac{\sigma^{2}}{\tau^{2}}$ and focusing on $(\sigma , \theta)$ as parameters.

\begin{figure}[btp]
  \centering

  \begin{subfigure}[h]{\textwidth}
  \centering
    \includegraphics[ width = \textwidth ,  height = 0.23\textheight]{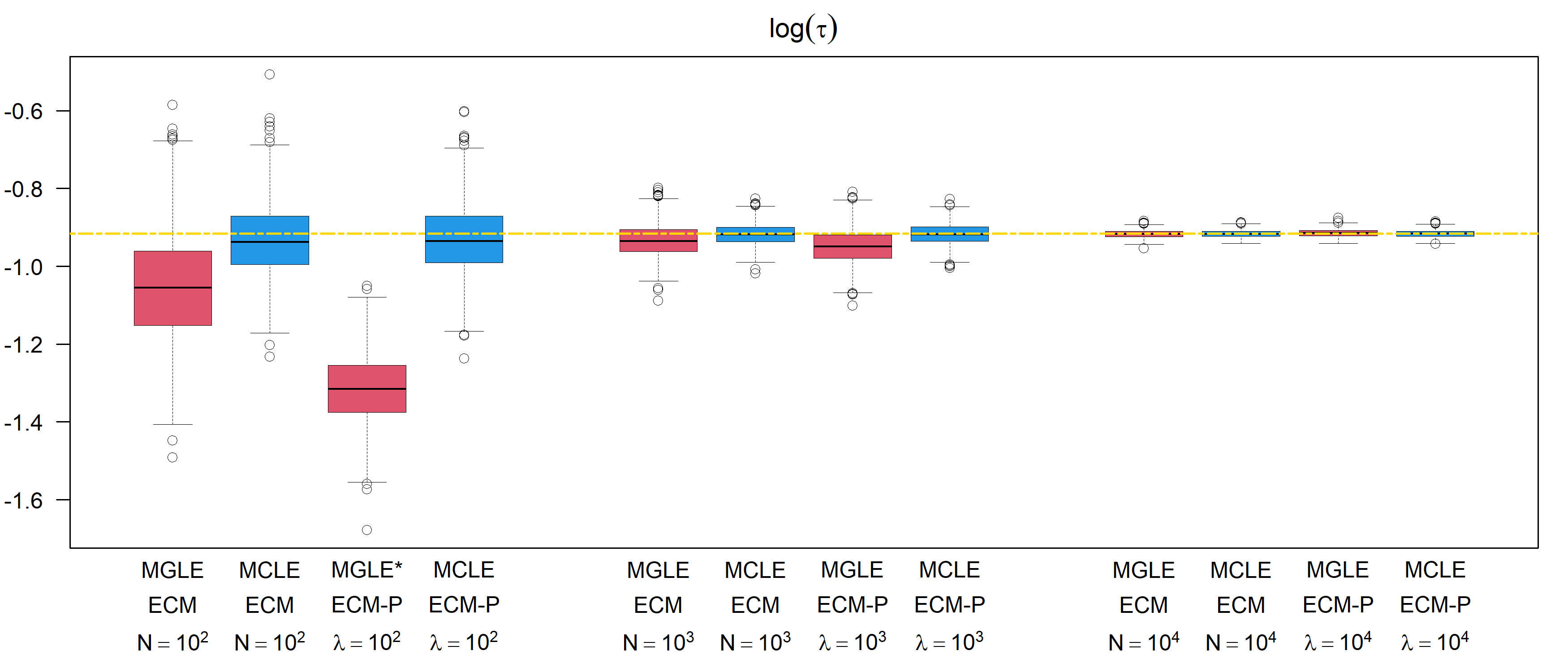}
  \end{subfigure}

  \begin{subfigure}[h]{\textwidth}
  \centering
    \includegraphics[  width = \textwidth , height = 0.23\textheight ]{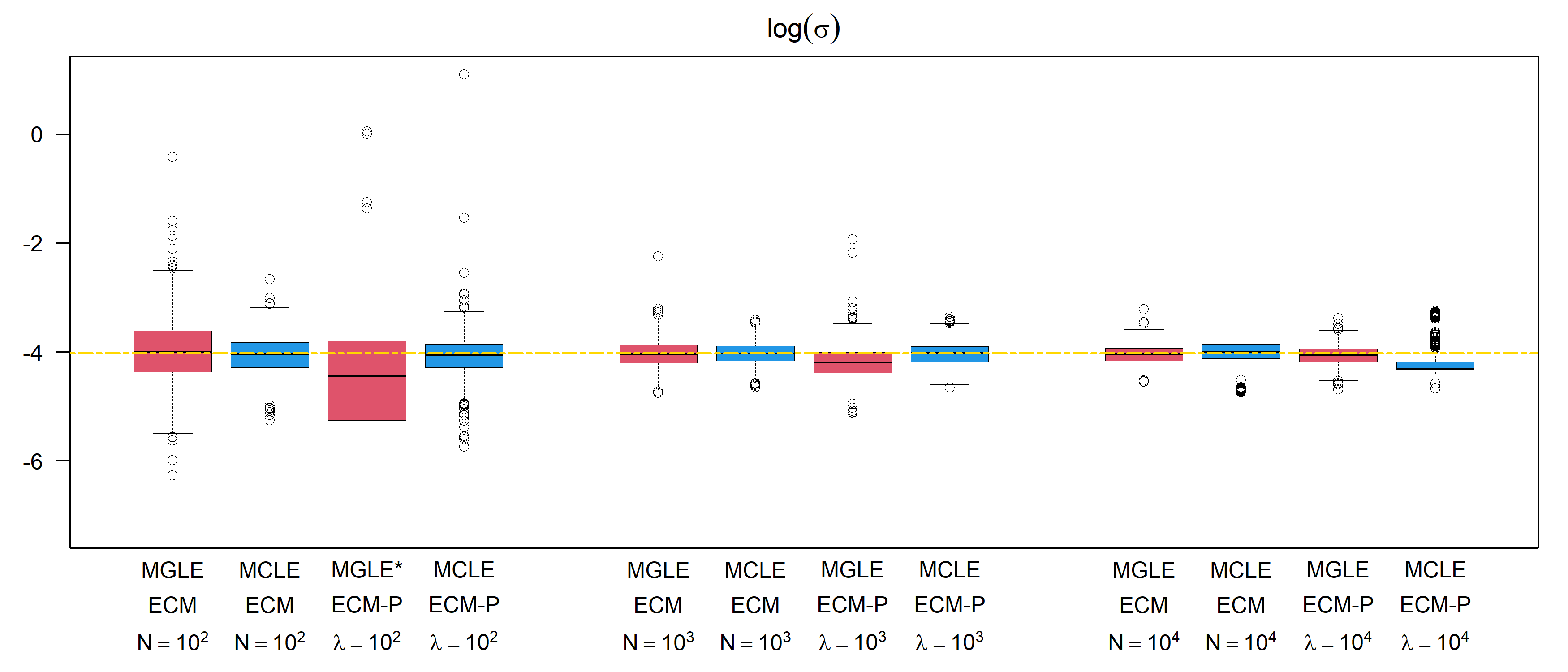}
  \end{subfigure}

  \begin{subfigure}[h]{\textwidth}
  \centering
    \includegraphics[  width = \textwidth , height = 0.23\textheight]{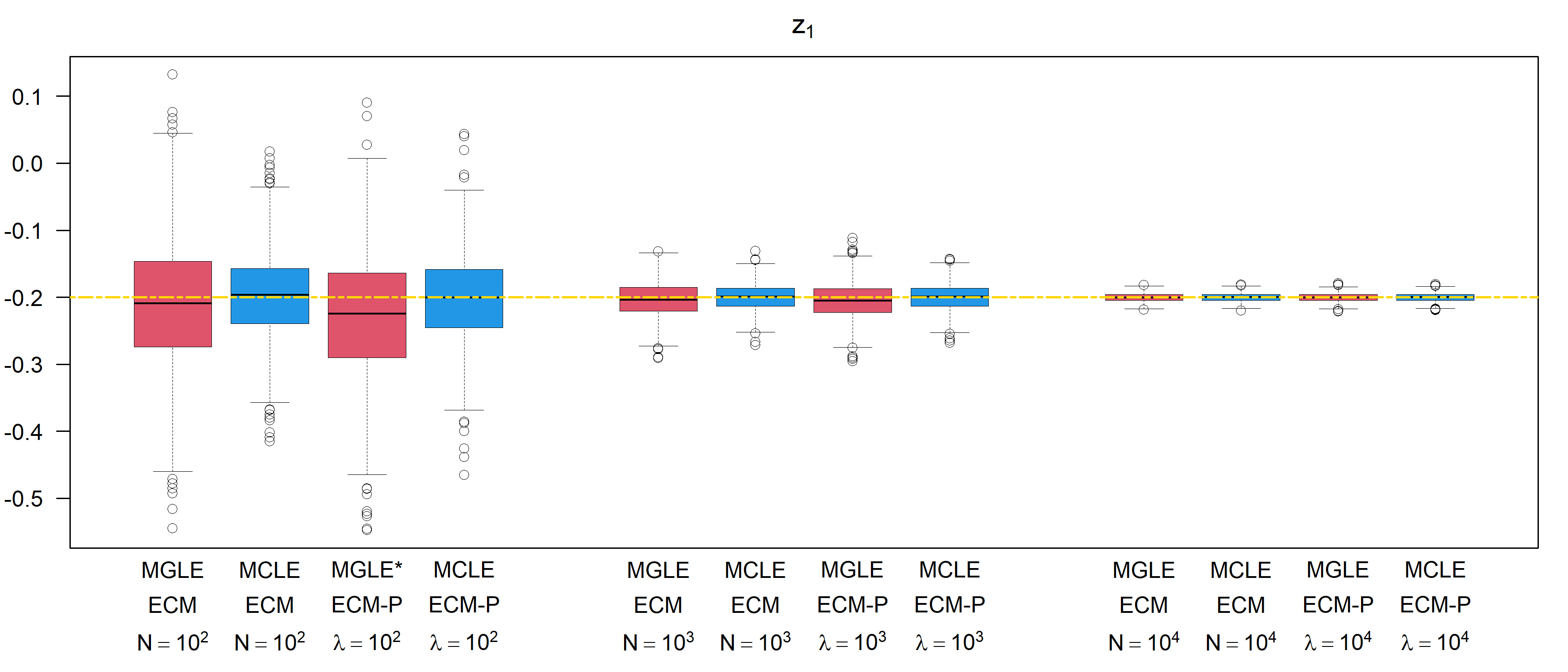}
  \end{subfigure}

  \begin{subfigure}[h]{\textwidth}
  \centering
    \includegraphics[ width = \textwidth , height = 0.23\textheight]{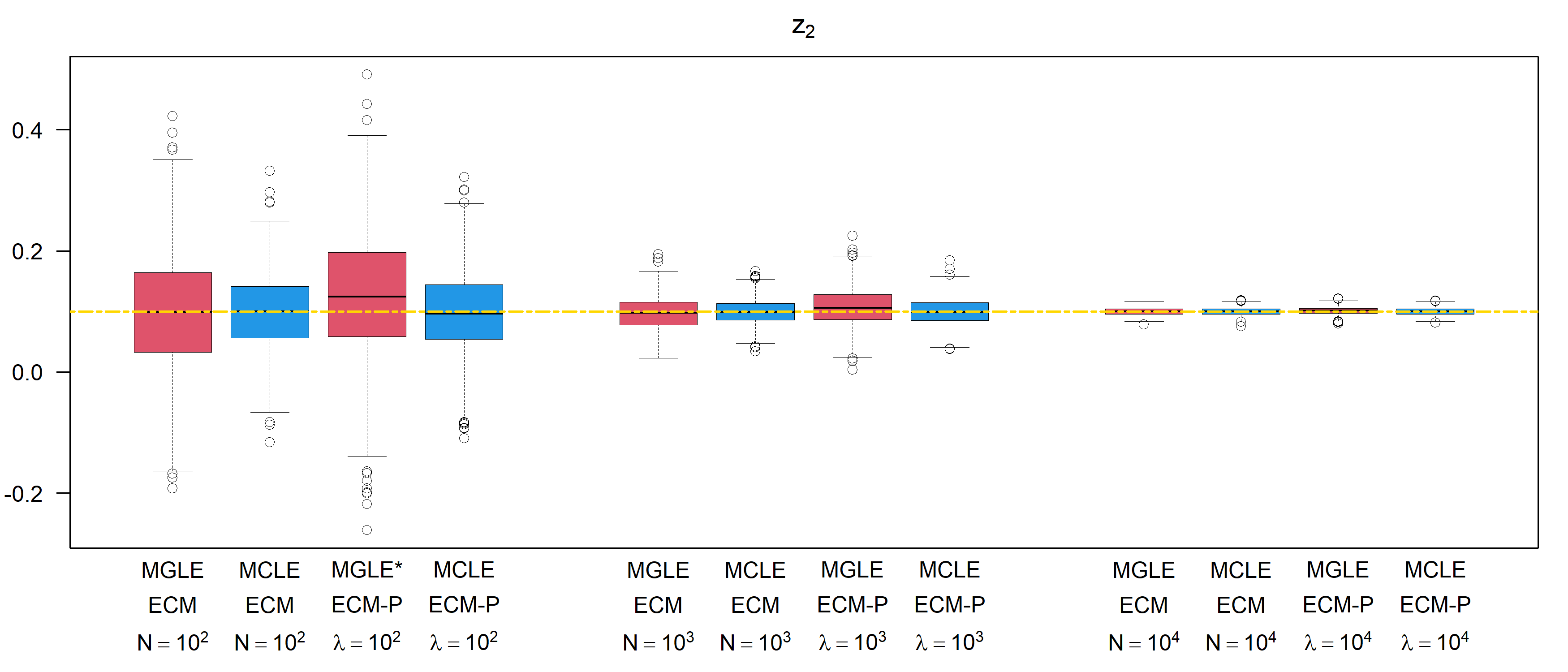}
  \end{subfigure}

  \caption{Estimates of steady-state OU parameters $(\tau,\sigma,z_{1},z_{2})$ in simulation study, with variable quantity of individuals $N$ and size rate $\lambda$ for ECM and ECM-Poisson (ECM-P) simulations respectively. Red colored box-plots correspond to MGLE, blue colored ones to MCLE. True parameter value indicated with a doted gold line. Sample size $1000$ in every setting (except MGLE ECM-P $\lambda = 10^{2}$, see Appendix \ref{App:DetailsSimu}).}
  \label{Fig:ResultsSimuStudy}
\end{figure}

\begin{figure}[btp]
  \centering

\begin{subfigure}[h]{\textwidth}
\centering
    \includegraphics[  width = 0.85\textwidth , height = 0.35\textheight]{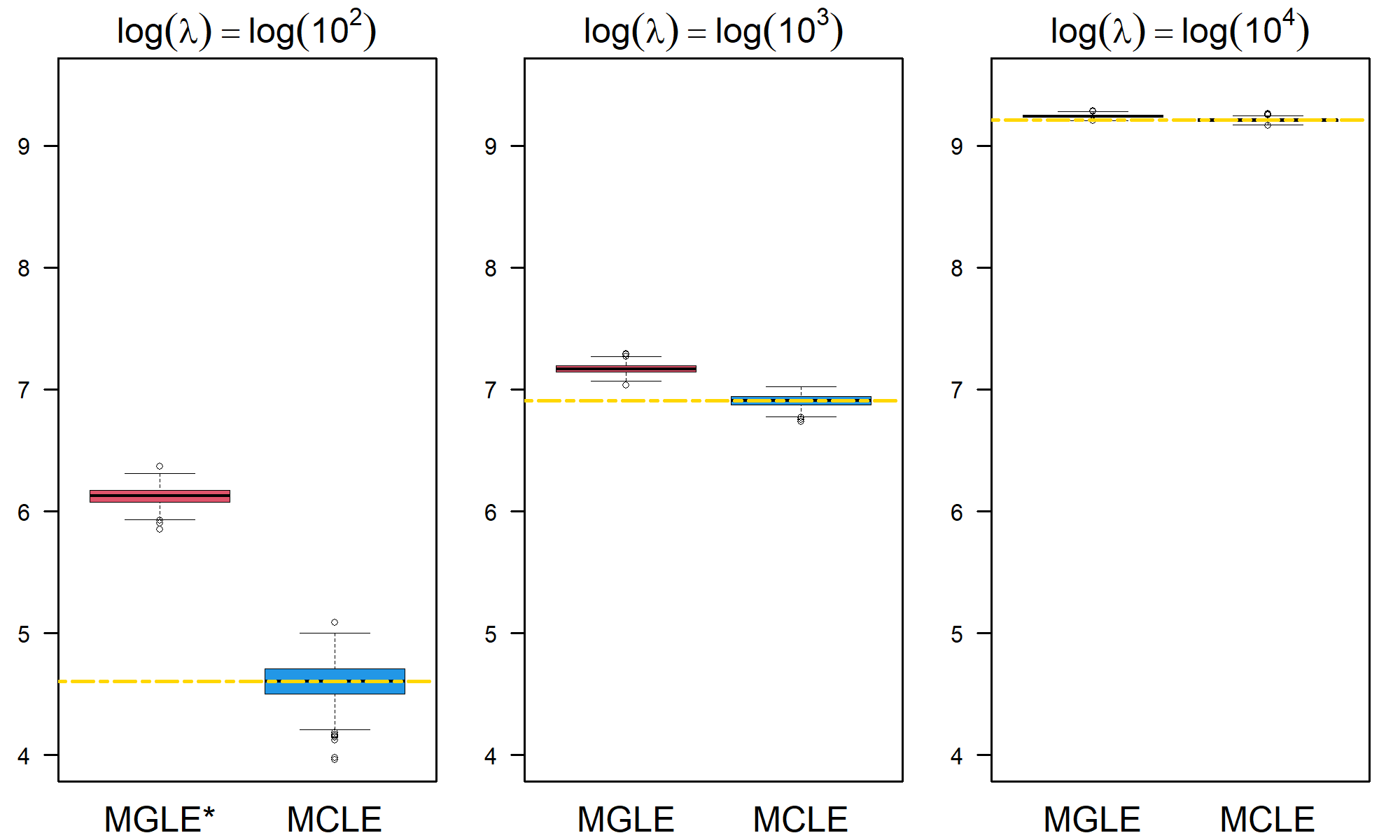}
  \end{subfigure}

  \caption{Estimations of size rates $\lambda$ in ECM-Poisson simulation studies with steady-state OU underlying movement. Red colored box-plots correspond to MGLE, blue colored ones to MCLE. True parameter value indicated with a doted gol line. Sample size $1000$ in every setting (except MGLE $\lambda = 10^{2}$, see Appendix \ref{App:DetailsSimu}).}
  \label{Fig:ResultsSimuStudyLambda}
\end{figure}

We consider a toy space-time setting in $2D$. We set $t_{0}=0$ and we consider $n=10$ count times taken randomly uniformly on the interval $[0,10]$. At each time, a random quantity of between $10$ and $50$ disjoint squares of length $\Delta x = 0.1$, randomly placed in the domain $[-1,1]^{2}$ are used as survey areas, see Figure \ref{Fig:ToySetting2D} in Appendix \ref{App:DetailsSimu}. We simulate steady-state OU moving individuals with parameters $\tau = 0.4, \sigma = 0.4\sqrt{0.002} \approx 0.0179$ (that is $\theta = 0.001$), and $z = (-0.2 , 0.1)$ and we obtain the count fields. We simulate $1000$ data sets per setting and estimate parameters using MGLE and MCLE. We study asymptotics for the estimates as the quantity of individuals grows, with $N = 10^{2},10^{3}, 10^{4}$ for ECM and $\lambda = 10^{2},10^{3}, 10^{4}$ for ECM-Poisson.  The estimations are done in the $\log$ scale for $\tau, \sigma$ and $\lambda$. Table \ref{tab:ResultsSimuStudyOU} contains a bias analysis of the results. Figures \ref{Fig:ResultsSimuStudy} and  \ref{Fig:ResultsSimuStudyLambda} contain Box-plots of the estimations of movement parameters and size rates (in the ECM-Poisson case) respectively. Estimates' correlograms are presented in Appendix \ref{App:DetailsSimu}.

\begin{table}[h!]
\centering
\resizebox{\textwidth}{!}{
\begin{tabular}{|c||c||c||c|c|c|c|c|}
\hline
\textbf{Distribution} & \textbf{Method} & \textbf{Size} & \makecell{ $\bm{\log(\hat{\tau})}$ \\ $\log(\tau_{theo}) \approx -0.916$ } & \makecell{ $\bm{\log(\hat{\sigma})}$ \\ $\log(\sigma_{theo}) \approx -4.024$   }   & \makecell{$\bm{\hat{z}_{1}}$ \\ $z_{1,theo} = -0.2$} & \makecell{$\bm{\hat{z}_{2}}$ \\ $z_{2,theo} = 0.1$} & $\bm{\log(\hat{\lambda})}$ \\
\hline\hline
\multirow{2}{*}{\makecell{\vspace{-3ex}  ECM}} & MGLE & \multirow{2}{*}{\makecell{\vspace{-3ex} $N=10^2$}} &
\makecell{$-1.053\ (-0.137)$\\\textit{0.196}} & 
\makecell{$-4.010\ (0.013)$\\\textit{0.617}} & 
\makecell{$-0.210\ (-0.010)$\\\textit{0.100}} & 
\makecell{$0.099\ (-0.001)$\\\textit{0.099}} & 
\makecell{$-$\\} \\
\cline{2-2} \cline{4-8}
 & MCLE & &
\makecell{$-0.934\ (-0.018)$\\\textit{0.092}} & 
\makecell{$-4.063\ (-0.039)$\\\textit{0.352}} & 
\makecell{$-0.198\ (0.002)$\\\textit{0.065}} & 
\makecell{$0.098\ (-0.002)$\\\textit{0.063}} & 
\makecell{$-$\\} \\
\hline
\multirow{2}{*}{\makecell{\vspace{-3ex}  ECM-P}} & MGLE* & \multirow{2}{*}{\vspace{-3ex}\makecell{ $\lambda_{theo}=10^2$ \\ $\log(\lambda_{theo}) \approx 4.605$ }} &
\makecell{$-1.315\ (-0.399)$\\\textit{0.409}} & 
\makecell{$-4.488\ (-0.464)$\\\textit{1.184}} & 
\makecell{$-0.226\ (-0.026)$\\\textit{0.102}} & 
\makecell{$0.125\ (0.025)$\\\textit{0.110}} & 
\makecell{$6.126\ (1.521)$\\\textit{1.523}} \\
\cline{2-2} \cline{4-8}
 & MCLE & &
\makecell{$-0.930\ (-0.014)$\\\textit{0.093}} & 
\makecell{$-4.075\ (0.009)$\\\textit{0.406}} & 
\makecell{$-0.201\ (-0.001)$\\\textit{0.066}} & 
\makecell{$0.098\ (-0.002)$\\\textit{0.067}} & 
\makecell{$4.599\ (-0.006)$\\\textit{0.151}} \\
\hline\hline
\multirow{2}{*}{\makecell{\vspace{-3ex}  ECM}} & MGLE & \multirow{2}{*}{\makecell{\vspace{-3ex} $N=10^3$}} &
\makecell{$-0.933\ (-0.017)$\\\textit{0.046}} & 
\makecell{$-4.043\ (-0.020)$\\\textit{0.259}} & 
\makecell{$-0.203\ (-0.003)$\\\textit{0.025}} & 
\makecell{$0.098\ (-0.002)$\\\textit{0.027}} & 
\makecell{$-$\\} \\
\cline{2-2} \cline{4-8}
 & MCLE & &
\makecell{$-0.918\ (-0.002)$\\\textit{0.028}} & 
\makecell{$-4.035\ (-0.011)$\\\textit{0.203}} & 
\makecell{$-0.201\ (-0.001)$\\\textit{0.020}} & 
\makecell{$0.099\ (-0.001)$\\\textit{0.021}} & 
\makecell{$-$\\} \\
\hline
\multirow{2}{*}{\makecell{\vspace{-3ex}  ECM-P}} & MGLE & \multirow{2}{*}{\vspace{-3ex}\makecell{ $\lambda_{theo}=10^3$ \\ $\log(\lambda_{theo}) \approx 6.908$ }} &
\makecell{$-0.949\ (-0.032)$\\\textit{0.056}} & 
\makecell{$-4.194\ (-0.171)$\\\textit{0.344}} & 
\makecell{$-0.205\ (-0.005)$\\\textit{0.028}} & 
\makecell{$0.107\ (0.007)$\\\textit{0.032}} & 
\makecell{$7.169\ (0.262)$\\\textit{0.264}} \\
\cline{2-2} \cline{4-8}
 & MCLE & &
\makecell{$-0.918\ (-0.001)$\\\textit{0.028}} & 
\makecell{$-4.033\ (-0.010)$\\\textit{0.213}} & 
\makecell{$-0.200\ (0.000)$\\\textit{0.020}} & 
\makecell{$0.100\ (0.000)$\\\textit{0.021}} & 
\makecell{$6.906\ (-0.001)$\\\textit{0.046}} \\
\hline\hline
\multirow{2}{*}{\makecell{\vspace{-3ex}  ECM}} & MGLE & \multirow{2}{*}{\makecell{\vspace{-3ex} $N=10^4$}} &
\makecell{$-0.917\ (-0.001)$\\\textit{0.010}} & 
\makecell{$-4.044\ (-0.020)$\\\textit{0.175}} & 
\makecell{$-0.200\ (-0.000)$\\\textit{0.006}} & 
\makecell{$0.100\ (-0.000)$\\\textit{0.006}} & 
\makecell{$-$\\} \\
\cline{2-2} \cline{4-8}
 & MCLE & &
\makecell{$-0.916\ (-0.000)$\\\textit{0.009}} & 
\makecell{$-4.020\ (-0.010)$\\\textit{0.266}} & 
\makecell{$-0.200\ (0.000)$\\\textit{0.006}} & 
\makecell{$0.100\ (-0.000)$\\\textit{0.006}} & 
\makecell{$-$\\} \\
\hline
\multirow{2}{*}{\makecell{\vspace{-3ex}  ECM-P}} & MGLE & \multirow{2}{*}{\vspace{-3ex}\makecell{ $\lambda_{theo}=10^4$ \\ $\log(\lambda_{theo}) \approx 9.210$ }} &
\makecell{$-0.915\ (-0.001)$\\\textit{0.010}} & 
\makecell{$-4.069\ (-0.045)$\\\textit{0.182}} & 
\makecell{$-0.200\ (-0.000)$\\\textit{0.006}} & 
\makecell{$0.101\ (0.001)$\\\textit{0.007}} & 
\makecell{$9.242\ (0.032)$\\\textit{0.035}} \\
\cline{2-2} \cline{4-8}
 & MCLE & &
\makecell{$-0.916\ (-0.000)$\\\textit{0.009}} & 
\makecell{$-4.137\ (-0.114)$\\\textit{0.379}} & 
\makecell{$-0.200\ (0.000)$\\\textit{0.006}} & 
\makecell{$0.100\ (-0.000)$\\\textit{0.006}} & 
\makecell{$9.211\ (0.001)$\\\textit{0.014}} \\
\hline
\end{tabular}
}
\caption{Results from simulation study to quantify bias in MGLE and MCLE of parameters of ECM and ECM-Poisson (ECM-P) counts induced by a steady-state OU movement process. The average of the estimates is presented for each parameter. In parentheses, the difference of this average with respect to the theoretical value. Down in italics, the square root of the mean squared error. Sample size is $1000$ in every setting (except for MGLE ECM-P $\lambda = 10^{2}$, see Appendix \ref{App:DetailsSimu}).}
\label{tab:ResultsSimuStudyOU}
\end{table}

Consistency is empirically supported: both MGLE and MCLE provide estimates which get closer to the true values as $N$ or $\lambda$ grow. This includes the estimate of $\lambda$ itself in the ECM-Poisson case and the autocorrelation parameter $\sigma$, which seems to be the most difficult to estimate in terms of reducing variance when adding extra information. In general, MCLE has less bias and variance than MGLE, especially for $\tau$ and $\lambda$, where the former is under- and the latter is overestimated.  MCLE presents few correlation among estimates. The Gaussian approximation in MGLE seems to be too crude for lower sizes, specially in the ECM-Poisson case for which the estimation was highly erratic, see more comments on this in Appendix \ref{App:DetailsSimu}. For $N,\lambda \geq 10^{3}$ both methods provide reasonable estimates for the movement parameters. For $N,\lambda = 10^{4}$ both are essentially unbiased and show similar variability, although with some undesired behavior for MCLE in the ECM-Poisson case, supposedly due to a multi-modal estimating function. While in general the knowledge of $N$ helps providing better estimates,
for $N,\lambda \geq 10^{3}$ the bias and variance are not importantly smaller when knowing $N$: as long as $N$ is big enough, the precise knowledge of $N$ is not required for a reasonable estimation. In summary: consistent estimation of movement parameters, including $\sigma$, is possible given the correct space-time count setting and with enough individuals; and MCLE performs better than MGLE in terms of bias and variability, but for large amount of individuals, MGLE provides adequate estimates with a less erratic and less computationally expensive estimating function.

\subsubsection{Illustration 2: are translocated lesser prairie-chickens behaving exploratory or sedentary?}
\label{Sec:AppChickens}

\begin{figure}[ ]
    \centering  \hspace{-0.5cm} \includegraphics[scale = 0.82]{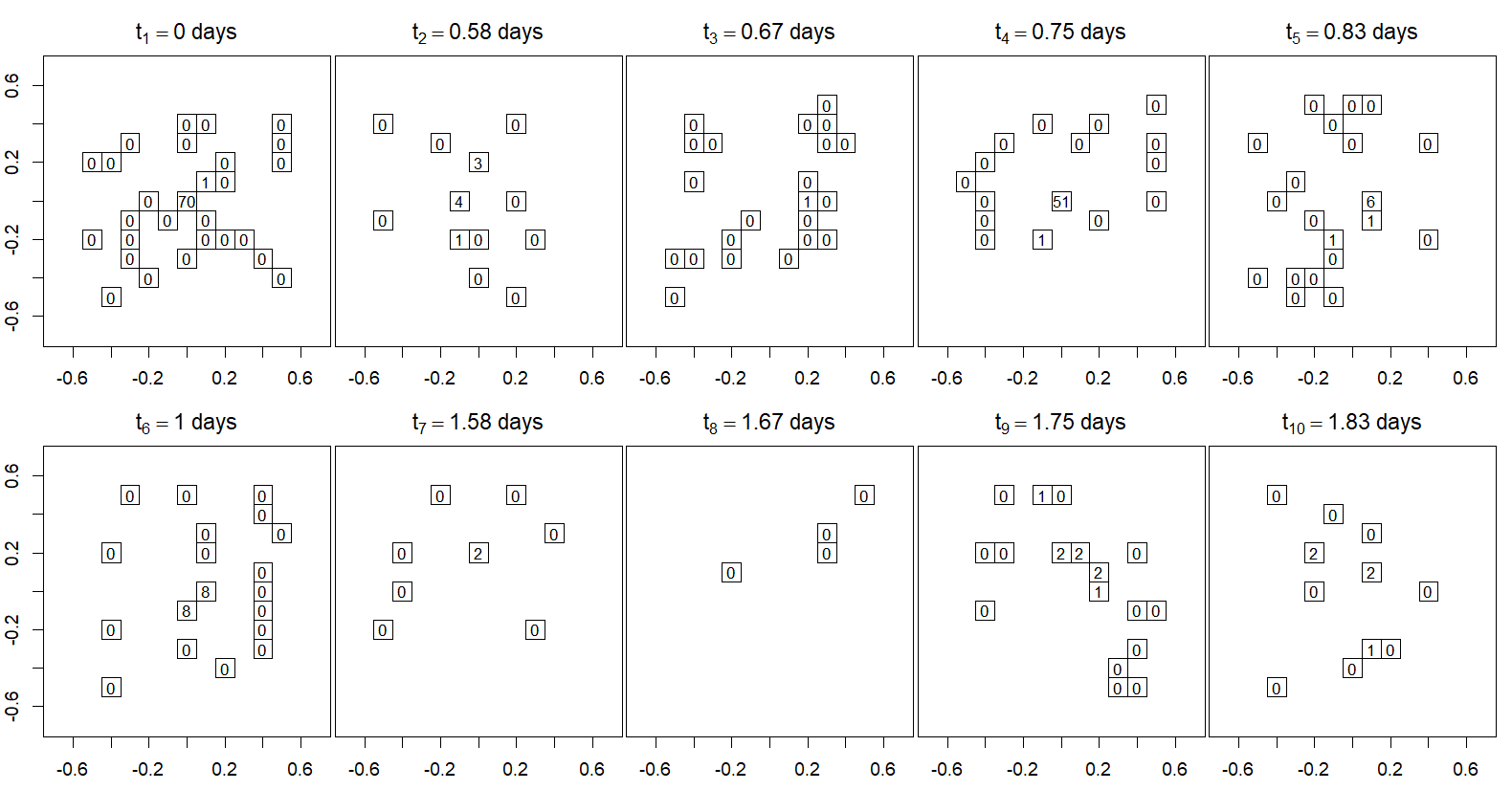}
    \caption{Lesser prairie-chicken abundance data constructed from telemetry data. Each plot shows the survey sub-squares of size $\Delta x = 0.1$ inside the domain $[-0.6,0.6]^{2}$ where the underlying moving individuals are counted at a given survey time $t_{k}$, $k=1,\ldots , 10$. Inside each sub-square, a number indicates the quantity of individuals there counted.}
\label{Fig:ChickensCountData}
\end{figure}

We consider data on $N=93$ lesser prairie-chickens, an endangered species in the United States. For conservation and study purposes, the individuals were translocated from northwest Kansas to southeastern Colorado and southwestern Kansas in 2016–2019 and telemetry devices were attached to each bird \citep{berigan2024lesser}. To show that our methodology allows to infer movement parameters even under loss of information, we transformed the tracking data into counts to be modeled with the ECM distribution. Although individuals were released at different places and on different days, the telemetry data were translated in space-time so all individuals start at an origin $x_{0} = (0,0)$ and at a common release time $t_{0}$. Due to information loss, $t_{0}$ is unknown and must be estimated. The individuals share $n=10$ location-record times, which are thus used as survey times, with $t_{1} = 0$ and the other $9$ times forming an irregular grid beneath a two-day horizon after $t_{1}$. The recorded positions are also scaled so they are contained in the domain $[-1,1]^{2}$. At each survey time, we emulate a count in $1$ to $30$ randomly selected squares of length $\Delta x = 0.1$ inside  $[-0.6 , 0.6]^2$. Since a central square around $x_{0}$ likely provides valuable information, we added such a square as survey count area for times $t_{1}$ and $t_{4}$. See Figure \ref{Fig:ChickensCountData} for the finally used count data.

One important biological question is whether this released population has an exploratory or sedentary movement behavior. Bird populations usually show a mixture of behaviors among individuals, sometimes related to sex \citep{berigan2024lesser}. Here we use the ECM distribution for inferring both movement parameters and the proportion of the population with exploratory movement type. A sedentary movement is modeled as a (non-steady state) OU process around $x_{0}$, with $\tau, \sigma$ to be estimated. The exploratory movement is modeled as a Brownian motion with standard deviation $\sigma$. Note that for a Brownian motion, formula \eqref{Eq:OUQuadraticvariationSigma} also holds, therefore it makes sense to interpret $\sigma$ as a ``speed'' parameter for either movement behavior. We introduce the parameter $\alpha \in [0,1]$, the probability of a randomly selected individual to be an explorer. The 
``trajectory'' model $X$ is therefore a mixture:
\begin{equation}
\label{Eq:XMixtureOUBrown}
    X \sim \begin{cases}
        X_{B} := Brown(\sigma), \ X_{B}(t_{0}) = x_{0} & \hbox{ with probability } \alpha \\
        X_{O} := OU( \tau , \sigma ), \ X_{O}(t_{0}) = x_{0} & \hbox{ with probability } 1 - \alpha.
    \end{cases}
\end{equation}
By conditioning, one obtains that the one-time and two-times marginals of the overall mixed trajectory model $X$ is given by the corresponding linear combination of the marginals associated to each movement:
\begin{equation}
\label{Eq:OneTimeMarginalMixture}
    \mathbb{P}\left( X(t) \in A  \right) = \alpha \mathbb{P}\left(X_{B}(t) \in A \right) + (1-\alpha) \mathbb{P}\left( X_{O}(t) \in A \right),
\end{equation}
\begin{equation}
\label{Eq:TwoTimesMarginalMixture}
        \mathbb{P}\left( X(t) \in A  , X(s) \in B \right) = \alpha \mathbb{P}\left( X_{B}(t) \in A \ , \ X_{B}(s) \in B  \right) + (1-\alpha) \mathbb{P}\left(  X_{O}(t) \in A \ , \ X_{O}(s) \in B \right).
\end{equation}
The one-time and two-times probabilities of the generated ECM counts then follow.

\begin{figure}[ht]
    \centering
        \includegraphics[width = \textwidth]{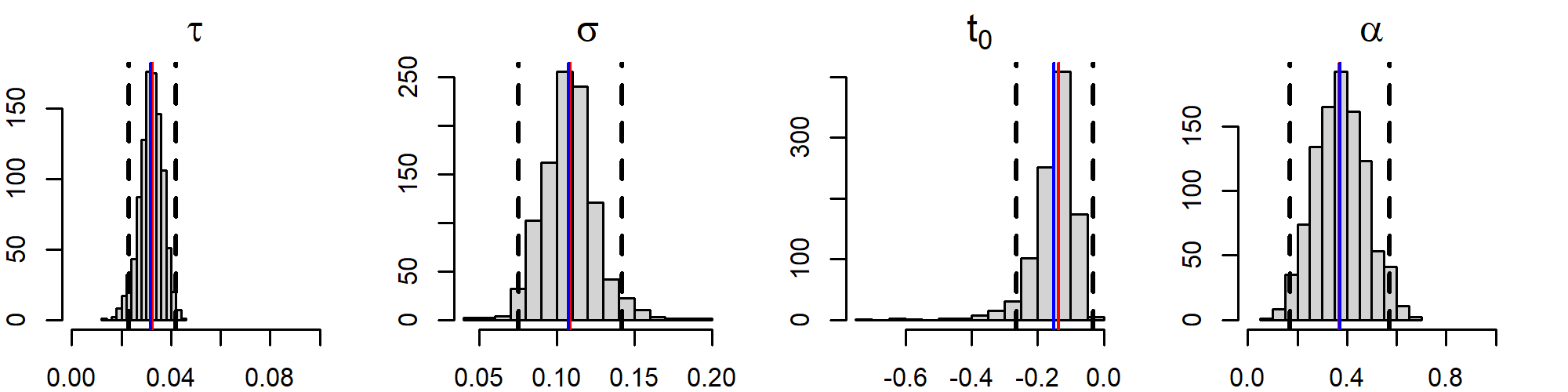} 
        \caption{Parametric bootstrap frequency histograms for estimates of $(\tau,\sigma,t_{0},\alpha)$ with $1000$ samples. Point estimates in red, sample mean in blue, $95\%$ confidence interval limits in dotted black.}
\label{Fig:ChickensHistograms}
\end{figure}

\begin{figure}[ ]
    \centering  \hspace{-0.5cm} \includegraphics[scale = 0.55]{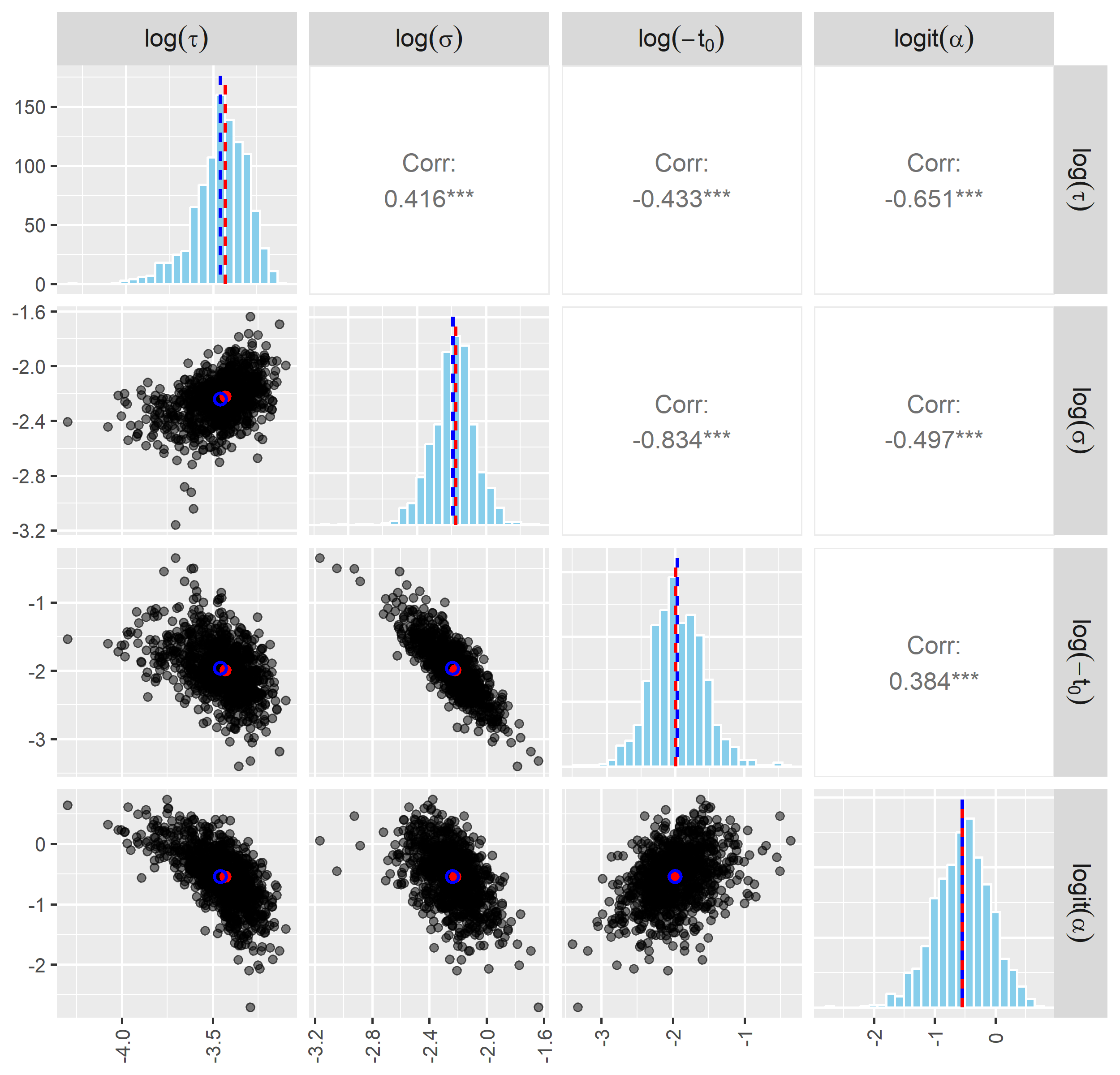}
    \caption{Correlogram of bootstrap samples of estimates of lesser prairie-chicken movement parameters $(\tau,\sigma,t_{0},\alpha)$. In histograms, true value marked with a dotted red, sample mean marked with dotted blue. In point clouds, true pair of values marked with a red point, sample mean marked with an empty-interior blue circle. Sample size of $1000$.}
\label{Fig:ChickensCorrelogram}
\end{figure}

From the simulation studies in Section \ref{Sec:SimuStudyOU}, we know that MCLE gives more adequate estimates than MGLE when $N \approx 10^{2}$, so we apply it to our case with $N=93$. We estimate the movement parameters $\sigma, \tau$, the initial time $t_{0}$ and the probability $\alpha$. Table \ref{tab:ChickenEstimates} presents the point estimates and $95\%$ confidence intervals obtained through parametric bootstrapping. Bootstrapped sampling histograms are presented in Figure \ref{Fig:ChickensHistograms}, and \ref{Fig:ChickensCorrelogram} contains the associated correlograms (with $\sigma,\tau,-t_{0}$ in  $\log$-scale, and $\alpha$ in $\logit$-scale).

\begin{table}[ht]
    \centering
    \resizebox{\textwidth}{!}{%
   \begin{tabular}{ |c||c|c| 
 c |c|c|c|c|  }
 \hline
 \multicolumn{6}{|c|}{Pairwise MCLE results} \\
 \hline
  &\begin{tabular}{@{}c@{}}   $\hat{\tau}$ \\ (dist-unit) 
 \end{tabular} & \begin{tabular}{@{}c@{}}   $\hat{\sigma}$ \\ (dist-unit/$\hbox{days}^{\frac{1}{2}}$) 
 \end{tabular} &  \begin{tabular}{@{}c@{}}   $\hat{t}_{0}$ \\ (days) 
 \end{tabular} & $\hat{\alpha}$ & $c\ell $  \\
 \hline
 Estimate & $0.0324 $ & $0.1085$ & $-0.1362$&  $0.3682 $  & $-8524.08$   \\
 $95\%$ Bootstrap CI   & $[0.0228 \ , \  0.0419]$ & $[ 0.0752 \ , \   0.1419]$ & $[-0.2658 \ , \  -0.0333]$ & $[ 0.1696 \ , \  0.5697 ]$ &  $-$  \\
 \hline
\end{tabular}
}
\caption{
Point estimates and $95\%$ confidence intervals (in brackets) for lesser prairie-chickens movement parameters, obtained with MCLE. Confidence intervals obtained from parametric bootstrapping with $1000$ samples. Composite likelihood at the estimate is shown in column  $c\ell$.}
\label{tab:ChickenEstimates}
\end{table}

The exploratory probability was estimated at $\hat{\alpha} \approx 0.37 $. There is some uncertainty on it, but nearly $90\%$ of bootstrap samples are less than $0.5$, suggesting a sedentary tendance during the initial two days. Parameters $\tau, \sigma, t_{0}$ are estimated with adequate uncertainty, specially $\tau$, whose low estimate suggests that the sedentary population tends to stay close to the release point. Correlograms \ref{Fig:ChickensCorrelogram} show strong correlations among estimates, suggesting that a suitable re-parametrization may be better for statistical purposes and that the knowledge of $t_{0}$ would further improve the estimation of other parameters.

\subsection{Estimating vote transfer}
\label{Sec:VoteTransfer}

One-final-option elections are usually held in at least two rounds. In the first, many options are available, while in the second only those candidates who satisfied a certain criterion in the first round can compete. One question that arises is if it is possible to infer, using only the vote counts from both rounds, the \textit{vote transfer}, that is, the proportions of voters preferring the different options in the second round given their preferences in the first. The problem of finding patterns of individual behavior given aggregate counts is known in sociology as the ecological inference problem \citep{schuessler1999ecological}. Many techniques have been developed for such problem. Commonly used methods are ecological regression  \citep{goodman1953ecological,goodman1959some}, where the objective proportions are linear regression coefficients, and King's method \citep{king2013solution}, where a truncated Gaussian prior distribution on the proportions is used for Bayesian analysis. A variation of King's method \citep{rosen2001bayesian,king2004ecological} appears to be the most used, and it has already been applied for inferring vote transfer \citep{audemard2024understanding}. Here we explain how to use the ECM distribution for this aim, providing a different methodology from the previously mentioned.\footnote{\textit{Practically} speaking, our methodology is an ecological regression where the variances of the residuals also depend on the parameters to estimate.}

We consider $n=2$ times representing the two election rounds. The categories at each time correspond to the possible options the voters have. Suppose there are $\tilde{m}_{1}$ candidates in the first round, $\tilde{m}_{2} = 2$ in the second, and that abstention is possible during both rounds, resulting in $m_{1} = \tilde{m}_{1}+1$ categories at time $1$ and $m_{2} = 3$ at time $2$. Assume there are $n_{D}$ electoral districts and that in each district $j \in \lbrace 1 , \ldots , n_{D} \rbrace$, there is a known number of voters, denoted $N_{j}$. Since we focus on vote transfer, one-time probabilities at first round are not important, the focus being on the conditional probabilities $p_{l'\mid l }^{(2 \mid 1)}$ of transition from one option to another (Eq. \eqref{Eq:ConditionalPathProbs}). Our main simplifying assumption is that in each district we have independent ECM vote counts with identical transition probabilities, and our objective is to estimate them. Since $\sum_{l'=1}^{3}p_{l'\mid l}^{(2 \mid 1)} = 1$, there are $2(\tilde{m}_{1}+1)$ different conditional probabilities to estimate. The independent replicates assumption makes this inference possible as long as $n_{D}$ is large enough.

Let us denote $\prescript{}{j}{\bm{Q}} = ( \prescript{}{j}{\vec{Q}}^{(k)}   )_{k=1,2}$ the ECM random arrangement containing the counts in district $j$. From Proposition \ref{Prop:TwoTimesDistQ}, at each district  the second-round count $\prescript{}{j}{\vec{Q}}^{(2)}$
is Poisson-Multinomial when conditioned on the first-round results $\prescript{}{j}{\vec{Q}}^{(1)}$. The log-likelihood of the transition probabilities $( p_{l' \mid l }^{(2 \mid 1)} )_{l',l}$ given a possible national-level result $\bm{q} = (  \prescript{}{j}{\bm{q}} )_{j=1,\ldots , n_{D}} = \left( ( \prescript{}{j}{\vec{q}}^{(k)}   )_{k=1,2} \right)_{j=1,\ldots , n_{D}}$ is given by
\begin{equation}
\label{Eq:LogLikeVoteTransfer}
    \ell\left(  (p_{l' \mid l }^{(2 \mid 1)})_{l',l }  \ \mid \ (\prescript{}{j}{\bm{q}} )_{j} \right) = \sum_{j=1}^{n_{D}}\ell_{SIM(m_{1} \ , \ \prescript{}{j}{\vec{q}^{(1)}} )}\left(  (p_{l' \mid l }^{(2 \mid 1)})_{l',l} \mid \prescript{}{j}{\vec{q}^{(2)}} \right),
\end{equation}
where $\ell_{SIM(m_{1} \ , \  \prescript{}{j}{\vec{q}^{(1)}} )}$ denotes the log-likelihood of the sum of $m_{1}$ independent multinomials with sizes given by the components of the vector $\prescript{}{j}{\vec{q}^{(1)}}$.

The Poisson-Multinomial likelihood has already been used in ecological inference problems in the context of predicting machine failure \citep{lin2023computing}. In our context, the explicit computation of the log-likelihood requires the evaluation of $m_{1}$ convolutions of dimension $2$, which is tractable for reasonable values of $m_{1}$. However, since the number of voters per district is usually large in national elections, we assume MGLE is adequate for estimation purposes. Hence, we replace the log-likelihood \eqref{Eq:LogLikeVoteTransfer} with a Gaussian pseudo-likelihood with corresponding mean and covariance.

\subsubsection{Illustration: Chilean 2021 presidential election}
\label{Sec:Chile}

The first round of the 2021 Chilean presidential election comprised $7$ candidates ranging from far-left to far-right. Since no candidate obtained $50\% + 1$ of valid votes in the first round, a second round was performed opposing left-winged candidate Gabriel Boric with far-right-winged José Kast. Two main questions arise in this context: did the vote transfer occur according to the usual left/right classification? Did the participation increase favor Boric, who later won the second round with $55.87\%$ of valid votes?  See Table \ref{tab:ResultsElectionChile2021} for the election results. The vote count information is publicly available in the Chilean election qualifying court's website.\footnote{See \url{https://tribunalcalificador.cl/resultados_de_elecciones/}, consulted for the last time the 1st May 2025.}

\begin{table}[ht]
\centering
\begin{tabular}{|l|l|c|c|}
\hline
\textbf{Option}              & \textbf{Political Tendance} & \textbf{Results 1st round} & \textbf{Results 2nd round} \\ \hline
\textbf{José Kast}              & Far-Right                   &        $ 1\ 961\ 779  $ ($  13.05\% $) &            $ 3\ 650\ 662 $  ($24.29\% $)  \\ \hline
\textbf{Sebastián Sichel}       & Center-Right                &        $898\ 635$  ($ 5.98\% $)              &                            \\ \hline
\textbf{Franco Parisi}          & Dissident/Right      &        $900\ 064$  ($ 5.99\% $)               &                            \\ \hline
\textbf{Yasna Provoste}         & Center-Left                 &        $815\ 563$   ($ 5.43\% $)     &                            \\ \hline
\textbf{Gabriel Boric}          & Left                        &        $1\ 815\ 024$   ($ 12.08\% $)    &        $ 4\ 621\ 231 $ ($ 30.74\% $) \\ \hline
\textbf{Marco Enríquez-Ominami} & Dissident/Left              &        $534\ 383$  ($ 3.56\% $) &                            \\ \hline
\textbf{Eduardo Artés}          & Far-Left                    &        $102\ 897$  ($ 0.68\% $)  &                            \\ \hline
\textbf{Abstention (incl. null/blank)}             &                             &                         $8\ 002\ 629$  ($ 53.24\% $)   &    $6\ 759\ 081$ ($ 44.97\% $)                     \\ \hline
\end{tabular}
\caption{Results of 2021 Chilean presidential election in quantity of votes (percentages in parenthesis).} 
\label{tab:ResultsElectionChile2021}
\end{table}

We applied our methodology with MGLE for inference. The number of people with right to vote is $N = 15\ 030\ 974$ for both rounds. As districts, we considered the territorial division of Chile into \textit{comunas}. Voters abroad were grouped into an extra fictional district. This resulted in $n_{D} = 347$ districts with number of voters ranging from $233$ to $403\ 129$. Null and blank votes are considered as abstention. We obtained estimates for vote transfer proportions and we quantified uncertainty with parametric bootstrapping. Results are presented in Table \ref{tab:ResultsChile}. Confidence intervals are quite narrow, presumably due to the strong and questionable iid assumption among and within districts.

\begin{table}[ht]
    \centering
    \resizebox{\textwidth}{!}{%
   \begin{tabular}{ |c||c|c|
 c |c|c|c|c|c|  }
 \hline
 \multicolumn{9}{|c|}{Vote transfer MGLE results} \\
 \hline
  &  Kast & Boric & Parisi & Sichel & Provoste & ME-O & Artés & Abstention  \\ 
  \hline 
  \hline 
  Boric &  \begin{tabular}{@{}c@{}}   $\approx 0 \% $ \\ $[0 \ , \  0.28]$\end{tabular}  & \begin{tabular}{@{}c@{}}  $\approx 100 \% $ \\ $[ 99.80 \ , \  100]$ \end{tabular}   & \begin{tabular}{@{}c@{}} $31.04 \% $ \\ $[30.41 \ , \  31.63 ]$ \end{tabular}   &\begin{tabular}{@{}c@{}}  $ 19.80 \% $ \\ $[19.21 \ , \  20.28 ]$ \end{tabular}  & \begin{tabular}{@{}c@{}} $52.13 \% $  \\ $[50.98 \ , \  53.29]$ \end{tabular}  &\begin{tabular}{@{}c@{}} $\approx 100 \% $ \\ $[99.25 \ , \  100]$ \end{tabular}  &\begin{tabular}{@{}c@{}}  $ \approx 100 \% $ \\ $[ 96.24 \ , \  100]$ \end{tabular}  & \begin{tabular}{@{}c@{}}  $ 16.34 \% $ \\ $[16.26 \ , \  16.57]$ \end{tabular}     \\
  \hline
  Kast &  \begin{tabular}{@{}c@{}}  $97.54 \% $ \\ $[ 97.15 \ , \  97.94]$ \end{tabular}    & \begin{tabular}{@{}c@{}} $\approx 0 \% $ \\ $[ 0 \ , \  0.01 ]$ \end{tabular}  &\begin{tabular}{@{}c@{}} $16.11 \% $ \\ $[15.57 \ , \  16.69]$ \end{tabular}   &\begin{tabular}{@{}c@{}}  $ 68.24 \% $ \\ $[67.55 \ , \  68.91]$ \end{tabular}  &\begin{tabular}{@{}c@{}} $ 30.18 \% $ \\ $[ 29.21 \ , \  31.18 ]$ \end{tabular}   & \begin{tabular}{@{}c@{}} $\approx 0 \% $ \\ $[ 0 \ , \  0.28 ]$ \end{tabular}   & \begin{tabular}{@{}c@{}} $\approx 0 \% $ \\ $[ 0 \ , \  2.69 ]$ \end{tabular}  & \begin{tabular}{@{}c@{}} $ 9.33 \% $  \\ $[ 9.19 \ , \  9.45 ]$ \end{tabular}  \\
  \hline
  Abstention  & \begin{tabular}{@{}c@{}} $2.47 \% $ \\ $[ 1.99 \ , \  2.84 ]$ \end{tabular} & \begin{tabular}{@{}c@{}}  $ \approx 0 \%$ \\ $[ 0 \ , \  0.17 ]$ \end{tabular} & \begin{tabular}{@{}c@{}}  $52.84 \% $ \\ $[52.16 \ , \  53.57]$ \end{tabular} & \begin{tabular}{@{}c@{}}  $ 11.96 \% $ \\ $[11.28 \ , \  12.77]$ \end{tabular} & \begin{tabular}{@{}c@{}} $17.69 \%  $ \\ $[16.42 \ , \  18.90]$ \end{tabular}   & \begin{tabular}{@{}c@{}} $ \approx 0 \% $ \\ $[0 \ , \  0.50]$ \end{tabular}  & \begin{tabular}{@{}c@{}}  $\approx 0 \% $ \\ $[0 \ , \  0.78]$ \end{tabular} & \begin{tabular}{@{}c@{}} $ 74.28 \% $  \\ $[74.12 \ , \  74.43]$ \end{tabular}  \\
  \hline
\end{tabular}
}
\caption{Point estimates and $95\%$ confidence intervals (in brackets) of vote transfer proportions during the 2021 Chilean presidential election, obtained with MGLE (in percentages). In rows, preferred option at second round. In columns, preferred option at first round. Confidence intervals obtained from parametric bootstrapping with $1000$ samples.}
    \label{tab:ResultsChile}
\end{table}

Under the assumed model, we learn, first, that the reduction in abstention  did favor Boric, but Kast also benefited from it. The most crucial vote transfer for Boric's victory may have been from Parisi voters. Parisi is commonly perceived as right or center-right \citep{quiroga2015debut}, but above all as a dissident from the traditional Chilean political system. His voters seem to have followed such tendance, with a huge abstention during the second round. But surprisingly, 
our estimated Parisi voters' support for leftist Boric was nearly twice their support for Kast. This suggests that the left/right classification of candidates may not be reasonable for classifying their voters. Results of more centrist Sichel and Provoste are also to be commented: while the vote transfer roughly respects the left/right tendance, there is a significant shift to the opposite side and to abstention in both cases.

\section{Comparison with other models of counts}
\label{Sec:Comparison}

The ECM distribution is connected to many models used in different fields. For sake of bibliographic completeness, this section is devoted to show the similarities with two widely used kinds of models: Poisson processes and Markov models.

\subsection{Count of Poisson point processes}
\label{Sec:PointProcesses}

Poisson point processes have applications in many fields \citep{ross1995stochastic,moller2003statistical,illian2008statistical}. Their simplicity stems from the sole need of modeling an intensity field, which gives the expected count over a space/time region. The counts are independent Poisson variables around such an average. The intensity can be modeled quite freely, sometimes relating it to covariates \citep{baddeley2005residual,thurman2014variable} or taken to be the solution of a suitable differential equation  \citep{wikle2003hierarchical,soubeyrand2014parameter,hefley2016hierarchical,zamberletti2022understanding}.

We can retrieve a structure of 
independent Poisson counts with the ECM-Poisson distribution as a particular scenario where the path probabilities satisfy
\begin{equation}
\label{Eq:FullPathIndependent}
    p_{l_{1},\ldots , l_{n}}^{(1,\ldots, n)} = p_{l_{1}}^{(1)} \cdots p_{l_{n}}^{(n)}.
\end{equation}
As a result, counts of space-time Poisson processes can be subsumed as a special case of this distribution. In such case, the time index $k = 1, \ldots , n$ refers to disjoint time intervals, and the categories at each $k$ represent disjoint spatial regions of count. The one-time probabilities govern the intensity.

If we apply this logic to moving individuals as in Section \ref{Sec:MovingIndividuals}, where the time index $k$ refers to a specific time point $t_{k}$, we obtain that the continuous-time stochastic processes satisfying \eqref{Eq:FullPathIndependent} for any possible choice of count times are those with independence at different times, no matter how small the time-lag. Such trajectory models have not continuous paths, thus they are not realistic for physical scenarios. As explained in Section \ref{Sec:MovingIndividuals}, the abundance random measure $\Phi$ in \eqref{Eq:DefPhit(A)Concept} is a continuous-time evolution of inhomogeneous spatial Poisson processes, but it is not a space-time point process.

One of the disadvantages of Poisson processes is the lack of autocorrelation in the count field. The most widely used point process model which includes dependence among counts is the Cox process \citep{moller1998log,diggle2010partial}. It is a Poisson process conditioned on a random intensity, which is often modeled phenomenologically in order to fit data or relate the counts to covariates, but we are unaware of a Cox-dependence modeling through movement of underlying individuals. Our approach tackles this problem by considering the natural autocorrelation induced by the movement.

In summary, ECM-related models here developed are arguably statistically richer than space-time Poisson processes. Their advantages include the explicit description of the temporal dependence between counts, which can be interpreted in a continuous-time setting, and that both intensity and space-time autocorrelation are related to individual movement, allowing to infer movement parameters which are unidentifiable in a purely intensity-based model.

\subsection{Markov models}
\label{Sec:MarkovCounts}

The ECM distribution does not assume any Markovian structure. Markov models can nonetheless be related to a particular case. Suppose the path probabilities can be written as
\begin{equation}
\label{Eq:PathProbMarkov}
    p_{l_{1},\ldots , l_{n}}^{(1,\ldots , n)} = p_{l_{n} \mid l_{n-1}}^{(n\mid n-1)}p_{l_{n-1} \mid l_{n-2}}^{(n-1\mid n-2)} \ldots p_{l_{2} \mid l_{1}}^{(2\mid 1)} p_{l_{1}}^{(1)}.
\end{equation}
In other words, the individuals follow a Markov chain across the categories. Then, the count vectors $\vec{Q}^{(1)} , \vec{Q}^{(2)} , \ldots , \vec{Q}^{(n)}$ follow a Hidden Markov Model (HMM) \citep{rabiner1986introduction,zucchini2009hidden}. The counts are not themselves Markov in general due to lumping
\citep{gurvits2005markov,kemeny1969finite}, but they are constructed from an underlying Markov chain, the vector $\vec{Q}^{(k)}$ containing information of the chain only at time $k$.

HMMs are very widely used in applied probability and are a reference in modeling counts in ecology \citep{knape2012fitting,glennie2023hidden}. Several techniques for treating the complicated likelihoods involved have been developed: the forward algorithm, the Viterbi algorithm and particular filtering techniques \citep{smyth1997probabilistic,doucet2001sequential}. Thus, the likelihood of the ECM distribution in the Markovian case \eqref{Eq:PathProbMarkov} can become tractable in some scenarios. Conversely, the fitting methods here presented may be added to the tool-box of inference methods for certain HMMs, which can also present intractable likelihoods \citep{dean2014parameter}.

One issue is still present when considering an underlying \textit{continuous} Markov process $X$ in the scenario of Section \ref{Sec:MovingIndividuals}. The Markovianity of $X$ does not imply the Markovianity of the counts. This is a form of continuous-to-discrete lumping, similar to encountered in discrete-valued Markov chains \citep{gurvits2005markov}. So, even with Markov trajectories such as Brownian motion, OU processes, or more generally Itô diffusions, the scenario is, strictly speaking, outside the scope of condition \eqref{Eq:PathProbMarkov}. Nonetheless, a Markov assumption on the path probabilities may still be used as a simplified ECM model, and the associated, more tractable likelihood, can be used as an estimating function or for Bayesian inference. The conditions in which this approximation produces adequate estimates have to be explored.

\section{Perspectives}
\label{Sec:Perspectives}

The properties of the ECM distribution here explored are modest in the sense that we only studied the minimum necessary to approach practical problems. Many additional properties such as higher-order marginals, Fisher information, Kullback-Leibler divergences with respect to distributions, or properties of diverse estimators, remain to be explored. A more theoretical study about the Gaussian approximation exploited in the MGLE estimator could be approached by using classical error-bounding results \citep{valiant2010clt,valiant2011estimating,daskalakis2015structure}. Asymptotic properties and uncertainty quantification for MGLE and MCLE may also be explored theoretically \citep{godambe1991estimating}.

One interesting question, especially in ecological applications, is the estimation of the number of individuals $N$ when it is unknown. We have shown that we can estimate $\lambda$ for the ECM-Poisson distribution if enough information is available, but an estimator of $\lambda$ is not an estimator of $N$. One could proceed studying the conditional distribution of $N$ given the information in $\bm{Q}$, which by Bayes' rule is given by
\begin{equation}
\label{Eq:NConditionalQ}
    \mathbb{P}\left( N = x \mid \bm{Q} = \bm{q} \right) = \frac{\mathbb{P}(\bm{Q} = \bm{q} \mid N = x)  \mathbb{P}\left( N = x \right) }{ \mathbb{P}\left( \bm{Q} = \bm{q} \right) }.
\end{equation}
Note that $\mathbb{P}(\bm{Q} = \bm{q} \mid N = x)$ is an ECM probability, $\mathbb{P}\left( N = x \right)$ is a Poisson probability and $ \mathbb{P}\left( \bm{Q} = \bm{q} \right)$ is an ECM-Poisson probability. Thus, one can either use point estimates for $\lambda$ and the other parameters and describe this distribution, or place a prior on $\lambda$ and derive the posterior of $N$ given $\bm{Q}$ and $\lambda$ using Bayesian inference. If $\lambda$ is big enough, the intractable probabilities in \eqref{Eq:NConditionalQ} can be approximated by Gaussian probabilities with  continuity correction. For small $\lambda$, the conditions under which the ECM and ECM-Poisson probabilities can be replaced by corresponding composite likelihoods may be studied  \citep{pauli2011bayesian,ribatet2012bayesian,syring2019calibrating}.

It is likely that future applications or variants of the ECM distribution will also present likelihood intractability. New ad-hoc fitting methods should be explored. Likelihood-free methods \citep{drovandi2022comparison} are an interesting option. Since simulating an ECM or ECM-Poisson arrangement is easy as long as the individual movement is easy to simulate, approximate Bayesian computation \citep{sisson2018handbook} can be used for Bayesian analyses, but case-specific summary statistics and discrepancies need to be chosen. Neural estimates \citep{lenzi2023neural,zammit2024neural} could also be explored, since they are versatile enough to be adapted to more complex but easily simulated scenarios such as individuals reproducing and dying or non-independent movement.

The iid assumption among individuals is clearly a strong hypothesis for biological or sociological applications. However, the ECM distribution can still be used as a basis for building more complex models. For example, counts of individuals with two or more different behaviors (such as in Section \ref{Sec:AppChickens} but \textit{knowing} the number of exploratory individuals) could be modeled as the sum of independent ECM arrangements, one for each type of movement. This avoids the identically distributed assumption, but requires to study the properties of sums of independent ECM arrangements. To avoid the independence assumption, one could try to condition on some external variable that influences individuals' behavior and then model their movement by invoking conditional independence. Then, counts are conditionally ECM, but not marginally. This also applies to the case where some individuals have a big influence on others. For example, a leader could follow a certain movement, and other individuals would follow conditionally independent movements given the leader's. This idea can be applied to either physical movement as in Section \ref{Sec:MovingIndividuals} or to sociological scenarios as in Section \ref{Sec:VoteTransfer}, in which case it would refer to opinion leader's effects.

\section*{Acknowledgments}

We warmly thank Liam A. Berigan and David Haukos for allowing us to use their data on lesser prairie-chickens. We also thank Denis Allard for his suggestions on the use of composite likelihood estimators and their application.

This research has been founded by the Swiss National Science Foundation grant number $310030\_207603$.

\begin{appendices}

\section{Proofs}
\label{App:Proofs}

\subsection{Proof of Proposition \ref{Prop:CharFunctionQ}}
\label{Proof:CharFunctionQ}

Let us define the $n$-dimensional random array $\bm{X} = ( X_{l_{1}, \ldots , l_{n}})_{l_{1},\ldots , l_{n}}$ with dimensions $m_{1}\times \ldots \times m_{n}$, in which each $X_{l_{1}, \ldots , l_{n}}$ means, for every $(l_{1} , \ldots , l_{n}) \in \{ 1 , \ldots , m_{1} \}\times \{ 1 , \ldots , m_{n}\}$,
\begin{equation}
    X_{l_{1},\ldots, l_{n}} := \hbox{``number of individuals doing the full-path  $l_{1},\ldots , l_{n}$''}.
\end{equation}
Since the $N$ individuals follow independently the same dynamics across categories, the array $\bm{X}$ can be wrapped into a random vector which follows a classical multinomial distribution, each component with success probability given by the corresponding full-path probability. The characteristic function of $\bm{X}$ is thus a multinomial one:
\begin{equation}
\label{Eq:CharX}
    \varphi_{\bm{X}}(\bm{\eta}) = \left[ \sum_{l_{1}=1}^{m_{1}} \ldots \sum_{l_{n}=1}^{m_{n}} p_{l_{1},\ldots, l_{n}}^{(1,\ldots , n)} e^{  i\eta_{l_{1},\ldots , l_{n}} }
 \right]^{N},
\end{equation}
for every consistent $n$-dimensional array of real values $\bm{\eta} = (\eta_{l_{1},\ldots , l_{n}})_{l_{1},\ldots , l_{n}}$.

Now, the variables in the ECM random arrangement $\bm{Q}$ are, by definition, the marginalization of the array $\bm{X}$ over each category and time, that is,
\begin{equation}
\label{Eq:QmarginX}
    Q_{l}^{(k)} = \sum_{\substack{l_{1},\ldots , l_{n} \\
    l_{k} = l} } X_{l_{1},\ldots , l_{n}}, \quad \forall k, l.
\end{equation}
Therefore, we can compute $\varphi_{\bm{Q}}$ from $\varphi_{\bm{X}}$. Let $\bm{\xi} = \left( \xi_{l}^{(k)} \right)_{l,k}$ be an arrangement in $\R^{m_{1} + ... + m_{n}}$. Using \eqref{Eq:QmarginX}, we have 
\begin{equation}
\begin{aligned}
    \varphi_{\bm{Q}}(\bm{\xi}) &= \mathbb{E}\left( e^{i \sum_{k=1}^{n}\sum_{l=1}^{m_{k}} Q_{l}^{(k)}  \xi_{l}^{(k)}  } \right) \\
    &= \mathbb{E}\left( e^{i\sum_{k=1}^{n}\sum_{l=1}^{m_{k}} \sum_{l_{1}=1}^{m_{1}} ... \sum_{l_{k-1}=1}^{m_{k-1}}\sum_{l_{k+1}=1}^{m_{k+1}} ... \sum_{l_{n}=1}^{m_{n}} X_{l_{1} , ... , l_{k-1} , l , l_{k} , ... , l_{n}} \xi_{l}^{(k)}  } \right).  \\
    &= \mathbb{E}\left( e^{i\sum_{l_{1}=1}^{m_{1}} ...\sum_{l_{n}=1}^{m_{n}} \left( \sum_{k=1}^{n} \xi_{l_{k}}^{(k)} \right) X_{l_{1} , ... , l_{n}}  } \right) \\
    &= \varphi_{\bm{X}}( \bm{\eta}),
    \end{aligned}
\end{equation}
where $\bm{\eta} = (\eta_{l_{1},\ldots , l_{n}})_{l_{1},\ldots , l_{n}}$ is the $n$-dimensional array given by $\eta_{l_{1},\ldots , l_{n}} = \sum_{k=1}^{n}\xi_{l_{k}}^{(k)}$. From \eqref{Eq:CharX} it follows 
\begin{equation}
     \varphi_{\bm{Q}}(\bm{\xi}) = \left[ 
 \sum_{l_{1}=1}^{m_{1}} ... \sum_{l_{n}=1}^{m_{n}} p_{l_{1} , ... , l_{n}}^{(1,...,n)} e^{i(\xi_{l_{1}}^{(1)} + ... + \xi_{l_{n}}^{(n)} )} \right]^{N}. \quad \blacksquare
\end{equation}

\subsection{Proof of Proposition \ref{Prop:OneTimeDistQ}}
\label{Proof:OneTimeDistQ}

Let $\vec{\xi} = (\xi_{1} , ... , \xi_{m_{k}} ) \in \R^{m_{k}}$. Let $\bm{\xi} = (\xi_{l}^{(k)})_{l,k}$ the arrangement in $\R^{m_{1}+\ldots+m_{n}}$ such that $\xi_{l}^{k'} = 0$ if $k' \neq k$ and $\xi^{(k)}_{l} = \xi_{l}$ for every $l \in \{ 1 , \ldots , m_{l} \}$. Then the characteristic function of the random vector $\vec{Q}^{(k)}$ satisfies
\begin{equation}
\label{Eq:CharacQkDevelopment}
\begin{aligned}
     \varphi_{\vec{Q}_{k}}(\vec{\xi}) &= \varphi_{\bm{Q}}(\bm{\xi}) \\
        &=\left[ \sum_{l_{1}=1}^{m_{1}} ... \sum_{l_{n}=1}^{m_{n}}  p_{l_{1} , ... , l_{n}}^{(1,...,n)}e^{i(\xi_{l_{1}}^{(1)} + ... + \xi_{l_{n}}^{(n)}   ) } \right]^{N} \\
        &= \left[ \sum_{l_{1}=1}^{m_{1}} ... \sum_{l_{n}=1}^{m_{n}} p_{l_{1} , ... , l_{n}}^{(1,...,n)}e^{i\xi_{l_{k}} }  \right]^{N} \\
        &= \left[ \sum_{l_{k}=1}^{m_{k}}p_{l_{k}}^{(k)} e^{i\xi_{l_{k}} } \right]^{N},
\end{aligned}
\end{equation}
which is the characteristic function of a multinomial random vector. Note that we have used marginalization for the one-time probabilities
\begin{equation}
\label{Eq:ProbMargin1Time}
    p_{l}^{(k)} = \sum_{\substack{l_{1},\ldots , l_{n} \\
    l_{k} = l}} p_{l_{1},\ldots , l_{n}}^{(1,\ldots , n)}. \quad \blacksquare
\end{equation}

\subsection{Proof of Proposition \ref{Prop:TwoTimesDistQ}}
\label{Proof:TwoTimesDistQ}

We remind that for two random vectors $\vec{X},\vec{Y}$ with values in $\R^{m}$ and $\R^{n}$ respectively, the conditioned characteristic function $\varphi_{\vec{X} | \vec{Y} = \vec{y}} $ satisfies
\begin{equation}
\label{Eq:ConnectionCharacPairConditionalDist}
\varphi_{\vec{X},\vec{Y}}(\vec{\xi},\vec{\eta}) = \int_{\R^{n}} \varphi_{\vec{X} | \vec{Y} = \vec{y}}(\vec{\xi})e^{i\vec{\eta}^{T}\vec{y}} d\mu_{\vec{Y}}(\vec{y}), \quad \vec{\xi} \in \R^{m}, \vec{\eta} \in \R^{n}, 
\end{equation}
where $\mu_{\vec{Y}}$ denotes the distribution measure of $\vec{Y}$. In our case, let $k' \neq k$. Since $\vec{Q}^{(k)}$ is multinomial (Proposition \ref{Prop:OneTimeDistQ}), we use \eqref{Eq:ConnectionCharacPairConditionalDist} and conclude 
\begin{equation}
\label{Eq:JointQkQtilde{k}UsingQkMultinomial}
\begin{aligned}
    \varphi_{\vec{Q}^{(k')} , \vec{Q}^{(k)}}(\vec{\xi} , \vec{\eta} ) &= \sum_{\substack{\vec{q} \in \lbrace 0 , ... , N \rbrace^{m_{k}} \\
    sum(\vec{q}) = N}} \varphi_{\vec{Q}^{(k')} | \vec{Q}^{(k)} = \vec{q} }(\vec{\xi} ) e^{i\vec{\eta}^{T}\vec{q}} \mathbb{P}\left( \vec{Q}^{(k)} = \vec{q} \right) \\
    &= \sum_{\substack{\vec{q} \in \lbrace 0 , ... , N \rbrace^{m_{k}} \\
    sum(\vec{q}) = N}} \varphi_{\vec{Q}^{(k')} | \vec{Q}^{(k)} = \vec{q} }(\vec{\xi} ) e^{i\vec{\eta}^{T}\vec{q}} \frac{N!}{q_{1}! ... q_{m_{k}}!} \left( p_{1}^{(k)} \right)^{q_{1}} ... \left( p_{l_{m_{k}}}^{(k)} \right)^{q_{m_{k}}}, 
    \end{aligned}
\end{equation}
for every $ \vec{\xi} = (\xi_{1} , ... , \xi_{m_{k'}} ) \in \R^{m_{k'}}$, and every $\vec{\eta} = (\eta_{1} , ... , \eta_{m_{k}} ) \in \R^{m_{k}}$. Here we denoted $sum(\vec{q}) = q_{1} + ... + q_{m_{k}}$. On the other hand,  $\varphi_{\vec{Q}^{(k')} , \vec{Q}^{(k)}}$ can be computed similarly as the marginal characteristic function $\varphi_{\vec{Q}^{(k)}}$ obtained in the proof of Proposition \ref{Prop:OneTimeDistQ}, by considering $\varphi_{\bm{Q}}$ evaluated at an arrangement with convenient null components and using marginalization for the two-times probabilities
 \begin{equation}
 \label{Eq:ProbMargin2times}
     p_{l,l'}^{(k,k')} = \sum_{\substack{l_{1},\ldots , l_{n} \\ 
     l_{k} = l , l_{k'} = l'}} p_{l_{1},\ldots , l_{n}}^{(1,\ldots , n)}.
 \end{equation}
The final result is (details are left to the reader)
\begin{equation}
\label{Eq:JointCharacQkQtilde{k}}
    \varphi_{\vec{Q}^{(k')} , \vec{Q}^{(k)}}(\vec{\xi} , \vec{\eta} ) = \left[ \sum_{l_{k'}=1}^{m_{k'}} \sum_{l_{k}=1}^{m_{k}} p_{l_{k},l_{k'}}^{(k,k')}e^{i(\xi_{l_{k'}} + \eta_{l_{k}} )}  \right]^{N}.
\end{equation}
 Using the factorisation $ p_{l_{k},l_{k'}}^{(k,k')}  =  p_{l_{k'} | l_{k}}^{(k' | k)} p_{l_{k}}^{(k)}$ and the multinomial theorem, we obtain
\begin{equation}
\label{Eq:JointCharacQkQtilde{k}Development}
\begin{aligned}
    \varphi_{\vec{Q}^{(k')} , \vec{Q}^{(k)}}(\vec{\xi} , \vec{\eta} ) &= \left[ \sum_{l_{k}=1}^{m_{k}} p_{l_{k}}^{(k)} e^{i\eta_{l_{k}}} \sum_{l_{k'}=1}^{m_{k'}} p_{l_{k'} | l_{k}}^{(k' | k)} e^{i\xi_{l_{k'}}} \right]^{N}  \\
    &= \sum_{\substack{\vec{q} \in \lbrace 0 , ... , N \rbrace^{m_{k}} \\
    sum(\vec{q}) = N}} \frac{N!}{q_{1}! ... q_{m_{k}}!} e^{i \vec{\eta}^{T}\vec{q} } \left( p_{1}^{(k)} \right)^{q_{1}} ... \left( p_{l_{m_{k}}}^{(k)} \right)^{q_{m_{k}}} \prod_{l_{k}=1}^{m_{k}} \left[ \sum_{l_{k'}=1}^{m_{k'}}p_{l_{k'} | l_{k}}^{(k' | k)}  e^{i\xi_{l_{k'}}}   \right]^{q_{l_{k}}}.
\end{aligned}
\end{equation}
Comparing term-by-term expressions  \eqref{Eq:JointQkQtilde{k}UsingQkMultinomial} and \eqref{Eq:JointCharacQkQtilde{k}Development}, we conclude
\begin{equation}
\label{Eq:ConditionalCharFunctionQtildekQk}
    \varphi_{\vec{Q}^{(k')} | \vec{Q}^{(k)} = \vec{q} }(\vec{\xi} ) = \prod_{l_{k}=1}^{m_{k}} \left[ \sum_{l_{k'}=1}^{m_{k'}} p_{l_{k'} | l_{k}}^{(k' | k)}e^{i\xi_{l_{k'}}}    \right]^{q_{l_{k}}},
\end{equation}
which is the product of $m_{k}$ multinomial characteristic functions, each one with desired size and probabilities vector.
Note that the term-by-term comparison is justified by the linear independence of the functions $\eta \in \R^{m_{k}} \mapsto e^{i\eta^{T}\vec{q}}$ for $\vec{q} \in \lbrace 0 , ... , N \rbrace^{m_{k}}$.  $\blacksquare$

\subsection{Proof of Proposition \ref{Prop:MomentsECM}}
\label{Proof:MomentsECM}

Since $\vec{Q}^{(k)}$ is multinomial (Proposition \ref{Prop:OneTimeDistQ}), one has immediately $\mathbb{E}(Q_{l}^{(k)}) = Np_{l}^{(k)}$. Using the covariance structure of a multinomial vector, we obtain when $k=k'$
\begin{equation}
    \label{Eq:InProofSecondMomQ}
    \mathbb{E}\left( Q_{l}^{(k)} Q_{l'}^{(k)} \right) = \delta_{l,l'}Np_{l}^{(k)}  -Np_{l}^{(k)}p_{l'}^{k'}  + N^{2}p_{l}^{(k)}p_{l'}^{(k)} = \delta_{l,l'}Np_{l}^{(k)}  + N(N-1)p_{l}^{(k)}p_{l'}^{(k)}.
\end{equation}
For the case $k \neq k'$, we use Proposition \ref{Prop:TwoTimesDistQ} and the properties of conditional expectations:
\small
\begin{equation}
\label{Eq:SecondMomentECMDevelop}
    \begin{aligned}
        \mathbb{E}(Q_{l}^{(k)}Q_{l'}^{(k')}) &= \mathbb{E}\left( Q_{l}^{(k)}\mathbb{E}\left( Q_{l'}^{(k')} \ \mid \ \vec{Q}^{(k)} \right)   \right) \\
        &= \mathbb{E}\left(  Q_{l}^{(k)} \sum_{l''=1}^{m_{k}} Q_{l''}^{(k)} p_{l' \mid l''}^{(k' \mid k)}   \right) && \substack{\hbox{(expectation of sum of indep.} \\ \hbox{multinomials)} } \\
        &= \sum_{l''=1}^{m_{k}} \left( \delta_{l,l''} N p_{l}^{(k)} + N(N-1)p_{l}^{(k)} p_{l''}^{(k)} \right) p_{l' \mid l''}^{(k' \mid k )}  && \hbox{(expression \eqref{Eq:InProofSecondMomQ})} \\
        &= N p_{l}^{(k)}p_{l'|l}^{(k'|k)} + N(N-1) p_{l}^{(k)}\sum_{l''=1}^{m_{k}} p_{l',l''}^{(k' , k)} && \substack{\hbox{(split the sum and use conditional } \\ \hbox{probability products)} } \\
        &= N p_{l , l'}^{(k,k')} + N(N-1)p_{l}^{(k)} p_{l'}^{(k')}. && \substack{\hbox{(conditional probability products } \\ \hbox{and marginalization \eqref{Eq:ProbMargin2times} )}}
    \end{aligned}
\end{equation}
\normalsize
With the non-centered second moments being computed, the covariance \eqref{Eq:CovQ} follows. $\blacksquare$

\subsection{Proof of Proposition \ref{Prop:CharFunctionQPoisson}}
\label{Proof:CharFunctionQPoisson}

Using that  $\bm{Q} \mid N  \sim ECM$ and the properties of conditional expectation we have
\begin{equation}
    \begin{aligned}
        \varphi_{\bm{Q}}(\bm{\xi} ) &= \mathbb{E}\left( e^{i \bm{\xi}^{T}\bm{Q} }  \right) \\
        &= \mathbb{E}\left(   \mathbb{E}\left(  e^{i \bm{\xi}^{T}\bm{Q} }  \mid  N \right)  \right)  \\
        &= \mathbb{E}\left(  \left[   \sum_{l_{1}=1}^{m_{1}} \ldots \sum_{l_{n}=1}^{m_{n}}  e^{i(\xi_{l_{1}}^{(1)} + \ldots + \xi_{l_{n}}^{(n)})} p_{l_{1},\ldots , l_{n}}^{(1,\ldots , n)}  \right]^{N}  \right), 
    \end{aligned}
\end{equation}
where we have used the characteristic function of an ECM random arrangement \eqref{Eq:CharFunctionQ}. Using the probability generating function of a Poisson random variable $G_{N}(z) = \mathbb{E}(z^{N}) = e^{\lambda(z-1)}$, we have
\begin{equation}
    \varphi_{\bm{Q}}(\bm{\xi} ) = \exp\Big\lbrace \lambda \left(  \sum_{l_{1}=1}^{m_{1}} \ldots \sum_{l_{n}=1}^{m_{n}}  p_{l_{1},\ldots , l_{n}}^{(1,\ldots , n)}e^{i(\xi_{l_{1}}^{(1)} + \ldots + \xi_{l_{n}}^{(n)})}     \  - 1  \right)  \Big\rbrace. \quad \blacksquare
\end{equation}

\subsection{Proof of Proposition \ref{Prop:OneTimeDistQPoisson}}
\label{Proof:OneTimeDistQPoisson}

Take $\bm{\xi} = (\xi_{l}^{(k)})_{l,k}$ be such that $\xi_{l}^{(k')} = 0$ for every $k' \neq k$, and set $\vec{\xi} = (\xi_{1}^{(1)}, \ldots , \xi_{m_{k}}^{(k)})$. Then,
\begin{equation}
    \begin{aligned}
        \varphi_{\vec{Q}^{(k)}}(\vec{\xi}) &= \varphi_{\bm{Q}}(\bm{\xi}) \\
        &= \exp\Big\lbrace  \lambda\left(\sum_{l_{1}=1}^{m_{1}} \ldots\sum_{l_{n}=1}^{m_{n}}  p_{l_{1},\ldots , l_{n}}^{(1,\ldots , n)}e^{i\xi_{l_{k}}^{(k)}}    \  - 1   \right)   \Big\rbrace \\
        &= \exp\Big\lbrace  \lambda\left(\sum_{l_{k}=1}^{m_{k}}  p_{l_{k}}^{(k)} e^{i\xi_{l_{k}}^{(k)}}     \  - 1   \right)   \Big\rbrace,
    \end{aligned}
\end{equation}
where we have used the marginalization \eqref{Eq:ProbMargin1Time}. Considering that $\sum_{l=1}^{m_{k}}p_{l}^{(k)} = 1$, we obtain
\begin{equation}
    \varphi_{\vec{Q}^{(k)}}(\vec{\xi}) = \prod_{l=1}^{m_{k}}  \exp\Big\{ \lambda( e^{i \xi_{l}^{(k)}} - 1 )   \Big\},  
\end{equation}
which is a product of Poisson characteristic functions. $\blacksquare$

\subsection{Proof of Proposition \ref{Prop:TwoTimesDistQPoisson}}
\label{Proof:TwoTimesDistQPoisson}

Let $\vec{\xi} = (\xi_{1} , \ldots , \xi_{m_{k'}}) \in \R^{m_{k'}}$ and $\vec{\eta} = (\eta_{1} , \ldots , \eta_{\tilde{m}_{k}}) \in \R^{\tilde{m}_{k}}$. We consider an arrangement $\bm{\xi} = (\xi_{l}^{(k)})_{l \in \{ 1 , \ldots , m_{k}\},\ k \in \lbrace 1 , \ldots , n \rbrace }$ such that
\begin{equation}
    \xi_{l}^{(k'')} = \begin{cases}
        \xi_{l} & \hbox{ if } k'' = k' \\
        \eta_{l} & \hbox{ if } k'' = k \hbox{ and } l \leq \tilde{m}_{k} \\
        0 & \hbox{otherwise}.
    \end{cases}
\end{equation}
Then, the joint characteristic function of $( \vec{Q}^{(k')} , \vec{\tilde{Q}}^{(k)} )$ is given by
\begin{equation}
    \label{Eq:VarphiQkQk'Poisson}
    \begin{aligned}
        \varphi_{\vec{Q}^{(k')} , \vec{\tilde{Q}}^{(k)}}( \vec{\xi} , \vec{\eta} ) = \varphi_{\bm{Q}}(\bm{\xi} ) &=\exp\Big\lbrace \lambda\left( \sum_{l_{1}=1}^{m_{1}}\ldots \sum_{l_{n}=1}^{m_{n}}p_{l_{1},\ldots,l_{n}}^{(1,\ldots , n)}e^{ i( \xi_{l_{k'}}^{(k')} + \xi_{l_{k}}^{(k)} ) }  \ - 1  \right)   \Big\rbrace \\
        &= \exp\Big\lbrace \lambda\left( \sum_{l'=1}^{m_{k'}} \sum_{l=1}^{m_{k}} p_{l,l'}^{(k,k')}e^{ i( \xi_{l}^{(k)} + \xi_{l'}^{(k')}) }  \ - 1  \right)   \Big\rbrace \\
        &= \exp\Big\lbrace \lambda\left( \sum_{l'=1}^{m_{k'}} \sum_{l=1}^{\tilde{m}_{k}} p_{l,l'}^{(k,k')}e^{ i( \eta_{l} + \xi_{l'}  ) }  +  \sum_{l'=1}^{m_{k'}} p_{m_{k} , l'}^{(k,k')}e^{i\xi_{l'}}  \ - 1  \right)   \Big\rbrace \\
        &= \exp\Big\lbrace \lambda\left(   \sum_{l'=1}^{m_{k'}} p_{m_{k} , l'}^{(k,k')}e^{i\xi_{l'}}  \ - 1  \right)   \Big\rbrace  \prod_{l=1}^{\tilde{m}_{k}} \exp\Big\lbrace e^{i\eta_{l}} \lambda\sum_{l'=1}^{m_{k'}} p_{l,l'}^{(k,k')}e^{ i \xi_{l'}   }    \Big\rbrace. 
    \end{aligned}
\end{equation}
On the other hand, using the same arguments as in the proof \ref{Proof:TwoTimesDistQ} (Eq. \eqref{Eq:ConnectionCharacPairConditionalDist}), considering that $\vec{\tilde{Q}}^{(k)}$ has Poisson independent components, we have
\begin{equation}
       \varphi_{\vec{Q}^{(k')} , \vec{\tilde{Q}}^{(k)}}( \vec{\xi} , \vec{\eta} ) = \sum_{q_{1}=0}^{\infty}\ldots \sum_{q_{\tilde{m}_{k}} = 0}^{\infty} \varphi_{\vec{Q}^{(k')} \mid \vec{\tilde{Q}} = (q_{1},\ldots , q_{\tilde{m}_{k}})}(\vec{\xi}) e^{i \vec{\eta}^{T}\vec{q} } \prod_{l=1}^{\tilde{m}_{k}} e^{ -\lambda p_{l}^{(k)} }\frac{[ \lambda p_{l}^{(k)} ]^{q_{l}}}{q_{l}!}.
\end{equation}
Now, the product of exponentials in the last line in \eqref{Eq:VarphiQkQk'Poisson} can be expressed as
\begin{equation}
\begin{aligned}
     \prod_{l=1}^{\tilde{m}_{k}} \exp\Big\lbrace e^{i\eta_{l}} \lambda \sum_{l'=1}^{m_{k'}} p_{l,l'}^{(k,k')}e^{ i \xi_{l'}   }    \Big\rbrace  &= \prod_{l=1}^{\tilde{m}_{k}} 
    \sum_{q=0}^{\infty}\frac{ \left[  e^{i\eta_{l}} \lambda \sum_{l'=1}^{m_{k'}} p_{l,l'}^{(k,k')}e^{ i \xi_{l'}   }  \right]^{q}  }{q!}   \\
    &= \sum_{q_{1}=0}^{\infty} \ldots \sum_{q_{\tilde{m}_{k}} = 0}^{\infty} \prod_{l=1}^{\tilde{m}_{k}} \frac{ \left[  e^{i\eta_{l}} \lambda\sum_{l'=1}^{m_{k'}} p_{l,l'}^{(k,k')}e^{ i \xi_{l'}   }  \right]^{q_{l}}  }{q_{l}!} \\
    &= \sum_{q_{1}=0}^{\infty} \ldots \sum_{q_{\tilde{m}_{k}} = 0}^{\infty} e^{i\vec{\eta}^{T}\vec{q}}\prod_{l=1}^{\tilde{m}_{k}} \frac{ \left[ \lambda \sum_{l'=1}^{m_{k'}} p_{l,l'}^{(k,k')}e^{ i \xi_{l'}   }  \right]^{q_{l}}  }{q_{l}!}. 
\end{aligned}
\end{equation}
We have therefore the equalities
\small
\begin{equation}
    \begin{aligned}
        \varphi_{\vec{Q}^{(k')} , \vec{\tilde{Q}}^{(k)}}( \vec{\xi} , \vec{\eta} ) &= \sum_{q_{1}=0}^{\infty}\ldots \sum_{q_{\tilde{m}_{k}} = 0}^{\infty} \varphi_{\vec{Q}^{(k')} \mid \vec{\tilde{Q}} = (q_{1},\ldots , q_{\tilde{m}_{k}})}(\vec{\xi}) e^{i \vec{\eta}^{T}\vec{q} } \prod_{l=1}^{\tilde{m}_{k}} e^{ -\lambda p_{l}^{(k)} }\frac{[ \lambda p_{l}^{(k)} ]^{q_{l}}}{q_{l}!} \\
        &= \sum_{q_{1}=0}^{\infty} \ldots \sum_{q_{\tilde{m}_{k}} = 0}^{\infty} \exp\Big\lbrace \lambda\left(   \sum_{l'=1}^{m_{k'}} p_{m_{k} , l'}^{(k,k')}e^{i\xi_{l'}}  \ - 1  \right)   \Big\rbrace e^{i\vec{\eta}^{T}\vec{q}}\prod_{l=1}^{\tilde{m}_{k}} \frac{ \left[ \lambda\sum_{l'=1}^{m_{k'}} p_{l,l'}^{(k,k')}e^{ i \xi_{l'}   }  \right]^{q_{l}}  }{q_{l}!}.
    \end{aligned}
\end{equation}
\normalsize
Comparing term-by-term, we must have for every $\vec{q} \in \mathbb{N}^{\tilde{m}_{k}}$
\small
\begin{equation}
    \varphi_{\vec{Q}^{(k')} \mid \vec{\tilde{Q}} = \vec{q} }(\vec{\xi})  \prod_{l=1}^{\tilde{m}_{k}} e^{ -\lambda p_{l}^{(k)} }\frac{[ \lambda p_{l}^{(k)} ]^{q_{l}}}{q_{l}!} = \exp\Big\lbrace \lambda\left(   \sum_{l'=1}^{m_{k'}} p_{m_{k} , l'}^{(k,k')}e^{i\xi_{l'}}  \ - 1  \right)   \Big\rbrace \prod_{l=1}^{\tilde{m}_{k}} \frac{ \left[ \lambda\sum_{l'=1}^{m_{k'}} p_{l,l'}^{(k,k')}e^{ i \xi_{l'}   }  \right]^{q_{l}}  }{q_{l}!}.
\end{equation}
\normalsize
Isolating $ \varphi_{\vec{Q}^{(k')} \mid \vec{\tilde{Q}} = \vec{q}}(\vec{\xi})$ we obtain
\small
\begin{equation}
\label{Eq:CondVarphiQk'Qkalmost}
\begin{aligned}
        \varphi_{\vec{Q}^{(k')} \mid \vec{\tilde{Q}} = \vec{q}}(\vec{\xi}) &= \exp\Big\lbrace \lambda\left(   \sum_{l'=1}^{m_{k'}} p_{m_{k} , l'}^{(k,k')}e^{i\xi_{l'}}  \ - 1  \right)   \Big\rbrace \prod_{l=1}^{\tilde{m}_{k}} \frac{ \left[ \lambda\sum_{l'=1}^{m_{k'}} p_{l,l'}^{(k,k')}e^{ i \xi_{l'}   }  \right]^{q_{l}}  }{q_{l}!}e^{\lambda p_{l}^{(k)}} \frac{q_{l}!}{[\lambda p_{l}^{(k)}]^{q_{l}}}  \\
        &= \exp\Big\lbrace  \lambda\left(   \sum_{l'=1}^{m_{k'}} p_{m_{k} , l'}^{(k,k')}e^{i\xi_{l'}} \ - 1 +   \sum_{l=1}^{\tilde{m}_{k}}p_{l}^{(k)} \right) \Big\rbrace \prod_{l=1}^{\tilde{m}_{k}}  \left[ \sum_{l'=1}^{m_{k'}} p_{l' \mid l}^{(k'\mid k )}e^{ i \xi_{l'}   }  \right]^{q_{l}}. 
\end{aligned}
\end{equation}
\normalsize
Finally, we consider the sum-up-to-one condition on (all) the one-time probabilities at time $k$ and by marginalization, we have
\begin{equation}        \sum_{l=1}^{\tilde{m}_{k}} p_{l}^{(k)} = 1 - p_{m_{k}}^{(k)} = 1  - \sum_{l'=1}^{m_{k'}} p_{m_{k},l'}^{(k,k')}.
\end{equation}
Putting this in \eqref{Eq:CondVarphiQk'Qkalmost} we conclude
\begin{equation}
\label{Eq:CharQk'CondQk}
\begin{aligned}         \varphi_{\vec{Q}^{(k')} \mid \vec{\tilde{Q}} = \vec{q}}(\vec{\xi}) &= \exp\Big\lbrace  \lambda \sum_{l'=1}^{m_{k'}} p_{m_{k},l'}^{(k,k')}\left( e^{i\xi_{l'}} - 1 \right)  \Big\rbrace \prod_{l=1}^{\tilde{m}_{k}}  \left[ \sum_{l'=1}^{m_{k'}} p_{l' \mid l}^{(k'\mid k )}e^{ i \xi_{l'}   }  \right]^{q_{l}} \\
&= \Bigg\{ \prod_{l'=1}^{m_{k'}} e^{  \lambda p_{m_{k},l'}^{(k,k')}\left( e^{i\xi_{l'}} - 1 \right) }  \Bigg\}\prod_{l=1}^{\tilde{m}_{k}}  \left[ \sum_{l'=1}^{m_{k'}} p_{l' \mid l}^{(k'\mid k )}e^{ i \xi_{l'}   }  \right]^{q_{l}}, 
\end{aligned}
\end{equation}
which is indeed the product between a characteristic function of an independent-Poisson-components random vector and a characteristic function which is the product of $\tilde{m}_{k}$ multinomial characteristic functions, all of them with their desired parameters. $\blacksquare$

\subsection{Proof of Proposition \ref{Prop:MeanCovQPoisson}}
\label{Proof:MeanCovQPoisson}

The mean and the covariance at same time $k=k'$ come directly from Proposition \ref{Prop:OneTimeDistQPoisson}. For $k\neq k'$, the non-centered second order moment can be obtained by conditioning with respect to $N$ and use the ECM second moment formula \eqref{Eq:SecondMomentECMDevelop}:
\begin{equation}
\begin{aligned}
        \mathbb{E}\left( Q_{l}^{(k)}Q_{l'}^{(k')}\right) &= \mathbb{E}\left( \mathbb{E}\left( Q_{l}^{(k)}Q_{l'}^{(k')} \mid N \right)  \right) \\
        &= \mathbb{E}\left( Np_{l,l'}^{(k,k')}  + (N^2 - N)p_{l}^{(k)}p_{l'}^{(k')} \right) \\
        &= \lambda p_{l,l'}^{(k,k')} + (\lambda^{2} + \lambda - \lambda)p_{l}^{(k)}p_{l'}^{(k')} \\
        &= \lambda p_{l,l'}^{(k,k')} + \lambda^{2}p_{l}^{(k)}p_{l'}^{(k')},
\end{aligned}
\end{equation}
where we have used that $N \sim Poisson(\lambda)$. The covariance follows immediately. $\blacksquare$

\subsection{Proof of Proposition \ref{Prop:PairsECMBivariateBinomial}}
\label{Proof:PairsECMBivariateBinomial}

We give reminders on how to construct a bivariate binomial distribution \citep{kawamura1973structure}. Consider a random vector $(X,Y)$ with binary components (bivariate Bernoulli). Its distribution can be parametrized by three quantities: the joint success probability $p_{X,Y} = \mathbb{P}( X = 1 , Y = 1)$, and the marginal success probabilities $p_{X} = \mathbb{P}( X = 1 )$, $p_{Y} = \mathbb{P}(Y = 1)$. In such case we have
\begin{equation}
\begin{aligned}
        \mathbb{P}( X = 1 , Y = 1) = p_{X,Y} \quad &; \quad \mathbb{P}( X = 1 , Y = 0) = p_{X} - p_{X,Y} \\
        \mathbb{P}( X = 0 , Y = 1) = p_{Y} - p_{X,Y} \quad &; \quad \mathbb{P}( X = 0 , Y = 0) = 1 - p_{X} - p_{Y} + p_{X,Y}.
\end{aligned}
\end{equation}
The characteristic function of $(X,Y)$ is easily computed
\begin{equation}
    \varphi_{X,Y}(\xi,\eta) = e^{i(\xi+\eta)}p_{X,Y} + e^{i\xi}( p_{X} - p_{X,Y} ) + e^{i\eta}( 
p_{Y} - p_{X,Y} ) + 1 - p_{X} - p_{Y} + p_{X,Y}.
\end{equation}
Suppose now $(X,Y)$ is the sum of $N$ iid bivariate Bernoulli random vectors with joint success probability $p_{X,Y}$ and marginal success probabilities $p_{X}$, $p_{Y}$. Then, $(X,Y)$ is said to follow a bivariate binomial distribution. Its characteristic function is then given by
\begin{equation}
    \varphi_{X,Y}(\xi,\eta) = \left[ e^{i(\xi+\eta)}p_{X,Y} + e^{i\xi}( p_{X} - p_{X,Y} ) + e^{i\eta}( 
p_{Y} - p_{X,Y} ) + 1 - p_{X} - p_{Y} + p_{X,Y} \right]^{N}.
\end{equation}

Consider the components $(Q_{l'}^{(k')} , Q_{l}^{(k)} )$ of an ECM random arrangement, with $k \neq k'$. Its characteristic function $\varphi_{Q_{l'}^{(k')} , Q_{l}^{(k)}}$ can be obtained by evaluating the characteristic function $\varphi_{\vec{Q}^{(k')} , \vec{Q}^{(k)}}$ (Eq. \eqref{Eq:JointCharacQkQtilde{k}}) at $\vec{\xi}$ and $\vec{\eta}$ such that $\xi_{l'}=\xi$, $\eta_{l} = \eta$ and the rest of their components are null. We obtain
\small
\begin{equation}
\begin{aligned}
        \varphi_{Q_{l'}^{(k')} , Q_{l}^{(k)}}( \xi , \eta ) &= \left[ \sum_{l_{k'}=1}^{m_{k'}} \sum_{l_{k}=1}^{m_{k}} e^{i \xi \delta_{l',l_{k'}} } e^{i \eta \delta_{l,l_{k}} } p_{l_{k},l_{k'}}^{(k,k')} \right]^{N} \\
        &= \left[ e^{i(\xi+\eta)}p_{l,l'}^{(k,k')} + e^{i \eta} \sum_{ \substack{l_{k'}=1 \\l_{k'} \neq l' }}^{m_{k'}}  p_{l,l_{k'}}^{(k,k')} \ + \ e^{i\xi} \sum_{ \substack{ l_{k}=1 \\ l_{k} \neq l }}^{m_{k}} p_{l_{k},l'} ^{(k,k')} \ + \ \sum_{ \substack{l_{k},l_{k'} \\ l_{k} \neq l , l_{k'} \neq l' } } p_{l_{k},l_{k'}}^{(k,k')} \right]^{N} \\
        &= \left[ e^{i(\xi+\eta)}p_{l,l'}^{(k,k')} + e^{i \eta}( p_{l}^{(k)} - p_{l,l'}^{(k,k')}) +  e^{i\xi} ( p_{l'}^{(k')} - p_{l,l'}^{(k,k')} )   +  1 - p_{l}^{(k)} - p_{l'}^{(k')} + p_{l,l'}^{(k,k')} \right]^{N}, 
\end{aligned}
\end{equation}
\normalsize
which is indeed the characteristic function of a bivariate binomial random vector. The formula for the distribution of the pairs \citep{kawamura1973structure} is given in Eq. \eqref{Eq:BivBinomialProb}. We use the notations $a \vee b = \max\{ a , b \}$ and $a \wedge b = \min\{ a , b \}$.

\small
\begin{equation}
\label{Eq:BivBinomialProb}
        \mathbb{P}( Q_{l}^{(k)} = q , Q_{l'}^{(k')} = q'  ) = \sum_{j = 0 \vee (q+q'-N)}^{q \wedge q' } \begin{aligned}
            &\frac{N!}{( N - q - q' + j )!  (q-j)! (q'-j)! j! }\left[ p_{l,l'}^{(k,k')} \right]^{j}\left[ p_{l}^{(k)} - p_{l,l'}^{(k,k')} \right]^{q-j} \\
            &\left[ p_{l'}^{(k')} - p_{l,l'}^{(k,k')} \right]^{q'-j}\left[ 1 - p_{l}^{(k)} -p_{l'}^{(k')} + p_{l,l'}^{(k,k')} \right]^{N - q - q' + j}
\end{aligned}.\blacksquare 
\end{equation}
\normalsize

\subsection{Proof of Proposition \ref{Prop:ECMPoissonPairBivariatePoisson}}
\label{Proof:ECMPoissonPairBivariatePoisson}

We give reminders on how to construct a bivariate Poisson distribution \citep{kawamura1973structure}. Consider three independent Poisson random variables $U,V,W$ with rates $\lambda_{U}, \lambda_{V}, \lambda_{W}$. Construct a random vector $(X,Y)$ as
\begin{equation}
    \begin{aligned}
        &X = W + U \\
        &Y = W + V.
    \end{aligned}
\end{equation}
We say then that $(X,Y)$ follows a bivariate Poisson distribution. This distribution can be parametrized with three quantities, the joint rate $\lambda_{X,Y} = \lambda_{W}$, and the marginal rates $\lambda_{X} = \lambda_{W} + \lambda_{U}$ and $\lambda_{Y} = \lambda_{W} + \lambda_{V}$. Each component of $(X,Y)$ is Poisson and $\Cov(X,Y) = \lambda_{X,Y}$. The characteristic function can be obtained using the independent sum $(X,Y) = (W,W) + (U,V)$:
\small
\begin{equation}
\begin{aligned}
     \varphi_{X,Y}(\xi,\eta) &= \varphi_{W,W}(\xi,\eta)\varphi_{U,V}(\xi,\eta) \\
     &= \varphi_{W}(\xi+\eta) \varphi_{U}(\xi)\varphi_{V}(\eta) \\
     &= \exp\Big\{ \lambda_{X,Y} (e^{i(\xi+\eta)} - 1)+(\lambda_{X}-\lambda_{X,Y})(e^{i\xi} - 1)+(\lambda_{Y} - \lambda_{X,Y} )(e^{i\eta} - 1)  \Big\}.
\end{aligned}
\end{equation}
\normalsize

Now we proced analogously as in the Proof \ref{Proof:PairsECMBivariateBinomial} of Proposition \ref{Prop:PairsECMBivariateBinomial}. The components $(Q_{l'}^{(k')} , Q_{l}^{(k)} )$ of an ECM-Poisson random arrangement, with $k \neq k'$ have joint characteristic function (Eq. \eqref{Eq:VarphiQkQk'Poisson})  
\small
\begin{equation}
\begin{aligned}
        \varphi_{Q_{l'}^{(k')} , Q_{l}^{(k)}}( \xi , \eta ) &=  \exp\Big\{  \lambda \sum_{l_{k'}=1}^{m_{k'}} \sum_{l_{k}=1}^{m_{k}} \left( e^{i \xi \delta_{l',l_{k'}} } e^{i \eta \delta_{l,l_{k}} } - 1\right)p_{l_{k},l_{k'}}^{(k,k')}  \Big\} \\
        &= \exp\Big\{ (e^{i(\xi+\eta)} - 1)\lambda p_{l,l'}^{(k,k')} + (e^{i \eta} - 1)\lambda \sum_{ \substack{l_{k'}=1 \\l_{k'} \neq l' }}^{m_{k'}}  p_{l,l_{k'}}^{(k,k')}  +  (e^{i\xi} -1 )\lambda\sum_{ \substack{ l_{k}=1 \\ l_{k} \neq l }}^{m_{k}} p_{l_{k},l'} ^{(k,k')} \Big\} \\
        &= \exp\Big\{ (e^{i(\xi+\eta)}-1)\lambda p_{l,l'}^{(k,k')} + (e^{i \eta}-1)\lambda( p_{l}^{(k)} - p_{l,l'}^{(k,k')}) + \lambda (e^{i\xi} -1)( p_{l'}^{(k')} - p_{l,l'}^{(k,k')} )  \Big\}, 
\end{aligned}
\end{equation}
\normalsize
which is indeed the characteristic function of a bivariate Poisson random vector with joint rate $\lambda p_{l,l'}^{(k,k')}$ and marginal rates $\lambda p_{l'}^{(k')} $ and $\lambda p_{l}^{(k)} $. The expression for the distribution of the pairs \citep{kawamura1973structure} is given by
\small
\begin{equation}
\label{Eq:BivPoissonProb}
\mathbb{P}\left( Q_{l}^{(k)} = q , Q_{l'}^{(k')} = q' \right) = \sum_{j=0}^{q \wedge q'}    e^{-\lambda (p_{l}^{(k)} + p_{l'}^{(k')}- p_{l,l'}^{(k,k')} ) } \frac{ [ \lambda (p_{l}^{(k)} - p_{l,l'}^{(k,k')} ) ]^{q-j}  }{(q-j)!} \frac{ [ \lambda (p_{l'}^{(k')} - p_{l,l'}^{(k,k')} ) ]^{q'-j}  }{(q'-j)!} \frac{ [ \lambda p_{l,l'}^{(k,k')} ]^{j}}{j!}.\blacksquare
\end{equation}
\normalsize

\section{Further details on simulation studies}
\label{App:DetailsSimu}

The simulation studies were implemented in \texttt{R}, using the \texttt{L-BFGS-B} method in the \texttt{optim} function for optimization. 
The optimization domain for the movement parameters is $(\log(\tau) , \log(\sigma) , z_{1} , z_{2}) \in [-8,6]\times[-8,10]\times[-1,1]\times[-1,1]$. In each optimization, three different initial parameter settings (\texttt{par0}) were provided towards ensuring a good optimization in the given domains. We set $\tau_{0} = \tau_{theo}/2$, $z_{1,0} =  0$, $z_{2,0} = 0$, and we try three different initial values for the autocorrelation parameter $\sigma$, by doing  $\theta_{0} \in \{ \frac{1}{10}\frac{\theta_{theo}}{2} ,  \frac{\theta_{theo}}{2}  ,  10\frac{\theta_{theo}}{2} \}$, and then $\sigma_{0} = \tau_{0}\sqrt{2\theta_{0}}$ (see the $(\theta,\sigma)$-parametrization of OU processes explained in Section \ref{Sec:SimuStudyOU}). In the ECM-Poisson case, which is usually more complicated, we used four initial values for the autocorrelation parameter by adding an extra order of magnitude, $\theta_{0} \in \{ \frac{1}{10}\frac{\theta_{theo}}{2} ,  \frac{\theta_{theo}}{2}  ,  10\frac{\theta_{theo}}{2} , 100\frac{\theta_{theo}}{2} \}$. In such case, the domain for $\log(\lambda)$ is $[\log(\lambda_{theo}/10) ,\log(10\lambda_{theo})]$ (except for the MGLE $\lambda = 10^2$ case, see further), and the initial value is $\lambda_{0} = \lambda_{theo}/2$.

For each $N$ and $\lambda$, and each method MGLE and MCLE, $1045$ simulations-estimations were performed, in order to take out potential erratic scenarios and still aiming for a total of $1000$ samples. By ``erratic'' we mean the following. The hessians at the optimum values proposed by \texttt{optim} are computed using the \texttt{hessian} function from the library \texttt{numDeriv}. The scenarios where the minimum eigenvalue of the hessian is smaller than $10^{-12}$ are considered erratic in terms of optimization (minimization), and are taken out from the samples. It turns out that almost none erratic scenarios were present for most settings, the only big exception being the case of MGLE applied to the ECM-Poisson distribution when $\lambda=10^{2}$. In such case, more than $50\%$ of samples were erratic. This required special attention. The domain for $\log(\lambda)$ had to be reduced to $[ \log(\lambda_{theo}/5) , \log(6\lambda_{theo}) ]$, and $2090$ simulation-estimations were tried-out, yielding to only $800$ non-erratic scenarios. Those $800$ samples are the ones presented in Section \ref{Sec:SimuStudyOU} and in the correlograms showed in this Appendix. The erratic estimates often lied on the border of optimization domains. This erratic behavior is likely due to the crude approximation of the Gaussian distribution when replacing the ECM-Poisson for small $\lambda$. We strongly suggest not to use MGLE for such settings.

\begin{figure}[htbp]
  \centering
  \begin{subfigure}[b]{0.49\textwidth}
    \includegraphics[width=\textwidth , height=0.38\textheight]{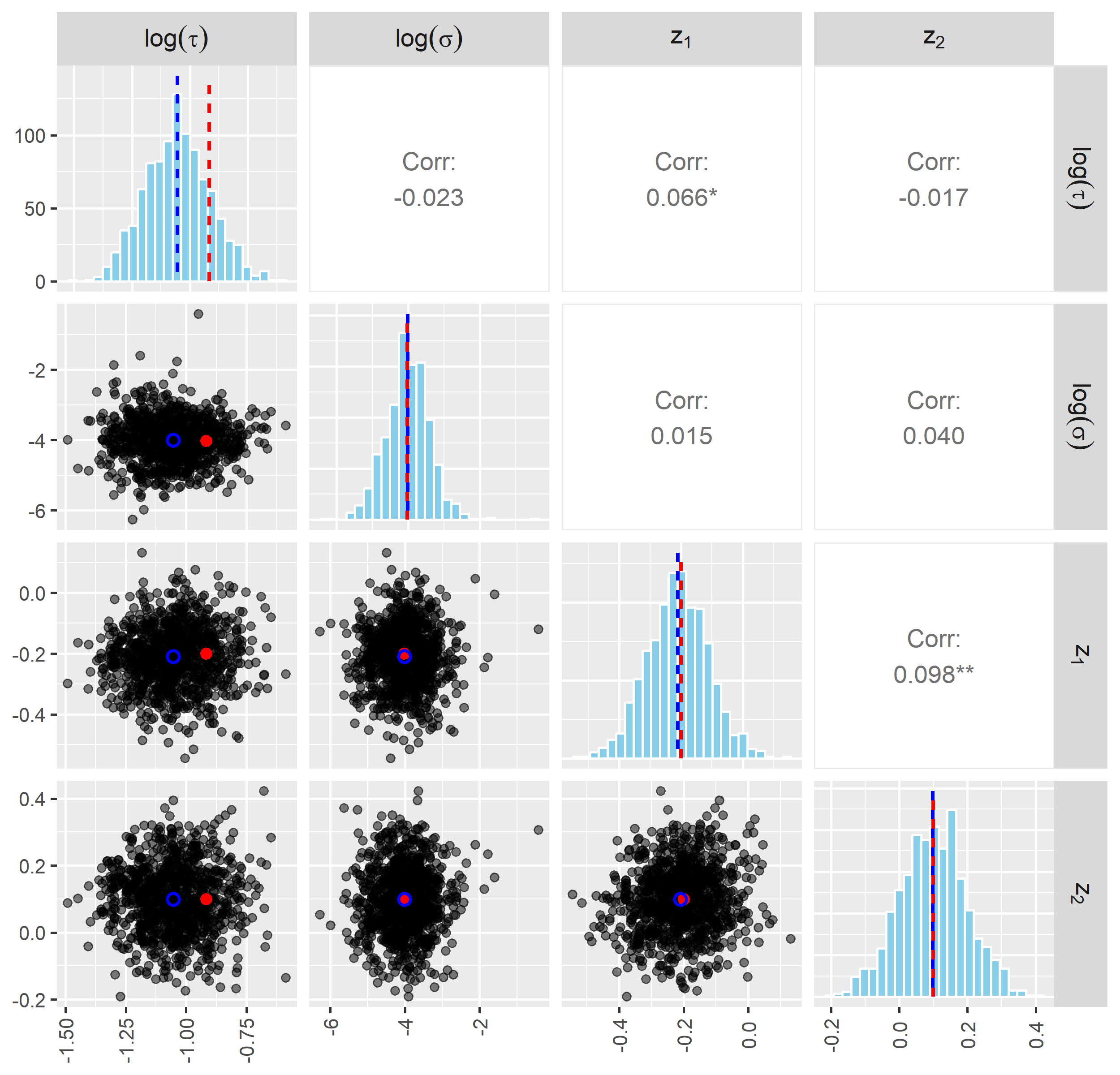}
    \caption{MGLE for ECM, $N = 10^2$}
  \end{subfigure}
  \hfill
  \begin{subfigure}[b]{0.49\textwidth}
    \includegraphics[width=\textwidth, height=0.38\textheight]{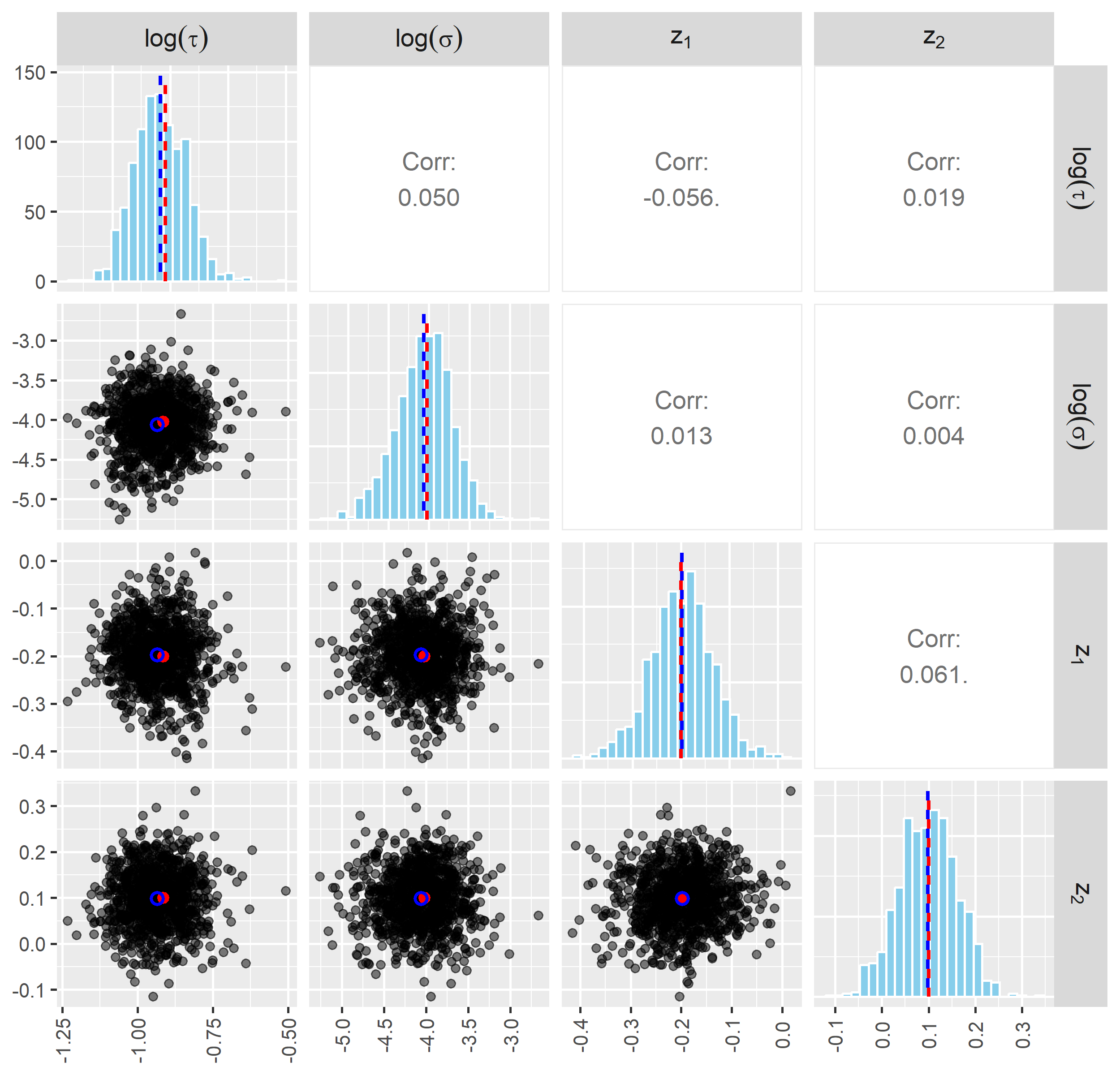}
    \caption{MCLE for ECM, $N = 10^2$}
  \end{subfigure}
  \hfill
  \begin{subfigure}[b]{0.49\textwidth}
    \includegraphics[width=\textwidth, height=0.38\textheight]{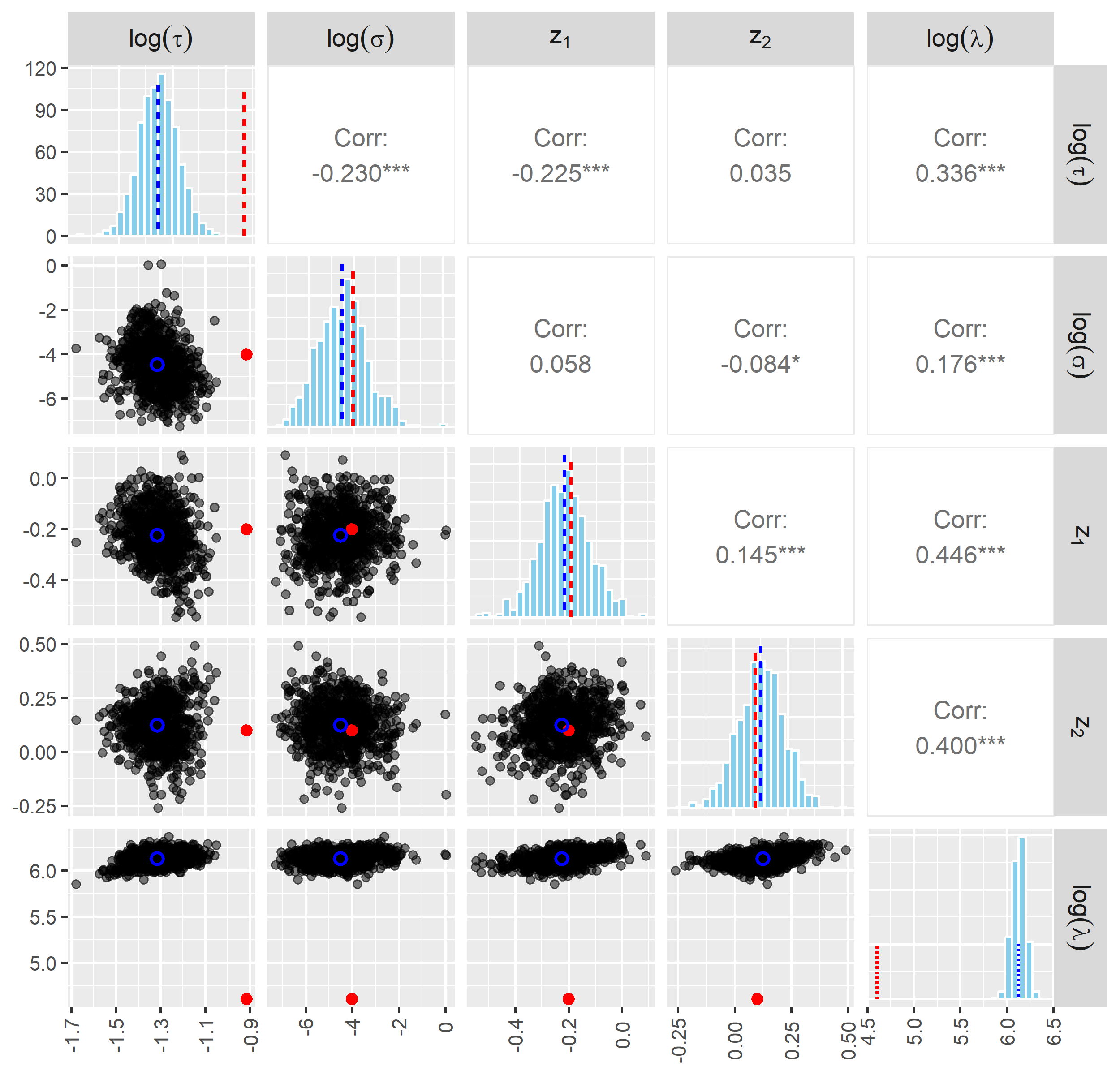}
    \caption{MGLE for ECM-Poisson*, $\lambda = 10^2$}
  \end{subfigure}
  \hfill
  \begin{subfigure}[b]{0.49\textwidth}
    \includegraphics[width=\textwidth, height=0.38\textheight]{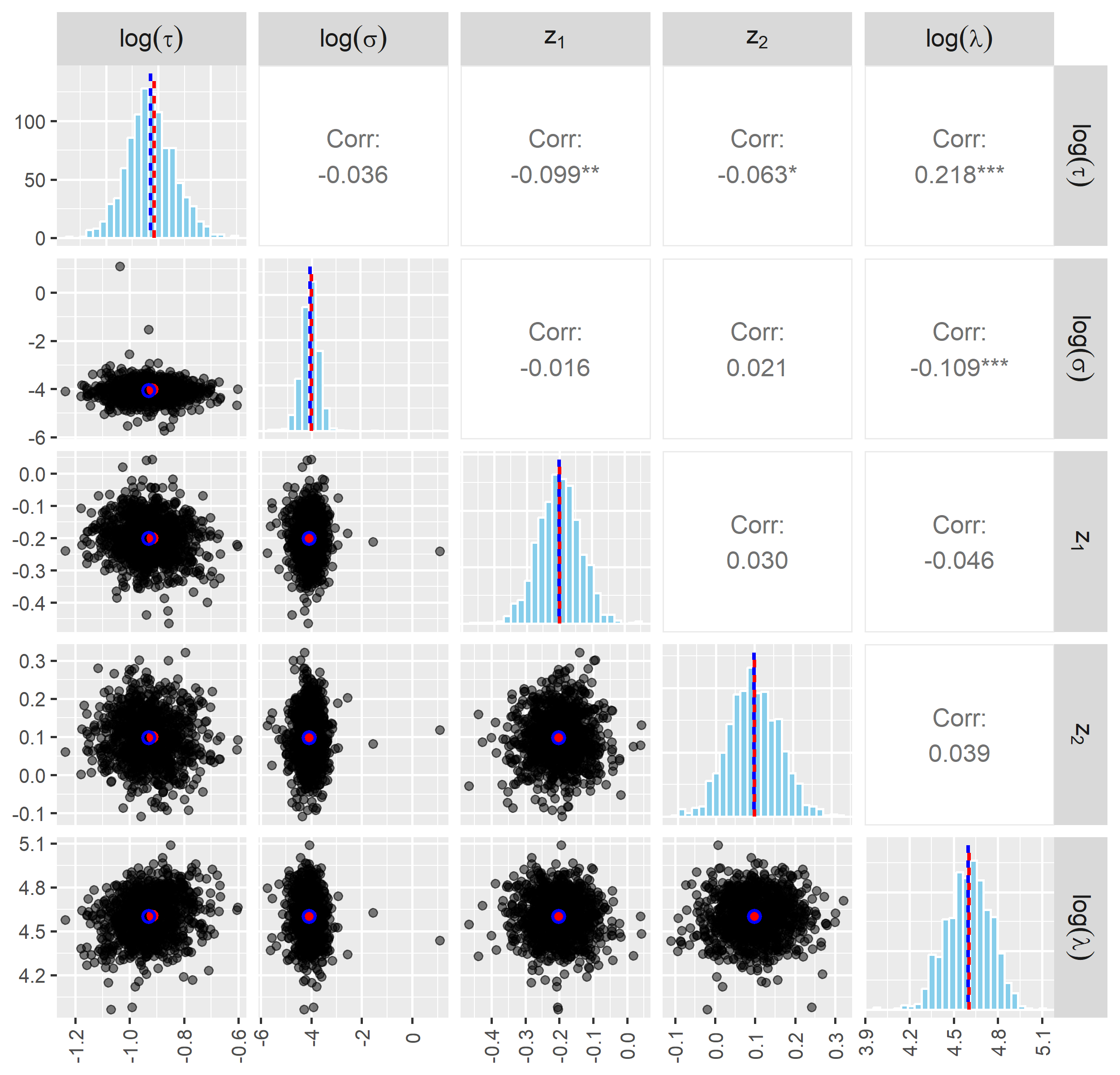}
    \caption{MCLE for ECM-Poisson, $\lambda = 10^2$}
  \end{subfigure}

  \caption{Correlograms of MGLE and MCLE estimates of steady-state OU parameters $(\tau,\sigma,z_{1},z_{2})$ in simulation studies. $N=10^2$ for ECM, $\lambda = 10^2$ for ECM-Poisson. In histograms, true parameter value marked with a dotted red line, sample mean marked with a dotted blue line. In point clouds, theoretical point value is indicated with a red point, the sample mean is indicated with an empty-interior blue circle. Sample size $1000$ in all settings, except MGLE ECM-Poisson, in which it is $800$ (see Appendix \ref{App:DetailsSimu}).}
  \label{Fig:Correlograms100}
\end{figure}

\begin{figure}[htbp]
  \centering
  \begin{subfigure}[b]{0.49\textwidth}
    \includegraphics[width=\textwidth , height=0.38\textheight]{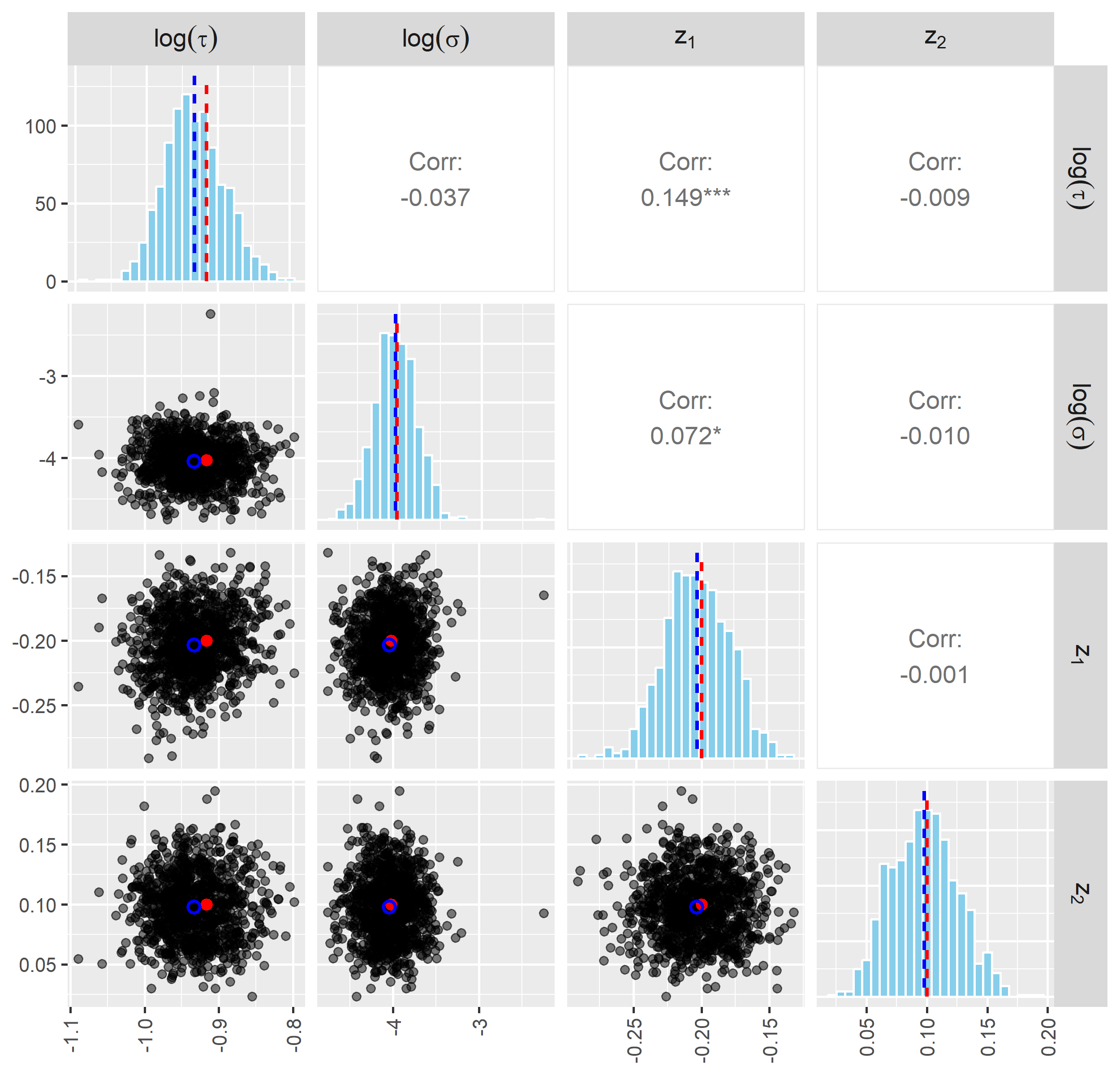}
    \caption{MGLE for ECM, $N = 10^3$}
  \end{subfigure}
  \hfill
  \begin{subfigure}[b]{0.49\textwidth}
    \includegraphics[width=\textwidth, height=0.38\textheight]{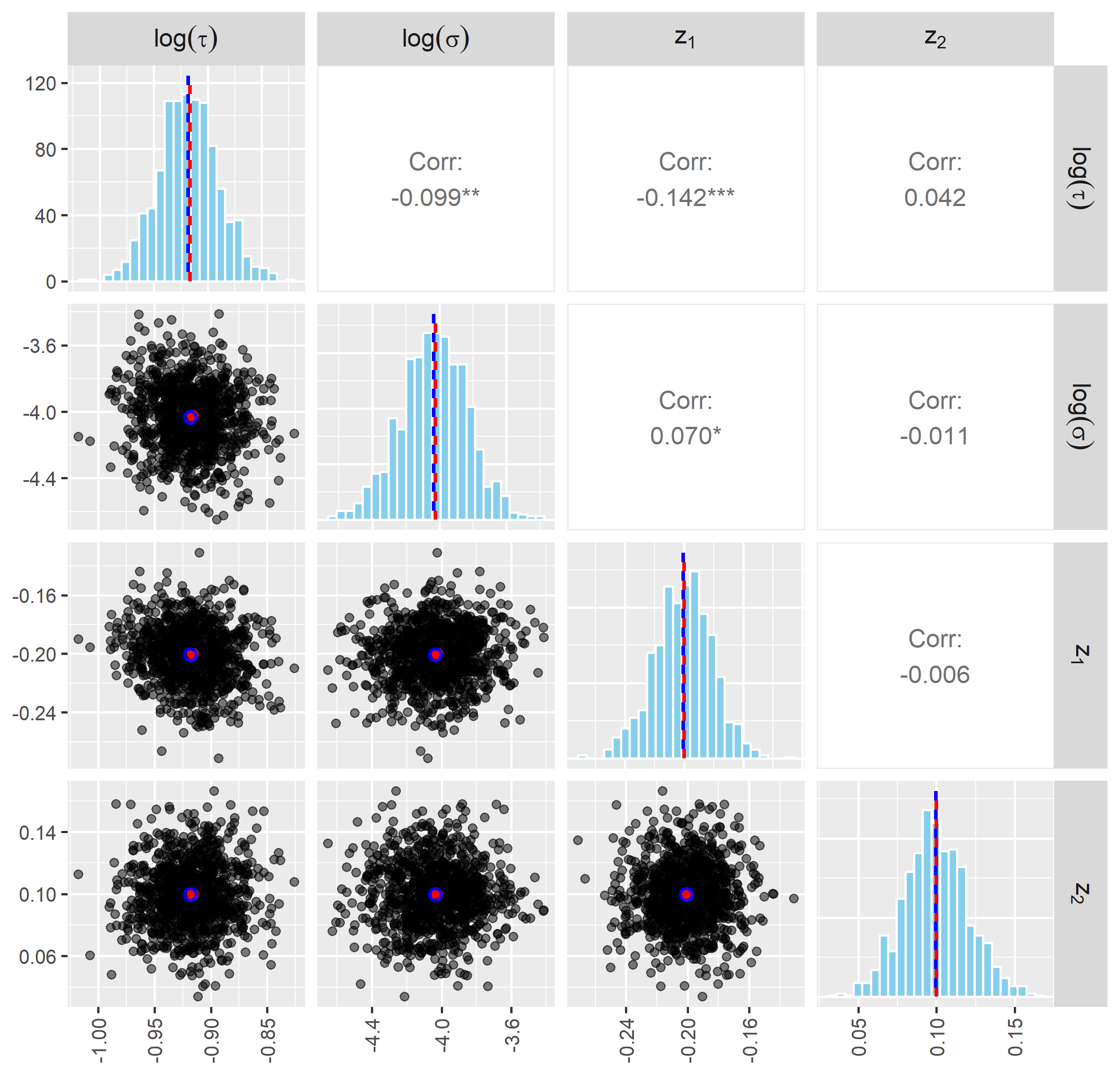}
    \caption{MCLE for ECM, $N = 10^3$}
  \end{subfigure}
  \hfill
  \begin{subfigure}[b]{0.49\textwidth}
    \includegraphics[width=\textwidth, height=0.38\textheight]{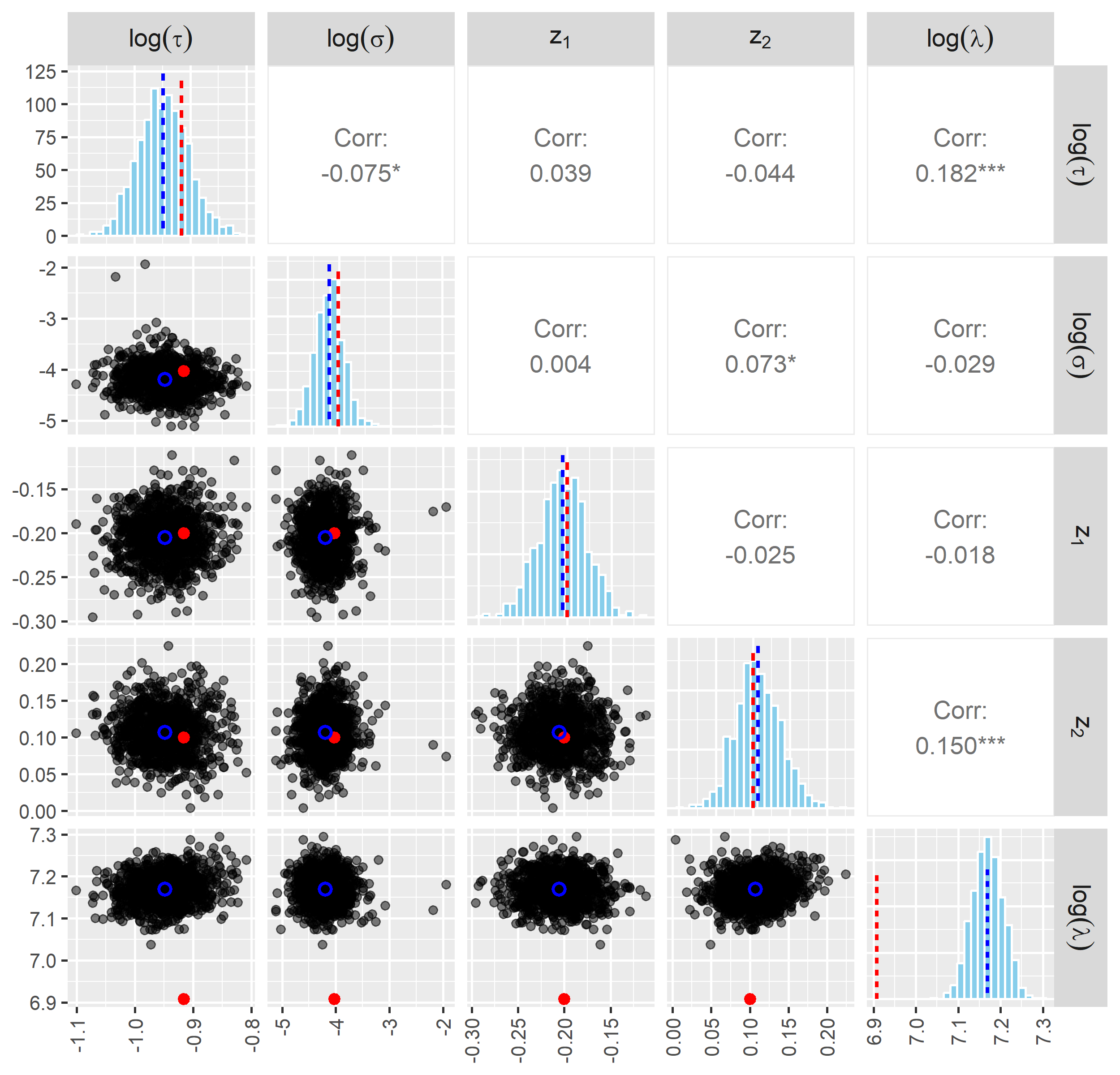}
    \caption{MGLE for ECM-Poisson, $\lambda = 10^3$}
  \end{subfigure}
  \hfill
  \begin{subfigure}[b]{0.49\textwidth}
    \includegraphics[width=\textwidth, height=0.38\textheight]{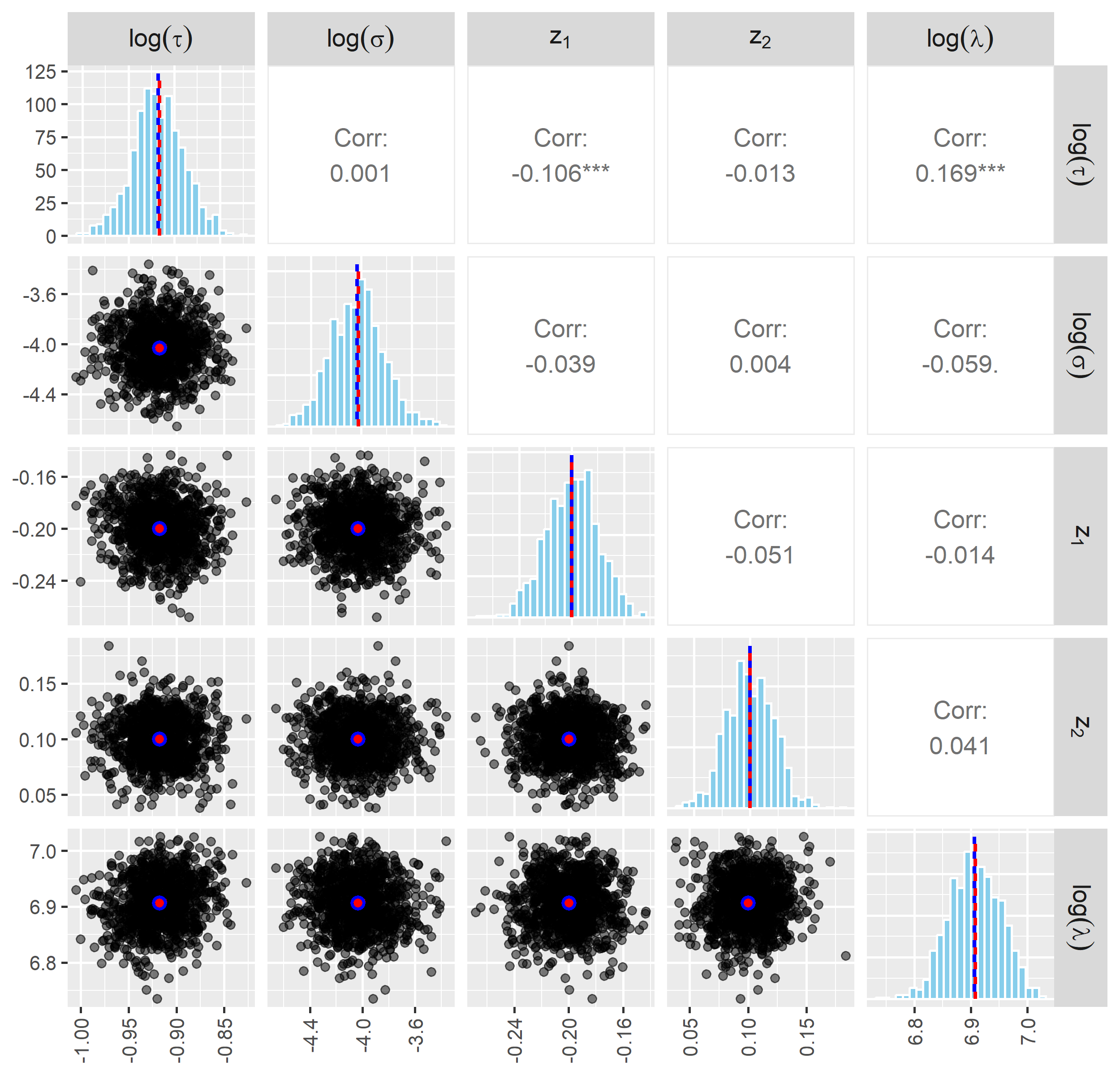}
    \caption{MCLE for ECM-Poisson, $\lambda = 10^3$}
  \end{subfigure}

  \caption{Correlograms of MGLE and MCLE estimates of steady-state OU parameters $(\tau,\sigma,z_{1},z_{2})$ in simulation studies. $N=10^3$ for ECM, $\lambda = 10^3$ for ECM-Poisson. In histograms, true parameter value marked with a dotted red line, sample mean marked with a dotted blue line. In point clouds, theoretical point value is indicated with a red point, the sample mean is indicated with an empty-interior blue circle. Sample size $1000$ in all settings.}
  \label{Fig:Correlograms1000}
\end{figure}

\begin{figure}[htbp]
  \centering
  \begin{subfigure}[b]{0.49\textwidth}
    \includegraphics[width=\textwidth , height=0.38\textheight]{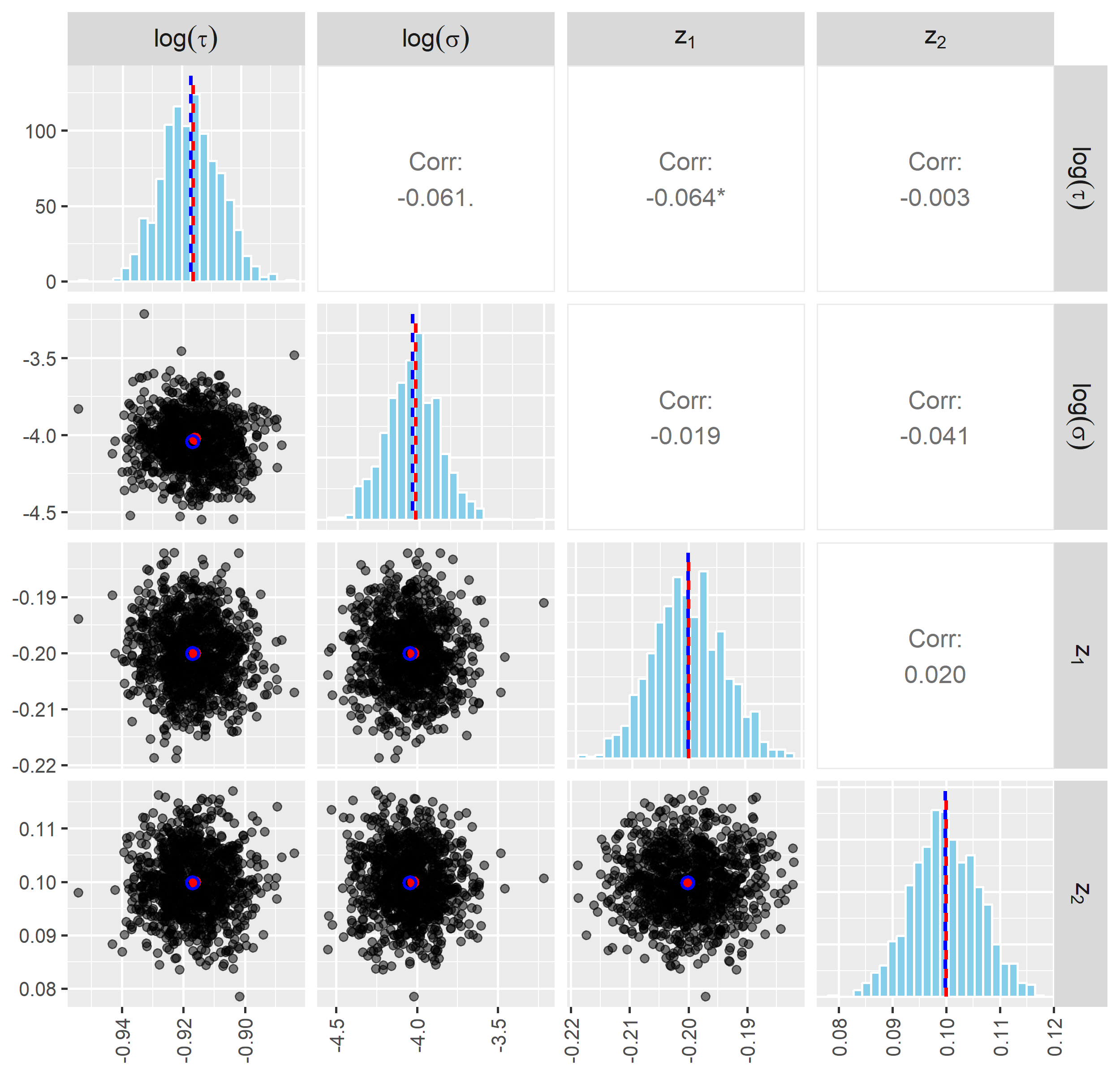}
    \caption{MGLE for ECM, $N = 10^4$}
  \end{subfigure}
  \hfill
  \begin{subfigure}[b]{0.49\textwidth}
    \includegraphics[width=\textwidth, height=0.38\textheight]{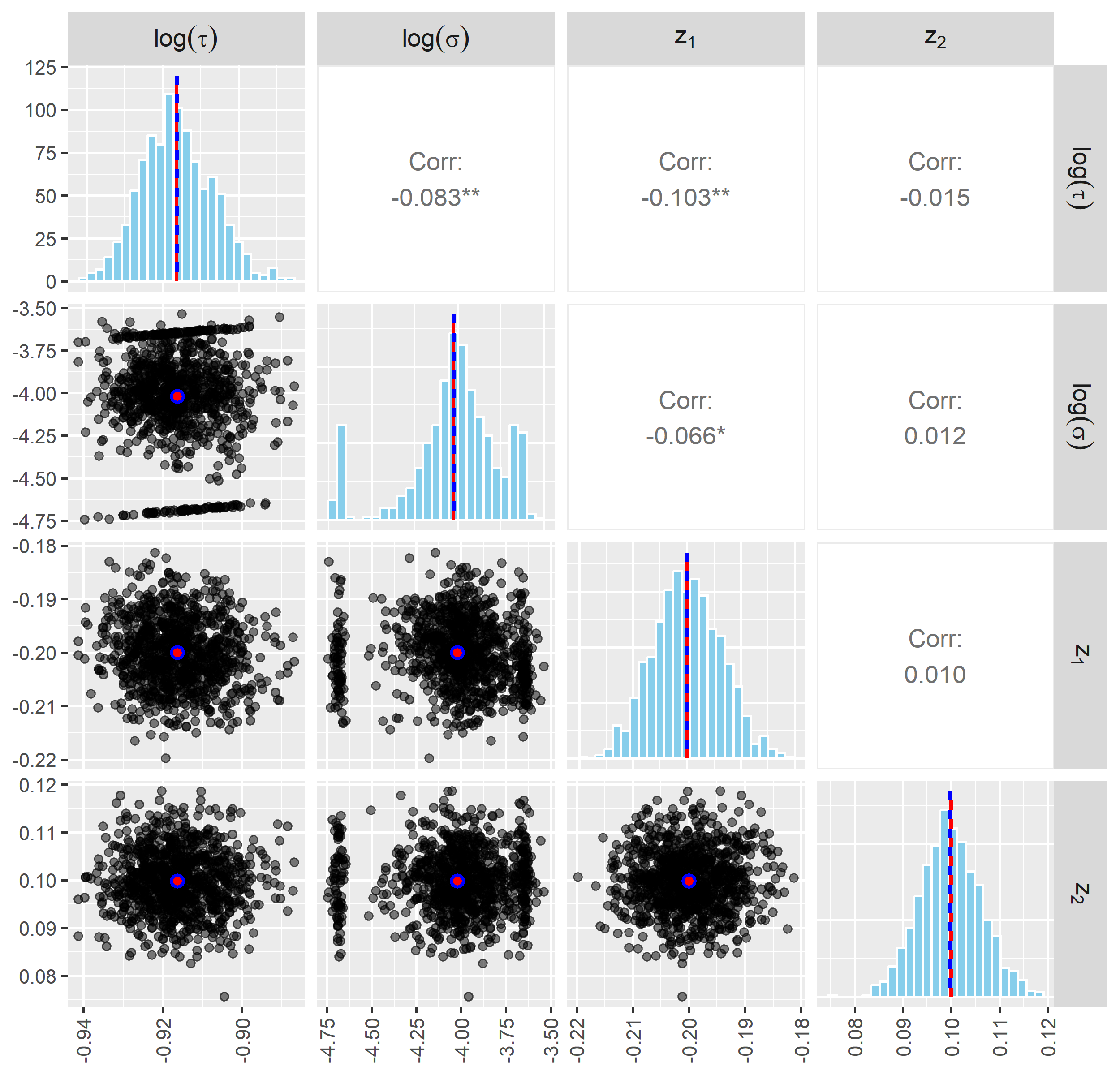}
    \caption{MCLE for ECM, $N = 10^4$}
  \end{subfigure}
  \hfill
  \begin{subfigure}[b]{0.49\textwidth}
    \includegraphics[width=\textwidth, height=0.38\textheight]{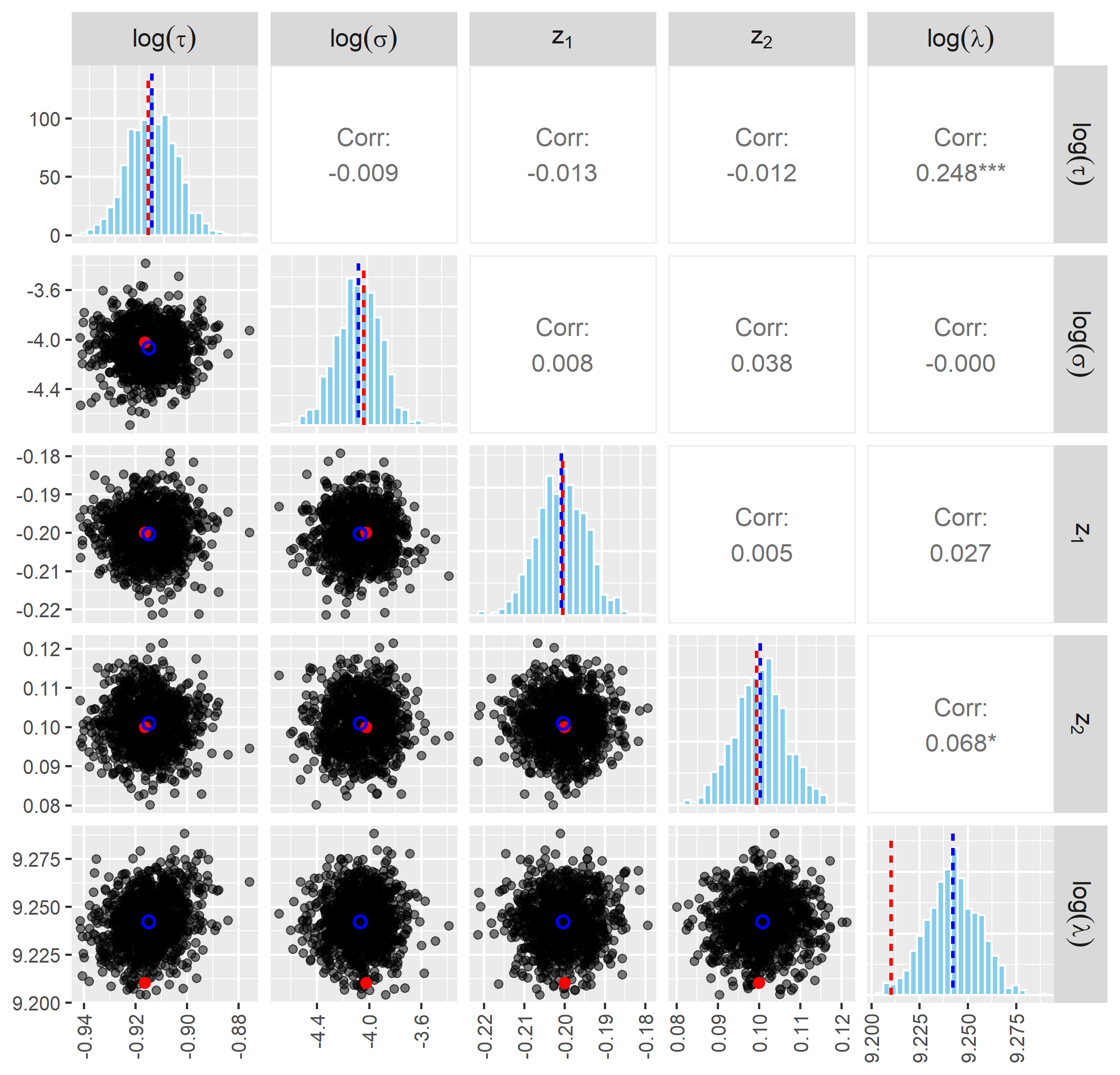}
    \caption{MGLE for ECM-Poisson, $\lambda = 10^4$}
  \end{subfigure}
  \hfill
  \begin{subfigure}[b]{0.49\textwidth}
    \includegraphics[width=\textwidth, height=0.38\textheight]{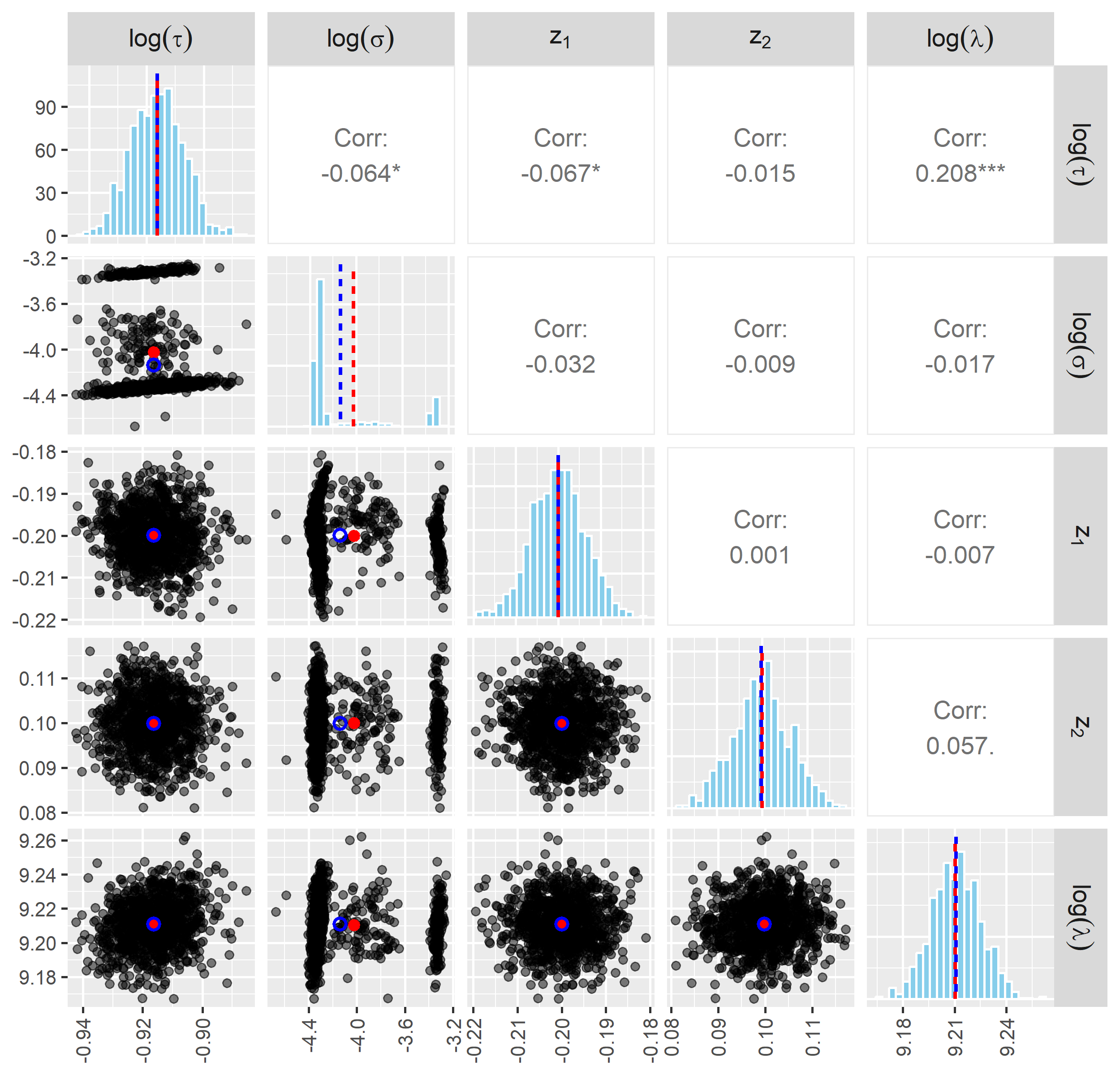}
    \caption{MCLE for ECM-Poisson, $\lambda = 10^4$}
  \end{subfigure}

  \caption{Correlograms of MGLE and MCLE estimates of steady-state OU parameters $(\tau,\sigma,z_{1},z_{2})$ in simulation studies. $N=10^4$ for ECM, $\lambda = 10^4$ for ECM-Poisson. In histograms, true parameter value marked with a dotted red line, sample mean marked with a dotted blue line. In point clouds, theoretical point value is indicated with a red point, the sample mean is indicated with an empty-interior blue circle. Sample size $1000$ in all settings.}
\label{Fig:Correlograms10000}
\end{figure}

The correlograms for all settings are presented in Figures \ref{Fig:Correlograms100}, \ref{Fig:Correlograms1000} and \ref{Fig:Correlograms10000}. A visual comparison of bias is accompanied in every case. There is an undesired complex scenario with MCLE for $N=10^{4}$ and $\lambda=10^{4}$, with multi-modal results. Contrarily to the MGLE ECM-Poisson $\lambda=10^{2}$ case, these are not erratic in the previously specified sense of optimization, but rather an indication of the multi-modality of the composite likelihood in this scenario. We remark, nonetheless, that the undesired lines formed around theoretical values in the point clouds are still inside a reasonable scale-range of the theoretical value. It is expected that for larger $\lambda$ or $N$, similar behavior will be present but inside a finer region around the theoretical value, respecting the asymptotic convergence.

\begin{figure}[ ]
    \centering  \hspace{-0.5cm} \includegraphics[scale = 0.28]{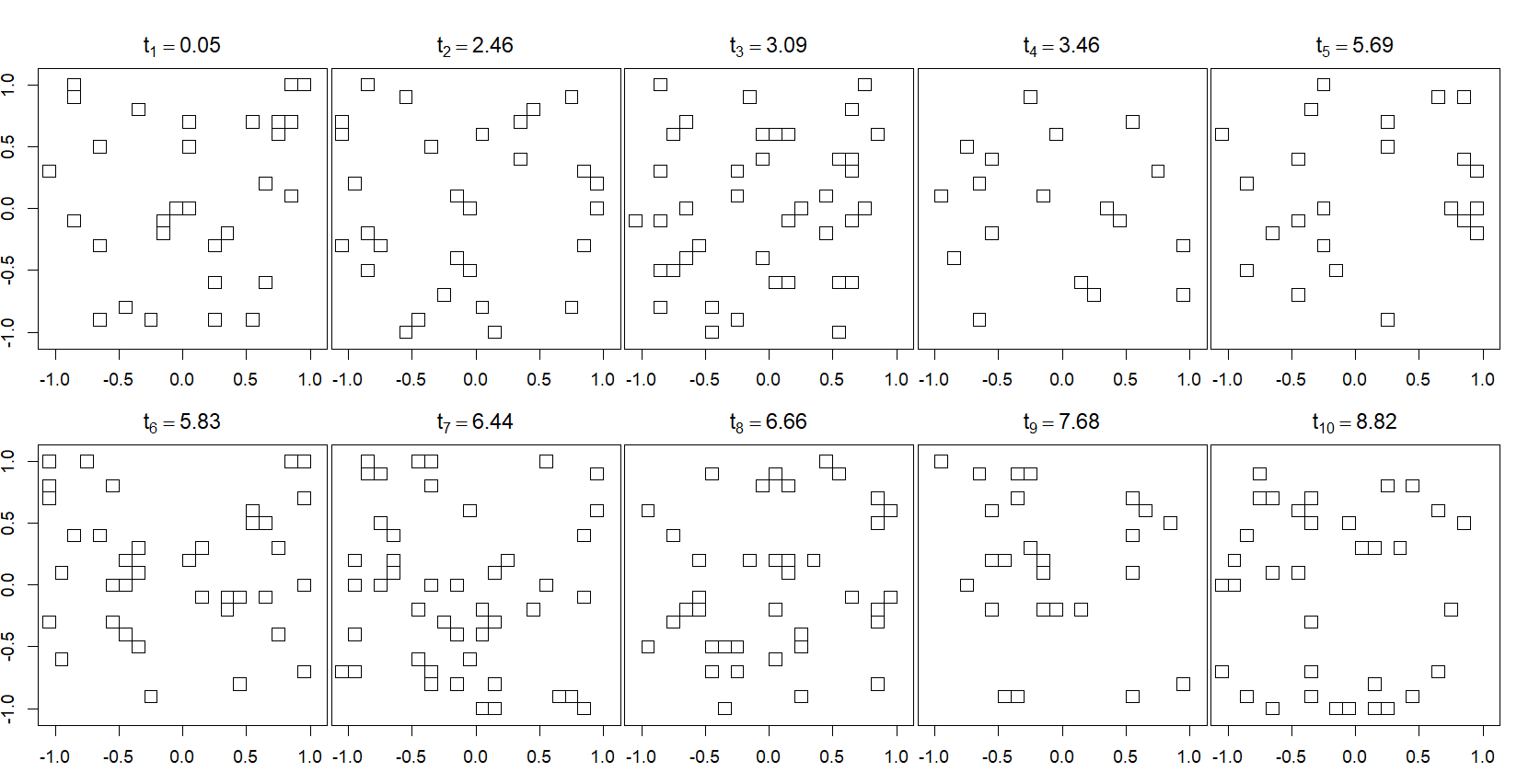}
    \caption{Survey squares and times used in simulation studies (Section \ref{Sec:SimuStudyOU}).}
\label{Fig:ToySetting2D}
\end{figure}

\end{appendices}

\newpage

\bibliography{mibib}
\bibliographystyle{apacite}

\end{document}